\documentclass[twoside,12pt,leqno]{article}
\advance\topmargin by -\headheight\advance\topmargin by -\headsep

\usepackage{amssymb,latexsym}
\usepackage{amsthm, amsmath, amsfonts}
\usepackage{hyperref}
\usepackage{empheq}
\usepackage{graphicx}
\usepackage{wrapfig}
\usepackage{slashed}
\usepackage{tikz} 
\usetikzlibrary{tikzmark}
\usepackage{tabu}
\usepackage{latexsym}
\usepackage{caption}
\usepackage{hyperref}

\def\blue{\color{blue}}

\def\magenta{\color{magenta}}

\hypersetup{
    colorlinks = true,
    colorlinks = false,
    linkcolor = blue,
    anchorcolor = red,
    citecolor = blue,
    filecolor = red,
    pagecolor = red,
    urlcolor = blue
}

\def\bz{\bar z}
\def\by{\bar y}

\def\bs{\bar s}

\def\p{\partial}
\def\bp{\bar\partial}

\def\b{\beta}

\def\bs{\bar s}

\def\ij{{\rm j}}

\def\bcal_k{\mathcal B_k}
\newcommand{\szego}{Szeg\"o\ }

\newcommand{\kahler}{K\"ahler }
\def\tr{{\rm Tr\,}}

\newcommand{\NPhi}{N_{\phi}}

\usepackage{amsmath}
\usepackage{amsthm}
\usepackage{amssymb}
\usepackage{amscd}
\usepackage{graphicx}
\usepackage{epsfig}
\usepackage{tikz} 
\usepackage{hyperref}

\numberwithin{equation}{section}
\newtheorem{theorem}{Theorem}[section]

\theoremstyle{definition}
\newtheorem{definition}[theorem]{Definition}

\numberwithin{equation}{section}

\allowdisplaybreaks
\begin{document}
\thispagestyle{empty}
\label{firstpage}




\markboth{Semyon Klevtsov }{Geometry and large N limits in Laughlin states}
$ $
\bigskip

\bigskip

\centerline{{\Large   Geometry and large N limits in Laughlin states}}

\bigskip
\bigskip
\centerline{{\large Semyon Klevtsov}}

\begin{center}
{\small\it
Institut f\"ur Theoretische Physik, Universit\"at zu K\"oln,\\ Z\"ulpicher Str. 77, 50937 K\"oln, Germany
}
\end{center}

\vspace*{.7cm}

\begin{abstract}
In these notes I survey geometric aspects of the lowest Landau level wave functions, integer quantum Hall state and Laughlin states on compact Riemann surfaces. In particular, I review geometric adiabatic transport on the moduli spaces, derivation of the electromagnetic and gravitational anomalies, Chern-Simons theory and adiabatic phase, and the relation to holomorphic line bundles, Quillen metric, regularized spectral determinants, bosonisation formulas on Riemann surfaces and asymptotic expansion of the Bergman kernel. 
 
Based on lectures given at the School on Geometry and Quantization, held at ICMAT, Madrid, September 7--11, 2015.
\end{abstract}
\newpage
\tableofcontents

\pagestyle{myheadings}
\section{Introduction}

Quantum Hall effect is observed in certain two-dimensional electron systems, such as GaAs heterostructures \cite{TSG1} and more recently in graphene \cite{Nov}, at large magnetic fields and low temperatures. In the most basic setup, the current $I_x$ is forced through a 2d sample in direction $x$ and the voltage $V_y$ is measured across the sample in $y$ direction, as shown on Fig.\ 1. The outcome of the measurement is that the Hall conductance $\sigma_H=I_x/V_y$ (Hall resistance shown on Fig.\ 1 is inverse Hall conductance, $R_{xy}=1/\sigma_H$) as a function of magnetic field strength at a fixed chemical potential undergoes a series of plateaux. There it takes on fractional values $\sigma_H\in\mathbb Q$, with small denominators, as measured in units of $e^2/h$. This effect is referred to as the "quantization" of Hall conductance. Even more remarkable, taking into account impurities of the samples, is the fact that quantization happens to a very high degree of accuracy, of the order $10^{-8}$. 

Quantum Hall effect comes in two varieties: integer QHE, where $\sigma_H\in\mathbb Z_+$, and fractional QHE with $\sigma_H$ rational and non-integer. On Fig.\ 1 the integer QHE plateaux are labelled as $1,2,3,4$ and all other plateaux corresponds the fractional QHE. The physics of integer and fractional QHE is very different: the former corresponds to non-interacting fermions, while the latter is a strongly interacting system (we refer to the classical survey Ref.\ \cite{Gir} for the introduction to the physics of QHE). However, there is a degree of similarity in that in both cases the mechanism behind the quantisation of Hall conductance alludes to Chern classes of certain vector bundles. 

\begin{figure}[t]
\begin{center}
\includegraphics[height=7cm]{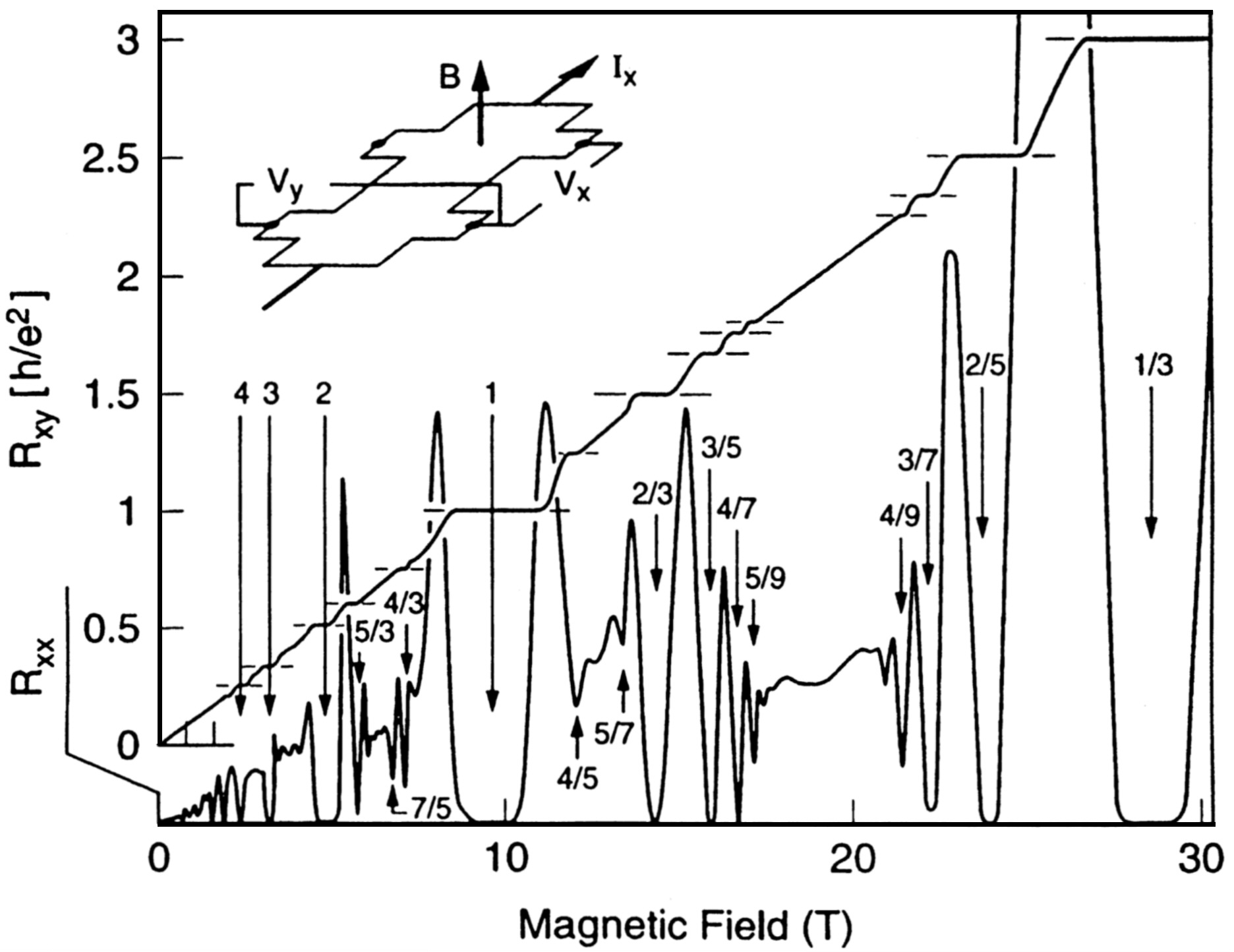}\\
{\small Figure 1.\, Hall resistance $R_{xy}$ and longitudinal resistance $R_{xx}$ vs. magnetic field $B$ (borrowed with permission from Ref.\ \cite{TSG1})}
\end{center}
\end{figure}

The standard approach to the theory of QHE is to assign a collective multi-particle electron wave function, or "state" $\Psi(z_1,...,z_N)$ to each plateaux, and then test its various properties against the experiment or numerics. In the integer QHE one could argue that the physics is captured by the first several energy levels in the tower of Landau levels of an electron in a strong magnetic field. The lowest Landau level (LLL) becomes highly degenerate at strong magnetic fields and is gapped due to the scale set by the cyclotron energy $\omega_c=eB/mc$. To satisfy the Pauli principle, the exact collective wave function of $N$ fermions on the fully filled LLL is completely antisymmetric and thus is given by the Slater determinant of one-particle wave functions, see Eq.\ \eqref{intplane}. One-particle wave functions are holomorphic functions on $\mathbb C$ and so is the integer QH state, apart from the overall non-holomorphic gaussian factor.

In the fractional QHE the many-body Hamiltonian contains interaction terms and thus it is hard to find an exact ground state analytically. Usually one proceeds by making educated guesses for the trial states. The most successful choice is the celebrated Laughlin state \cite{L}, corresponding to the values of Hall conductance given by simple fractions $\sigma_H=1/\beta,\,\beta\in Z_+$, see e.g.\ plateau labelled $1/3$ on Fig.\ 1. This wave function is not made out of one-particle states, although one could look at it as corresponding to a partially filled LLL (notation $\nu=1/\beta$ is also widely used, where $\nu$ is the "filling fraction" for the LLL). In particular, it is also a holomorphic function of coordinates times the gaussian factor, see Eq.\ \eqref{LS}. 

Another well-known state is Pfaffian state \cite{MR}, corresponding to $\sigma_H=5/2$ plateau, which is not pictured on Fig.\ 1. Other states were proposed to describe various other plateaux, as an incomplete list of best known examples we refer to the hierarchy states of Refs.\ \cite{H,Hal}, composite fermion states \cite{Jain}, Read-Rezayi states \cite{RR1999}, but our main focus here is on Laughlin states. Interestingly, the Laughlin states are also widely used in the contexts other than the FQHE, in particular we shall mention $d+id$-wave superconductors \cite{RG} and chiral spin liquids \cite{KL}.

The standard approach to the explanation of the quantization of Hall conductance, put forward in Ref. \cite{T1}, is inherently geometric (the first explanation of the quantization is Laughlin's transport of charge argument, which also invokes non-trivial geometry of annulus \cite{L1981}). There the integer QH effect was considered for a 2d electron gas in a periodic potential. Then the Hall conductance  was essentially described as the first Chern number of the line bundle over the Brillouin zone, which is a torus in the momentum space, see Ref.\ \cite{ASS} for this interpretation. We shall mention that this approach can be generalized to include impurities in the framework of Chern classes in non-commutative geometry \cite{BES}.

In these notes we follow a closely related, but not equivalent line of thought, known as the {\it geometric adiabatic transport}, see e.g.\ Ref.\ \cite{Av} for introduction. One considers a model 2d electron system either with periodic boundary conditions or on a compact Riemann surface $\Sigma$ \cite{AS,ASZ}. In the latter case the magnetic field is created by a configuration of $k$ magnetic monopoles inside the surface and is described by the line bundle $L^k$ of degree $k$ over $\Sigma$. When the Riemann surface has nontrivial topology, i.e., for genus $\rm g>0$, one can create magnetic field flux thought the holes of the surface using solenoids wrapped around the nontrivial one-cycles of $\Sigma$. On the surface this leads to the magnetic field acquiring flat gauge connection part, characterized by Aharonov-Bohm phases, e.g. in the case of torus, the phases are real numbers $\varphi_1,\varphi_2\in [0,1]^2$. The space of solenoid phases is itself a $2\rm g$-dimensional torus $T^{2\rm g}$, which is known to mathematicians as the moduli space of flat connections, or the Jacobian variety $Jac(\Sigma)$. The non-zero time derivative of phases $\dot\varphi$ gives rise to the electric current along the surface. 

The integer QH state is a section of the line bundle, called determinant line bundle, over $Jac(\Sigma)$. With the help of the Kubo's formulas Hall conductance can be expressed as the first Chern class of this line bundle, see e.g.\ \cite{AS,ASZ} for details. In particular, when the integer QH wave function is transported adiabatically along a smooth closed contour $\mathcal C\in Jac(\Sigma)$, it acquires an adiabatic Berry phase \cite{Berry}. For contractible contours it is equal to the integral of the Chern curvature of the canonical Berry connection, known in this context as adiabatic connection and curvature, over the area enclosed by the process. This argument can be extended to the Hall conductance in the fractional case \cite{TW,T2}, where the novel feature is topological degeneracy of the fractional QH states on the Riemann surfaces of genus $\rm g>0$. 

Studying QH states on a Riemann surface can also help uncover hidden properties of the QH states, such as e.g. the Hall viscosity. Apart from the moduli space of flat connections, complex structure moduli of Riemann surfaces $\mathcal M_{\rm g}$ provide another parameter space as a new arena for the geometric adiabatic transport. Avron-Seiler-Zograf  \cite{ASZ1} (see also \cite{L1}) considered the integer QH state as a section of the line bundle over the moduli space of complex structures of the torus $\mathcal M_1$. The latter is the fundamental domain of the action of $PSL(2,\mathbb Z)$ on the complex upper half plane, pictured on Fig.\ 4. They computed the curvature of the adiabatic connection on this line bundle over $\mathcal M_1$, which is proportional to the Poincar\'e metric on the upper half plane. The coefficient of proportionality was interpreted as the non-dissipative (anomalous, Hall) viscosity of the quantum Hall "electronic liquid", we refer to \cite{Ho} for the recent review. 

Geometric adiabatic transport was subsequently generalized to integer QH state on higher genus Riemann surfaces \cite{L2}. The Hall viscosity for various fractional QH states was derived Refs.\ \cite{TV1,TV2,R,RR,FHS}, see especially Ref.\ \cite{R} for the comprehensive study. In general, the coefficients entering the adiabatic curvature are called {\it adiabatic transport coefficients}. In this sense the Hall conductance $\sigma_H$ is a transport coefficient for the adiabatic transport on the moduli space of flat line bundles. To sum up, mathematically one would like to construct various QHE states a Riemann surface and describe the vector bundles arising on the moduli space of Riemann surface, for a large number of particles.

QH states can be also constructed on surfaces with curved metric and inhomogeneous magnetic field. This allows to determine the effect of gravitational anomaly in QHE starting directly from the wave functions. In Ref.\ \cite{K} the integer QH state was defined on a compact Riemann surface of any genus with an arbitrary metric $g_{z\bz}$ and for the constant magnetic field with the integer flux $k$ through the surface. The main task is to compute the asymptotics for large number of particles $N$ of the normalization factor of the integer QH state on curved backgrounds, which also has the meaning of the generating functional for the density-density correlation functions. The tool for the derivation of the asymptotics is the Bergman kernel expansion for large powers of the holomorphic line bundle $L^k$. Since the number of particles $N$ is of order $k$, this asymptotics is equivalent to large $N$ limit in the number of electrons. The gravitational anomaly appears as the order $\mathcal O(1)$ term in the $1/k$ expansion of the generating functional \cite{K}. This calculation was generalized to the case of inhomogeneous magnetic field in \cite{KMMW}. 

The advantage of the Bergman kernel method is that it is mathematically rigorous, and that its large $k$ expansion is very well understood, see \cite{Z,C,MM,Xu}. However, this method does not appear to be generalizable to the fractional QH case. There exists several physics methods to compute the asymptotics of the generating functional in this case. In Refs.\ \cite{CLW,CLW1,LCW} the generating functional and the gravitational anomaly for the Laughlin states was derived using the Ward identity method, developed in the important Ref.\ \cite{ZW}. In Ref.\ \cite{FK} an alternative derivation was given, based on the vertex operator construction of Laughlin states \cite{MR} adapted to curved backgrounds. This derivation, which we review here relies on path-integral arguments and does not directly refers to the standard "plasma screening" argument for large $N$ scaling limit in QH states \cite{L}, see also \cite{GN,BGN}. 

Asymptotic expansion of the generating functional for QH states consists of two parts: the anomalous part, where non-local functionals of the metric and magnetic field enter, and exact part, which includes an infinite series of local invariants of scalar curvature and its derivatives. 
The anomalous part consists of three terms corresponding to electromagnetic, mixed and gravitational anomalies, with three independent coefficients, see e.g.\ Eq.\ \eqref{logzh}. The coefficient in front of the electromagnetic anomaly is the Hall conductance $\sigma_H=1/\beta$ and the coefficient $\varsigma_H$ in front of the mixed gravitational-electromagnetic anomaly is related to Hall viscosity (for the Laughlin states on torus the Hall viscosity is $\eta_H=\NPhi/4$). Gravitational anomaly gives rise to a new adiabatic transport coefficient which was dubbed $c_H$, for "Hall central charge" in \cite{KW}, and "apparent central charge" in Ref.\ \cite{BR1}. For Laughlin states the Hall central charge is $c_H=1-3q^2$, where $q$ is background charge, see Eq.\ \eqref{q}. We shall stress that the theory is not conformally invariant, since there is a scale associated with the magnetic field $l^2_B=1/B$, and $c_H$ is one of infinitely many coefficients appearing in the asymptotic expansion (in $l_B$) of the generating functional. For a closely related point of view on the gravitational anomaly in QH states we refer to Ref.\ \cite{BR1}.

In parallel to these developments the gravitational anomaly in QHE was derived from the $2+1$d picture where $2$ stand for the space and $1$ for the time dimensions. Description of the $2+1$d long distance effective action in QHE in terms of Chern-Simons action goes back to Refs.\ \cite{FKer,FS,WZ}. There the non-relativistic Chern-Simons theory was constructed using $2+1$d gauge $A$ and spin $\omega$ connections and has schematic form $\int AdA+Ad\omega$, where the first term represents the electromagnetic anomaly and the second term  corresponds to the mixed anomaly. Recently the gravitational anomaly contribution  $\int \omega d\omega$ to this effective was computed starting from $2+1$d non-relativistic fermions in the integer QHE \cite{AG1}, see \cite{AG4} for the fractional case, and also Refs.\ \cite{Son,AG2,AG3,BR,CR}. These results are in complete agreement with the $2$d generating functional Eq.\ \eqref{logzh}, although the exact meaning of the matching between the terms in 2d and $2+1$d actions is not immediately clear. It was understood in Ref.\ \cite{KMMW} that the geometric part of the adiabatic phase for the integer QH state can be expressed in Chern-Simons form equivalent to that of Ref.\ \cite{AG1}, thus connecting 2d and $2+1$d approaches to QHE effective action in the integer QHE case. The derivation of the Chern-Simons action as adiabatic phase is based on Quillen theory \cite{Q} and on the Bismut-Gillet-Soul\'e formula \cite{BGS3} for the curvature of the Quillen metric for the holomorphic section of the determinant line bundle, i.e. integer QH state (Bismut-Gillet-Soul\'e formula was also invoked in QHE context in Ref.\ \cite{TP06}).

In these notes we review these developments and make an attempt to put them into a broader mathematical physics context. In particular, in the QHE context we cover such topics as holomorphic line bundles and $\bp$-operator, asymptotic expansion of the Bergman kernel, regularized spectral determinants and the Quillen metric, bosonisation formulas on Riemann surfaces, Bismut-Gillet-Soul\'e anomaly formula and Chern-Simons action. In this regard the lectures are aimed at both theoretical physicists working on QHE and mathematicians interested in learning about the subject.  

We begin in Section 2 with the one-particle states on Riemann surfaces, where we introduce the holomorphic line bundle, $\bp$-operator and describe the one-particle wave functions on the lowest Landau level as holomorphic sections of the magnetic monopole line bundle. Our main working example is the torus, where we construct the wave functions and study geometric adiabatic transport on the moduli space in detail. In section 3 we introduce the integer QHE wave function and the generating functional on Riemann surfaces with arbitrary magnetic field and Riemannian metric. We then use Bergman kernel expansion to compute the generating functional to all orders in $1/k$. We relate the non-local terms in the expansion to the gauge, mixed and gravitational anomalies and the local part of the expansion to the regularized determinant of the spectral laplacian for the line bundle. 

In Section 4 we define the Laughlin states, review their construction on the round sphere and flat torus.  We then review in detail the vertex operator construction of the Laughlin states and explain how it reduces to the bosonisation formulas on higher genus Riemann surfaces in the integer QHE case. 
We work out vertex operator construction on the torus, paying particular attention to the role of the spin structures and modular transformations. In Section 5 we proceed to the definition of the generating functional for the Laughlin states and review its asymptotic expansion for large number of  particles. Then we study geometric adiabatic transport and derive adiabatic curvature for the integer QH and for the Laughlin states. Finally we review the relation between the geometric part of the adiabatic phase and Chern-Simons action.

These lectures reflect the point of view on the geometry of Laughlin states taken in Refs. \cite{K,DK,FK,KW,KMMW}. In addition to the work already mentioned here, a number of exciting recent developments in geometry of QHE appeared in recent years, which regrettably are not surveyed here. These include emergent geometry approach to Laughlin states \cite{H1,H2}, Newton-Cartan approach to the geometry of QHE \cite{Son}, recently experimentally realized QH states on singular surfaces  \cite{JSimon} and emergent conformal symmetry \cite{CCLW}, genons \cite{BJQ,Gr}, Dehn twist process on the torus \cite{KVW,ZMP}, quantum Hall effect on \kahler manifolds \cite{KN,DK,B2,KN1,Zh}, to mention just a few. We also plan to provide a more detailed account of higher genus Laughlin states in a separate publication \cite{K2016}. Apart form that, we do not discuss a vast topic of quasi-hole excitations above the Laughlin states and their non-abelian statistics, although this is something which can be treated by methods reviewed here. We hope to address some of the aforementioned topics elsewhere. 

{\bf Acknowledgements.} I would like to thank the organizers of the summer School on Geometry and Quantization for the invitation to give this course and for their excellent job organizing the School and to participants for their interest and valuable feedback. I am indebted to my collaborators on the topics reviewed here: M. Douglas, F. Ferrari, X. Ma, G. Marinescu, P. Wiegmann and S. Zelditch. I have greatly benefitted from conversations with A.~Abanov, J.-M.~Bismut, T.~Can, J.~Dubail, B.~Estienne, A.~Gromov, B.~Hanin, V.~Pasquier, P~Zograf and I am especially grateful to P. Wiegmann for numerous enlightening discussions over recent years. I would like to thank A.~Gromov and B.~Hanin for reading the draft and suggesting improvements. 

This work was partially supported by the German Excellence Initiative at the University of Cologne,  DFG-grant ZI 513/2-1, grant NSh-1500.2014.2 and grant RFBR 14-01-00547. I would like to acknowledge support from the Simons Center for Geometry and Physics, Stony Brook University, where these notes were finalized.

\section{Lowest Landau level on Riemann surface}

\subsection{Background}
\label{background}

We consider a compact connected Riemann surface $\Sigma$ of genus $\rm g$, a positive Hermitian holomorphic line bundle $(L,h)$ of degree $\deg L=1$ and the $k$th tensor power $(L^k,h^k)$, where $h(z,\bz)$ is a Hermitian metric. Given some complex structure $J$ there exist local complex coordinates $z,\bz$ where the Riemannian metric on $\Sigma$ is diagonal $ds^2=2g_{z\bz}|dz|^2$, and the area of $\Sigma$ is normalized as $\int_\Sigma\sqrt g d^2z=2\pi$. The curvature two-form of the Hermitian metric $h^k(z,\bz)$,
\begin{align}\label{F}
F:=F_{z\bz}idz\wedge d\bz=-(\p_z\p_{\bz}\log h^k)  i dz\wedge d\bz,
\end{align}
where $ i=\sqrt{-1}$, defines the magnetic field strength two-form on $\Sigma$. We will mainly work with the scalar magnetic field density $B$, defined as follows
\begin{align}\label{magn}
B=g^{z\bz}F_{z\bz}.
\end{align}
The total flux $\NPhi$ of the magnetic field through the surface is an integer  
\begin{equation}\label{flux}
\NPhi=\frac1{2\pi}\int_\Sigma F
=\frac1{2\pi}\int_\Sigma B\sqrt gd^2z,
\end{equation} 
and it is equal to the degree $\deg L^k=k$ of the line bundle $L^k$, 
\begin{equation}\nonumber
\NPhi=k.
\end{equation}
In this section and in Sec.\ \ref{iqhe} we will use $k$ for the degree/flux of magnetic field, and reserve the notation $\NPhi$ for Sec.\ \ref{lsrs}.

The scalar curvature $R$ of the metric is given by
\begin{equation}\nonumber
R=2g^{z\bz}R_{z\bz}=-2g^{z\bz}\p_z\p_{\bz}\log\sqrt g
=-\Delta_g\log\sqrt g,
\end{equation}
where $\sqrt g=2g_{z\bz}$ and the scalar Laplacian is 
$\Delta_g=2g^{z\bz}\p_z\p_{\bz}$. The Euler characteristic of $\Sigma$ is the integral of the scalar curvature over the surface   
\begin{equation}\nonumber
\chi(\Sigma)=\frac1{4\pi}\int_\Sigma R\sqrt gd^2z=2-2{\rm g}.
\end{equation}
We will also introduce the gauge connection for the magnetic field $F=dA$ and spin connection for the curvature written in components as follows
\begin{align}\nonumber
&F_{z\bz} i dz\wedge d\bz= (\p_zA_{\bz}-\p_{\bz}A_z)dz\wedge d\bz,\\\nonumber
&R_{z\bz}\,  i dz\wedge d\bz=(\p_z\omega_{\bz}-\p_{\bz}\omega_z) dz\wedge d\bz.
\end{align}
Sometimes it will be convenient to use the symmetric gauge, where 
\begin{align}\label{symgauge}
&A_z=\frac12 i \p_z\log h^k,\quad A_{\bz}
=-\frac12  i \p_{\bz}\log h^k,\\\nonumber
&\omega_z=\frac12 i\p_z\log g_{z\bz},\quad\omega_{\bz}
=-\frac12  i\p_{\bz}\log g_{z\bz}.
\end{align}
Here $A_z$ and $A_{\bz}$ (id. $\omega_z$ and $\omega_{\bz}$) are complex conjugate and the components of the connections $A_x,A_y$ in real coordinates defined as $A_zdz+A_{\bz}d\bz=A_xdx+A_ydy$ are real-valued.

We will consider a more general choice of the line bundle which is the tensor product 
${\rm L}=L^k\otimes K^s\otimes L_{\varphi}$, where $K$ is the canonical line bundle and $L_{\varphi}$ is the flat line bundle, which has degree zero. This choice is motivated by different physical meaning of the components of ${\rm L}$. As we already mentioned, the line bundle $L^k$ corresponds to the magnetic field, created by a distribution of monopole charges inside the compact surface, with the total flux $k$ through $\Sigma$, Eq. \eqref{flux}. If $s>0$, the sections of $K^s$ correspond to tensors of weight $(s,0)$, i.e. invariant objects of the form $t_{zz...z}(dz)^s$ with $s$ holomorphic indices $z$ (for $s<0$ one considers holomorphic vector fields of weight $-s$), see e.g.\ \cite[\S II.E]{DP}. The Hermitian norm-squared $||t(z,\bz)||^2$ of a section $t(z,\bz)$ of $L^k\otimes K^s$ reads
\begin{equation}\nonumber
\|t(z,\bz)\|^2=|t(z,\bz)|^2h^k(z,\bz)g_{z\bz}^{-s}(z,\bz).
\end{equation}
Physically this means that parameter $s$ is the gravitational (or conformal) spin of the wave function. The curvature of the Hermitian metric on $K^s$ is thus $ i sR_{z\bz}dz\wedge d\bz$ and $\deg K^s=\frac1{2\pi}\int_\Sigma  i sR_{z\bz}dz\wedge d\bz=-s\chi(\Sigma)$. Since $\chi(\Sigma)$ is even this allows for half-integer values of the spin $s$.

Finally, the flat line bundle $L_{\varphi}$ takes into account the flat connection part of the gauge field of total flux zero (line bundle of degree zero), which nonetheless can have non-trivial monodromy around the 1-cycles of the Riemann surface. Physically this corresponds to the magnetic field, created by the solenoid coils wrapped around the 1-cycles of the surface and $\varphi$ labels the solenoid phases. Namely, let $(A_a,B_b)\in H_1(\Sigma,\mathbb Z),\;a,b=1,...,\rm g$ be a canonical basis of one-cycles in $\Sigma$ and $\alpha_a,\beta_b\in H^1(\Sigma,\mathbb Z)$ be the dual basis of harmonic one-forms 
\begin{align}\nonumber
&\int_{A_a}\alpha_b=\delta_{ab},\quad\int_{A_a}\beta_b=0\\\nonumber
&\int_{B_a}\alpha_b=0,\quad\int_{B_a}\beta_b=\delta_{ab}.
\end{align}
The space of phases $(\varphi_{1a},\varphi_{2b})\in[0,1]^{2\rm g}$ span $2\rm g$ dimensional torus $T^{2\rm g}_{[\varphi]}$, also known as the Jacobian variety $Jac(\Sigma)$. The flat connections can be explicitly parameterized as follows
\begin{equation}\label{flatconn}
A^{\varphi}=2\pi\sum_{a=1}^{\rm g}(\varphi_{1a}\alpha_a-\varphi_{2a}\beta_a).
\end{equation}
This gauge connection has the monodromy $e^{2\pi i\varphi_{1a}}$ around the cycle $A_a$ and $e^{-2\pi i\varphi_{2b}}$ around $B_b$ and so do the wave functions, which we will define in a moment. 

It will be useful to write the flat connection \eqref{flatconn} in terms of the basis of holomorphic differentials $\omega_a$, normalized as 
\begin{equation}\nonumber
\int_{A_a}\omega_b=\delta_{ab},\quad\int_{B_a}\omega_b=\Omega_{ab},
\end{equation}
where $\Omega$ is the period matrix of Riemann surface, which is a ${\rm g}\times{\rm g}$ complex symmetric matrix with ${\rm Im}\,\Omega>0$, see e.g.\ \cite[Ch.\ 2,\S 2]{M} and \cite{AMV}. Then the harmonic one-forms are related to the holomorphic differentials by the following linear transformation
\begin{align}\nonumber
&\alpha=-\bar\Omega(\Omega-\bar\Omega)^{-1}\omega+\Omega(\Omega-\bar\Omega)^{-1}\bar\omega,\\\nonumber
&\beta=(\Omega-\bar\Omega)^{-1}\omega-(\Omega-\bar\Omega)^{-1}\bar\omega,
\end{align}
where summation over matrix and vector indices is understood. We can now write the connection \eqref{flatconn} as
\begin{equation}\label{flatconn1}
A^{\varphi}=2\pi\varphi(\Omega-\bar\Omega)^{-1}\bar\omega-2\pi\bar\varphi(\Omega-\bar\Omega)^{-1}\omega,
\end{equation}
where 
\begin{equation}\label{varphic}
\varphi_a=\varphi_{2a}+\Omega_{ab}\varphi_{1b}
\end{equation}
is the complex coordinate on $Jac(\Sigma)$.

\subsection{Lowest Landau level, $\bp$-equation and holomorphic sections}
\label{holosec}

The Hamiltonian for a particle with gravitational spin $s$ on surface $(\Sigma,g)$ in the magnetic field $B$ can be written in complex coordinates as
\begin{equation}\label{H}
H=\frac1mD_z D_{\bz}+\frac{2-g_s}4e\hbar B-\frac s4R+cR,
\end{equation}
where $g_s$ is the Land\'e $g$-factor and in our conventions mass, charge and Planck's constant will be set to one. The derivative operator here is $D_{\bz}=g_{z\bz}^{-\frac12}(i\p_{\bz}-s\omega_{\bz}+A_{\bz})$. The additional term $cR$, where $c$ is a numerical constant, is sometimes added to the Hamiltonian depending on the quantization scheme, see e.g.\ \cite{Kir} for review. For the large flux $k$ of the magnetic field, the lowest energy level (lowest Landau level, LLL) for this hamiltonian is highly degenerate, provided the sum of the last three terms in \eqref{H} is constant
\begin{equation}\label{H1}
E_0=\frac{2-g_s}4 B-\frac s4R+cR=\frac{2-g_s}4k+\left(c-\frac s4\right)\frac{\chi(\Sigma)}2,
\end{equation}
and this constant is equal to the ground state energy $E_0$. Note that the constant magnetic field $B={\rm const}$ corresponds to $F_{z\bz}=kg_{z\bz}$, where $g_{z\bz}$ is not necessarily a constant scalar curvature metric. There exist various choices of constants when Eq.\ \eqref{H1} holds for inhomogeneous magnetic field and/or non-constant scalar curvature metrics. Since we are interested in the case when LLL is highly degenerate, we will ignore the curvature and magnetic field terms in \eqref{H} in what follows. In this case the LLL wave functions solve the first order PDE: 
\begin{equation}\label{lll}
D_{\bz}\Psi=0.
\end{equation} 

As we will see in a moment, equation \eqref{lll} is a local version of a globally defined (on $\Sigma$) $\bp$-equation,
\begin{equation}
\label{dbareq}
\bp_{\rm L} s(z)=0,
\end{equation}
for a line bundle ${\rm L}=L^k\otimes K^s\otimes L_{\varphi}$. Here the $\bp$-operator acts from $\mathcal C^\infty$ sections of $\rm L$ to (0,1) forms with coefficients in $\mathcal C^\infty$ sections,
\begin{equation}\nonumber
\bp_{\rm L}: \mathcal C^\infty(\Sigma,\rm L)
\to\Omega^{0,1}(\Sigma,\rm L).
\end{equation}
The global solutions to Eq.\ \eqref{dbareq} are called the holomorphic sections of $\rm L$, and these will be our LLL wave functions. The vector space of holomorphic sections of $\rm L$ is usually denoted as $H^0(\Sigma,\rm L)$.

By the Riemann-Roch theorem, the dimension of the space of holomorphic sections $\dim H^0(\Sigma,\rm L)$ satisfies the relation
\begin{equation}\nonumber
\dim H^0(\Sigma,{\rm L})-\dim H^1(\Sigma,{\rm L})=\deg(\rm L)+1-\mathrm{g},
\end{equation}
see e.g.\ \cite[p.\,245-6]{GH}. Here $H^1(\Sigma,\rm L)$ is the first Dolbeaut cohomology group, which by the Kodaira vanishing theorem \cite[p.\,154]{GH} vanishes $H^1(\Sigma,\rm L)=0$ for $k$ large enough, and the latter is exactly the large magnetic flux condition relevant for applications to QHE. The precise technical condition for vanishing is $\deg L^k\otimes K^s>\deg K$, i.e., $k+ 2 (\mathrm{g}-1)(s-1)>0$ (using $\deg L_{\varphi} =0$ and $\deg {\rm L}=\deg L^k+\deg K^s=k+2s({\rm g}-1)$), see \cite[p.\,215]{GH}. Hence,  we have for the total number of LLL wave functions
\begin{equation}\label{Nk}
N\equiv N_{k,s}=\dim H^0(\Sigma,{\rm L})=k+(1-{\rm g})(1-2s)\,.
\end{equation}
This formula again reminds us that we can allow for half-integer spins $s$, in which case the wave function is a spinor on $\Sigma$.

In quantum mechanics the wave functions are always normalized with respect to the $L^2$ norm. The $L^2$ inner product of sections reads
\begin{equation}\label{norm}
\langle s_1,s_2\rangle_{L^2}=\frac1{2\pi}\int_\Sigma \bs_1s_2\,
h^kg_{z\bz}^{-s}\sqrt gd^2z.
\end{equation}
One could also formally write the wave function in some local coordinate system as 
\begin{equation}\nonumber
\Psi_l(z,\bz)=s_l(z)h^{k/2}(z,\bz)g^{-s/2}(z,\bz),
\end{equation}
although other inequivalent choices for $\Psi$ will be considered in what follows. Then the inner product looks like the standard quantum-mechanical one
\begin{equation}\nonumber
\langle\Psi_1|\Psi_2\rangle_{L^2}
=\frac1{2\pi}\int_\Sigma \Psi_1^*\Psi_2\sqrt gd^2z.
\end{equation}
Now, in the symmetric gauge \eqref{symgauge}, the $\bp$-equation for
the locally defined wave functions $\Psi_l(z,\bz)$ reduces to the local form Eq. \eqref{lll} that we started with, $D_{\bz}\Psi=0$.
The operator $\bar D$ is a twisted version of the global operator $\bp_{\rm L}$ in \eqref{dbareq}, namely
\begin{equation}\label{Dbz}
D_{\bz}=  h^{\frac k2} g_{z\bz}^{-\frac{s+1}2}\circ
i\bp_L\bigl( h^{-\frac k2} g_{z\bz}^{\frac s2}\; \circ\bigr) 
=g_{z\bz}^{-\frac12}(i\p_{\bz}-s\omega_{\bz}+A_{\bz}),
\end{equation}
where spin $s$ couples to the spin-connection. 

Note that in the discussion above we did not impose any conditions on $B$ and $R$, and we can consider holomorphic sections on $\Sigma$ with an arbitrary metric and with inhomogeneous magnetic field. Suppose $g_0$ is constant scalar curvature metric and $B_0=g_0^{z\bz}F_{0z\bz}$ is constant magnetic field. Integrating magnetic field over $\Sigma$ we conclude that the constant equals the total flux $B_0=k$. A natural way to parameterize an arbitrary curved Riemannian metric $g$ and inhomogeneous magnetic field $B$ is via the \kahler potential $\phi(z,\bz)$ and magnetic potential $\psi(z,\bz)$,
\begin{align}\label{gmetric}
&g_{z\bz}=g_{0z\bz}+\p_z\p_{\bz}\phi,\\
&F_{z\bz}=F_{0z\bz}+k\p_z\p_{\bz}\psi,\quad h^k=h_0^ke^{-k\psi}.\label{magneticp}
\end{align}
Here $\phi$ is a scalar function, satisfying $\p_z\p_{\bz}\phi>-g_{0z\bz}$, so that the metric $g_{z\bz}$ is everywhere positive on $\Sigma$. In other words, the (1,1) forms $F$ and $F_0$ \eqref{F} (cf. $g_{z\bz}idz\wedge d\bz$ and $g_{0z\bz}idz\wedge d\bz$) are in the same \kahler class. We use lower-case $\psi$ for the magnetic potential and and upper-case $\Psi$ for the wave functions throughout the paper, which shall not be confused. 

The area of the surface computed with the metrics $g_0$ and $g$ is the same, which is a natural parameterization in view of incompressibility of QH electronic liquid. It follows from the formulas above and from the definition of magnetic field Eq.\ \eqref{magn}, that constant magnetic field implies magnetic and \kahler potentials being equal up to an irrelevant constant and vice versa,
\begin{equation}\label{constmagn}
B=k\iff \phi=\psi+{\rm const}.
\end{equation} 
We will now consider examples where we explicitly construct some reference orthonormal bases of the LLL states, for constant scalar curvature metric and constant magnetic field. We start here with the sphere, and in the next subsections review LLL on the torus. 

{\it Lowest Landau level on sphere}. In the complex coordinate $z$ induced from $\mathbb C$ by stereographic projection, the round metric on the sphere reads
\begin{equation}\label{roundsp}
g_{0z\bz}=\frac{1}{(1+|z|^2)^2}.
\end{equation}
It has constant scalar curvature $R(g_0)=4$ (the subscript $0$ will mostly be reserved for the metric of constant scalar curvature on the surface). For the constant magnetic field $F_{0z\bz}=kg_{0z\bz}$, and thus one can choose the Hermitian metric \eqref{F} as
\begin{equation}\label{roundsp1}
h^k_0(z,\bz)=\frac{1}{(1+|z|^2)^k}.
\end{equation}
There are no flat connections with nontrivial monodromy since there are no non-contractible cycles ($H^1(S^2,\mathbb Z)$ is trivial) and basis of holomorphic sections of $L^k\otimes K^s$ can be constructed from the polynomial-valued $(s,0)$-forms $(dz)^s$, $z(dz)^s$, $z^2(dz)^s$,... of the maximal degree $N-1=k-2s$ constrained by the condition of finiteness of the $L^2$-norm Eq. \eqref{norm}. Denoting the orthogonal basis as $s_l(z)=c_lz^{l-1}(dz)^s$, we can derive the normalization coefficients $c_l$ from orthogonality condition on the basis
\begin{equation}\nonumber
\langle s_l,s_m\rangle_{L^2}=\delta_{lm}c_l^2\frac1{2\pi}\int_{\mathbb C} \frac{|z|^{2l-2}}{(1+|z|^2)^{k-2s+2}}2d^2z=\frac{c_l^2}{(k-2s+1) C_{k-2s}^{l-1}}\delta_{lm},
\end{equation}
where $C_{k-2s}^{l-1}$ is the binomial coefficient. The integral above converges for $j\leqslant k-2s$. Thus the reference orthonormal basis of sections of $L^k\otimes K^s$ on sphere reads
\begin{equation}\label{spherebasis}
s_l(z)=\sqrt{k-2s+1}\sqrt{C_{k-2s}^{l-1}}\,z^{l-1}(dz)^s,\quad l=1,...,k-2s+1.
\end{equation}
The basis is non-empty when $k\geqslant2s$. For spin zero particles there are $k+1$ states on LLL, one extra normalizable state compared to the Landau problem in the planar domain. 

\subsection{Moduli spaces and adiabatic curvature}\label{at}

Torus is the first nontrivial example where moduli parameters appear. The flat torus can be represented as a quotient $T^2=\mathbb C/\Lambda$ of the complex plane by a lattice $\Lambda=m+n\tau,\;m,n\in\mathbb Z$ and $\tau\in\mathbb H$ ($\mathbb H$ is the complex upper-half-plane) is the complex structure modulus, see picture on Fig. 2 (left). 

In addition to $\tau$ there are moduli parameters associated with the moduli space of flat connections $Jac(T^2)=T_{[\varphi]}$ (moduli of the flat line bundle $L_{\varphi}$). These are two "solenoid phases" $(\varphi_1,\varphi_2)\in[0,1]^2=T_{[\varphi]}$. The corresponding flat connections, defined in Eq.\ \eqref{flatconn}, read in this case
\begin{align}\label{Atorus}
A^{\varphi}=2\pi(\varphi_1\alpha_1-\varphi_2\beta_1), \quad \alpha_1=\frac{\tau d\bz-\bar\tau dz}{\tau-\bar\tau},\quad \beta_1=\frac{dz-d\bz}{\tau-\bar\tau}.
\end{align}

The normalization is such that $\int_{A_1}A^{\varphi}=2\pi\varphi_1,\,\int_{B_1}A^{\varphi}=-2\pi\varphi_2$ for the 1-cycles $A_1=[0,1],\,B_1=[0,\tau]$.
We can give the Jacobian torus $T_{[\varphi]}$ a natural complex structure, defining the complex coordinate on  $T_{[\varphi]}$ according to Eq.\ \eqref{varphic},
\begin{equation}\label{comcoor}
\varphi=\varphi_2+\varphi_1\tau.
\end{equation}
In this complex coordinate $T_{[\varphi]}$ looks exactly like the coordinate space torus, see the picture on the right on Fig.\ 2. Of course, one should remember that these are two different spaces -- one is the physical space where particles live and the other one is the parameter space.

\begin{figure}[h]
\begin{center}
\raisebox{-.12cm}{\begin{tikzpicture}
\draw[-] (-.2,0) -- (3,0) node[below] {};
  \draw[-] (0,-.2) -- (0,2.4) node[left] {};
  \draw[scale=1,domain=0:.8,smooth,variable=\x,blue,very thick] plot ({\x},{2*\x});
    \draw[scale=1,domain=-0.01:2.02,smooth,variable=\x,blue,very thick] plot ({\x},{0});
    \draw[scale=1,domain=2:2.8,smooth,variable=\x,blue,very thick] plot ({\x},{2*\x-4});
  \draw[scale=1,domain=0.787:2.817,smooth,variable=\x,blue,very thick]  plot ({\x},{1.6});
          \draw[scale=1,domain=2.7:3,smooth,variable=\x,black] plot ({\x},{2.1});
        \draw[scale=1,domain=2.09:2.4,smooth,variable=\x,black] plot ({2.7},{\x});
  \draw[black] (2.87,2.5) node[below]{$z$};
    \draw[black] (0.8,2.0) node[below]{$\tau$};
    \draw[black] (2,0) node[below]{$\scriptstyle1$};
                \draw[black] (-0.07,-0.19) node[right]{${\scriptstyle 0}$};
  \end{tikzpicture}}\quad\quad\quad\quad\quad\quad
  \begin{tikzpicture}
\draw[-] (-.2,0) -- (3,0) node[below] {};
  \draw[-] (0,-.2) -- (0,2.4) node[left] {};
  \draw[scale=1,domain=0:.8,smooth,variable=\x,magenta,very thick] plot ({\x},{2*\x});
    \draw[scale=1,domain=-0.01:2.02,smooth,variable=\x,magenta,very thick] plot ({\x},{0});
    \draw[scale=1,domain=2:2.8,smooth,variable=\x,magenta,very thick] plot ({\x},{2*\x-4});
  \draw[scale=1,domain=0.787:2.817,smooth,variable=\x,magenta,very thick]  plot ({\x},{1.6});
          \draw[scale=1,domain=2.7:3,smooth,variable=\x,black] plot ({\x},{2.1});
        \draw[scale=1,domain=2.09:2.4,smooth,variable=\x,black] plot ({2.7},{\x});
  \draw[black] (2.87,2.5) node[below]{$\varphi$};
    \draw[black] (0.8,2.0) node[below]{$\tau$};
    \draw[black] (2,0) node[below]{$\scriptstyle1$};
                \draw[black] (-0.07,-0.19) node[right]{${\scriptstyle 0}$};
  \end{tikzpicture}\\
{\small Figure 2.\, Coordinate space torus $T^2$ (left) and Jacobian torus $T_{[\varphi]}=Jac(T^2)$ (right).}
\end{center}
\end{figure}

Our wave functions on $T^2$ will also depend on the moduli space coordinates, i.e., on two complex parameters $(\tau,\varphi)$, while the energy of the ground state is independent of $(\tau,\varphi)$. This situation falls into the broader context of Berry phases \cite{Berry} and, when the parameter space is the moduli space, this is also known as {\it geometric adiabatic transport}. Before constructing the wave functions on the torus explicitly, we briefly review adiabatic (Berry) connection and curvature on moduli spaces. Suppose the wave functions depend on a point in the moduli space, which is a complex space $Y$ of complex dimension $\dim Y$. In our case the parameter space will be 
\begin{equation}\label{parsp}
Y=\mathcal M_{\rm g}\times Jac(\Sigma), 
\end{equation}
where $\mathcal M_{\rm g}$ is the moduli space of complex structures of genus-{\rm g} Riemann surface and $Jac(\Sigma)$ is the moduli space of flat connections \eqref{flatconn}. To be more precise, Eq.\ \eqref{parsp} is not exactly a direct product, but a fibration, since $Jac(\Sigma)$ as a complex manifold varies over $\mathcal M_{\rm g}$ due to the choice of complex structure \eqref{comcoor}. This will not be important for us here, but should be kept in mind. We consider some local coordinates $(y^\mu,\bar y^{\bar\mu}), \mu=1,..,\dim Y$, and we will usually suppress the index $\mu$ for simplicity. All wave functions we will encounter here have the following schematic form
\begin{equation}\label{wf}
\Psi_l(x|y,\by)=\frac{F_l(x|y)}{\sqrt{Z(y,\bar y)}},
\end{equation}
where $F_l(x|y),\,l=1,..,n$ are locally holomorphic functions of $y$. The wave functions can be one-particle wave functions or multi-particle states, so $x$ schematically denotes coordinates of one or several particles on $\Sigma$. The normalization factor $Z$ is fixed by the condition that the wave functions are canonically normalized $\langle\Psi_l|\Psi_m\rangle=\delta_{lm}$, and crucially $Z$ is independent of the index $r$. The basis of wave functions on $\Sigma$ varies over $Y$ as a frame of certain Hermitian vector bundle $E$ of rank $n$ over $Y$. Moreover, in quantum mechanics there is a canonical Hermitian connection $\mathcal A$ on $E$, called Berry, or adiabatic connection \cite{Berry}, given by
\begin{align}\label{aconn}
&(\mathcal A_y)_{lm}= i\langle\Psi_l|\p_y\Psi_m\rangle_{L^2},
\\\label{acurv}
&\mathcal R_{lm}=d\mathcal A_{lm}+ (\mathcal A\wedge\mathcal A)_{lm}.
\end{align}
This is a Chern connection \cite{Simon} since its curvature is a $(1,1)$-form on $Y$.
Specifically for the wave functions of the form \eqref{wf} the formulas above simplify and the connection and curvature can be expressed via the normalization factor as 
\begin{align}\label{adconn}
&(\mathcal A_y)_{lm}= i\p_y\langle\Psi_l|\Psi_m\rangle_{L^2}- i\langle\p_y\Psi_l|\Psi_m\rangle_{L^2}=\delta_{lm}\frac i2\p_y\log Z,\\\label{adcurv}
&\mathcal R_{lm}=-\delta_{lm}(\p_y\p_{\by}\log Z)\, i dy\wedge d\by.
\end{align}
Adiabatic connection and curvature in Eq.\ \eqref{adcurv} are scalar matrices, i.e., $\mathcal A$ is the projectively flat connection, and normalization factor $Z$ plays the role of the Hermitian metric on $E$. The vector bundle $E$ appears to be very close to being a direct sum of line bundles for each degenerate state. However this will only be the case for the integer QH state (which is non degenerate), but not for the LLL states on the torus and not for the Laughlin states, because these states have non-abelian monodromy around 1-cycles of $Y$, as we will see later.

\subsection{Lowest Landau level on the torus}

For the discussion of various aspects of quantum mechanics of an electron in the magnetic field on the torus we refer to \cite{DN,Av,L1}. 

The metric on the flat torus reads
\begin{equation}\label{torusmetr}
g_{0z\bz}=\frac{2\pi i}{\tau-\bar\tau},
\end{equation}
and its area is normalized to be $2\pi$. The Hermitian metric on $L^k$, corresponding to the constant magnetic field can be chosen as 
\begin{equation}\label{tildeh0}
\tilde h_0^k(z,\bz)=\exp\left({\frac{\pi i k}{\tau-\bar\tau}(z-\bz)^2}\right),
\end{equation}
so that $F_0=kg_0$. The canonical bundle $K$ is a trivial holomorphic line bundle on the torus, so we set gravitational spin $s=0$ in this case. 

The transformation rules for the holomorphic parts of the wave functions under the lattice shifts can be derived from the condition that $\bar{\tilde s}(z)\tilde s(z)\tilde h_0^k(z,\bz)$ transforms as a scalar. Then the lattice shifts act on $s(z)$ via the automorphy factor
\begin{align}\label{torusmonod0}
\tilde s(z+t_1+t_2\tau)=(-1)^{2t_2\delta+2t_1\varepsilon}e^{-\pi i kt_2^2\tau-2\pi i kt_2z+2\pi i(t_1\varphi_1-t_2\varphi_2)}\tilde s(z),\quad t_1,t_2\in\mathbb Z.
\end{align}
Here in addition to the solenoid phases $\varphi_1,\varphi_2$ we introduced the spin structure parameters $\varepsilon,\delta\in\{0,\frac12\}$. These label periodic or anti-periodic boundary conditions for the wave function around the 1-cycles of the torus. The four spin structures are divided into two classes: $(0,0)$, $(0,{\scriptstyle\frac12})$, $({\scriptstyle\frac12},0)$ are called even and $({\scriptstyle\frac12},{\scriptstyle\frac12})$ is odd. At first glance introducing spin structures appears to be a redundancy since $\varepsilon,\delta$ correspond to points in the space of phases $\varphi_1,\varphi_2$, but as we will see later spin structures and phases transform in a different fashion under modular transformations, so it is instructive to keep track of the spin structure.  

The transformation rules \eqref{torusmonod0} are holomorphic in $\tau$, but not in $\varphi$. In order to stay in accordance with Eq.\ \eqref{wf} we slightly change the Hermitian metric \eqref{tildeh0} and consequently the automorphy factors as follows  
\begin{align}\label{hermtor}
&h_0^k(z,\bz)=\exp\left({\frac{\pi i k}{\tau-\bar\tau}(z-\bz)^2}+\frac{2\pi i}{\tau-\bar\tau}(z-\bz)(\varphi-\bar\varphi)\right),\\\label{torusmonod}
&s(z+t_1+t_2\tau)=(-1)^{2t_2\delta+2t_1\varepsilon}e^{-\pi i kt_2^2\tau-2\pi i kt_2z-2\pi it_2\varphi} s(z),
\end{align}
so that the transformations are holomorphic in $\varphi$. These are the transformation properties satisfied by theta functions with characteristics, see Eq.\ \eqref{theta} in the Appendix and Ref.\ \cite{M} for their definition and properties. A particularly convenient choice of the basis of theta functions is  
\begin{equation}\label{basis}
s_l^{\varepsilon,\delta}(z)=\vartheta\left[\begin{array}{c}\frac{\varepsilon+l}k \\{\scriptstyle\delta}\end{array}\right](kz+\varphi,k\tau), \quad l=1,...,k.
\end{equation}
Here we explicitly indicated the choice of spin-structure $\varepsilon,\delta$. We stress that $l$ labels the degenerate LLL states while $\varepsilon,\delta$ are treated as external quantum numbers, labelling the choice of boundary conditions. Thus we have constructed $k$ LLL states, and in accordance with the Riemann-Roch theorem \eqref{Nk}, $\dim H^0(T^2,L^k)=k$.  

Taking into account that the complex coordinate depends on the complex structure as  $z=x_1+\tau x_2$, in the basis Eq.\ \eqref{basis} we can write the wave functions on the torus as
\begin{equation}\label{holombasis}
\Psi_l^{\varepsilon,\delta}(x|\tau,\varphi)=\frac1{\sqrt{Z(\tau,\bar\tau,\varphi,\bar\varphi)}} \cdot e^{2\pi i\varphi x_2+\pi i kx_2^2\tau}\,\vartheta\left[\begin{array}{c} \frac{\varepsilon+l}k \\\scriptstyle\delta\end{array}\right](kz+\varphi,k\tau),
\end{equation}
which is consistent with the generic abstract form given in Eq.\ \eqref{wf} since $\varphi$ now enters holomorphically in the numerator. Computing the $L^2$ inner product \eqref{norm} for the basis \eqref{basis} 
\begin{equation}\label{L2s}
\langle s_l,s_m\rangle_{L^2}=\sqrt{\frac i{k(\tau-\bar\tau)}}\cdot e^{-\frac{\pi i}{k}\frac{(\varphi-\bar\varphi)^2}{\tau-\bar\tau}}\delta_{lm}
\end{equation}
allows us to determine the normalization $Z$-factor
\begin{equation}\label{Zfactor}
Z(\tau,\bar\tau,\varphi,\bar\varphi)=\sqrt{\frac i{k(\tau-\bar\tau)}}\cdot e^{-\frac{\pi i}{k}\frac{(\varphi-\bar\varphi)^2}{\tau-\bar\tau}},
\end{equation}
cf. Eq.\ \eqref{wf}.

\subsection{Modular group and geometric adiabatic transport}\label{monodrom}

In addition to the adiabatic connection and curvature \eqref{adcurv}, which will be determined momentarily from \eqref{Zfactor}, the vector bundle over $Y$ of ground states \eqref{basis} on $\Sigma$ is characterized by the monodromies around non-trivial one-cycles in the moduli space. The flat connection moduli $\varphi$ belong to the torus $T_{[\varphi]}=\mathbb C/\Lambda,\;\Lambda=t_1+t_2\tau,\,t_1,t_2\in\mathbb Z$ and the group $\mathbb Z+\mathbb Z$ of lattice shifts $\varphi\to\varphi+t_1+t_2\tau$ acts on the holomorphic wave functions \eqref{basis}, Hermitian metric \eqref{hermtor} and the $Z$-factor \eqref{Zfactor}. We have

\begin{align}\nonumber
&s_l^{\varepsilon,\delta}(z|\varphi+t_1+t_2\tau,\tau)=e^{-\frac{\pi i}kt_2^2\tau-\frac{2\pi i}k t_2(kz+\varphi)}\cdot \sum_{m=1}^kU_{lm} s_{m}^{\varepsilon,\delta}(z|\varphi,\tau),\\
\nonumber&\phantom{aaaaaaaaaaaaaaaaaaa} {\rm where}\quad U_{lm}=e^{\frac{2\pi i}k(t_1l+t_1\varepsilon-t_2\delta)}\delta_{l,m-t_2},\\\nonumber
&h_0^k(z,\bz|\varphi+t_1+t_2\tau,\tau)=e^{2\pi it_2(z-\bz)}h_0^k(z,\bz|\varphi,\tau),
\\
\nonumber
&Z(\varphi+t_1+t_2\tau,\bar\varphi+t_1+t_2\bar\tau,\tau,\bar\tau)=e^{-\frac{\pi i}k t_2^2(\tau-\bar\tau)-\frac{2\pi i}k t_2(\varphi-\bar\varphi)}\cdot Z(\varphi,\bar\varphi,\tau,\bar\tau).
\end{align}
We see that group group of lattice transformations on $T_{[\varphi]}$ acts on the basis of states in a unitary representation, given by the unitary matrix $U$, and preserves the spin structure. Note that while shifts by $t_1$ act diagonally on the basis, the action of shift in $t_2$ is non-diagonal and amounts to relabelling the basis $l\to l+t_2$. We will see shortly that the adiabatic curvature is given by scalar matrix \eqref{projphi}, yet due to the non-diagonal action of $\varphi_1$ shifts the vector bundle of LLL states is not a direct sum of one-dimensional bundles for each wave function.

Next we discuss the monodromies on the complex structure moduli space. The symmetry group of the lattice $\Lambda$ is the modular group $PSL(2,\mathbb Z)$, which preserves the torus. It consists of matrices
\begin{equation}\nonumber
\quad\left(\begin{matrix} a & b \\ c & d \end{matrix}\right),\quad a,b,c,d\in\mathbb Z,\quad ad-bc=1,
\end{equation}
defined up to an overall sign $a,b,c,d\to -a,-b,-c,-d$. Its action (modular transformation) on $\tau$, has to be accompanied by the action on $z=x+y\tau$ and $\varphi=\varphi_2+\varphi_2\tau$, since the latter depend on $\tau$. The full action is given by
\begin{equation}\nonumber
(\tau,\varphi,z)\to\left(\frac{a\tau+b}{c\tau+d},\frac{\varphi}{c\tau+d},\frac{z}{c\tau+d}\right),
\end{equation}
while $\varepsilon$ and $\delta$ do not transform (that is why their role is slightly different from solenoid phases $\varphi$, as was mentioned after Eq.\ \eqref{torusmonod0}).
The group $PSL(2,\mathbb Z)$ is generated by two elements $T:\tau\to\tau+1$ and $S:\tau\to-1/\tau$, subject to relations $S^2=1$ and $(ST)^3=1$. 

The action of the modular group on the wave functions \eqref{basis} can be derived using Eqns. (\ref{Ttr}, \ref{Str}) in the Appendix. We obtain
\begin{align}\label{T}
&T\circ s_l^{\varepsilon,\delta}(z|\varphi,\tau)=\sum_{m=1}^NU^T_{lm}s_m^{\varepsilon,\delta+\varepsilon-\lambda}(z|\varphi,\tau),\quad U^T_{lm}=e^{\frac{\pi i}k(l+\varepsilon)(l-\varepsilon+2\lambda)}\delta_{lm},\\\nonumber
&T\circ h_0^k(z,\bz)=h_0^k(z,\bz),
\\\nonumber
&T\circ Z(\varphi,\bar\varphi,\tau,\bar\tau)=Z(\varphi,\bar\varphi,\tau,\bar\tau),\\\label{S}
&S\circ s^{\varepsilon,\delta}_l(z|\varphi,\tau)=\sqrt{-i\tau}\cdot e^{\frac{\pi i}k\frac{(kz+\varphi)^2}{\tau}} \sum_{m=1}^kU^S_{lm} s^{\delta,\varepsilon}_{m}(z|\varphi,\tau),\\\nonumber
&{\rm where}\quad U^S_{lm}=\frac1{\sqrt k}e^{-\frac{2\pi i}k(\varepsilon\delta+l(m+2\varepsilon))},\\\nonumber
&S\circ h_0^k(z,\bz)=e^{-\pi ik\frac{z^2}\tau+\pi ik\frac{\bz^2}{\bar\tau}-2\pi i\frac{z\varphi}\tau+2\pi i\frac{\bz\bar\varphi}{\bar\tau}}h_0^k(z,\bz),\\\nonumber
&S\circ Z(\varphi,\bar\varphi,\tau,\bar\tau)=\sqrt{\tau\bar\tau}\cdot e^{\frac{\pi i}k\frac{\varphi^2}{\tau}-\frac{\pi i}k\frac{\bar\varphi^2}{\bar\tau}}\cdot Z(\varphi,\bar\varphi,\tau,\bar\tau).
\end{align}
Here we introduced the parity indicator constant for the degeneracy of LLL states
\begin{align}\label{lambda0}
\lambda=\frac k2-\left[\frac k2\right]=\begin{cases}
0, & \text{for }k\in{\rm even}\\
\frac12, & \text{for }k\in{\rm odd}.\\
\end{cases}
\end{align}
We see that the action of the modular group is given essentially by the unitary matrices $U^S$ and $U^T$, and all other factors appearing in transformation formulas cancel out between $s(z), h_0$ and $Z$. 

\begin{figure}
\begin{center}
  \begin{tabular}{c|c|c}
&\quad\,${\scriptstyle \lambda=0}$\quad\,&\quad\,${\scriptstyle\lambda=\frac12}$\quad\,\\[2pt] 
$(\varepsilon,\delta)$  & ${\blue T}$ \quad ${\magenta S}$ & ${\blue T}$ \quad ${\magenta S}$ \\[2pt]\hline
 &&\\[-4pt]
   $(0,0)$ & ${\blue\circlearrowleft}$ \;\; ${\magenta\circlearrowleft}$ & {\blue\tikzmark{a}\;\;\tikzmark{a1}} \;\;\; ${\magenta\circlearrowleft}$  \\[8pt] 
            $\bigl(0,\frac12\bigr)$ & ${\blue\circlearrowleft}$ \;\;\, \tikzmark{b}\;\;\tikzmark{b1} & \tikzmark{c}\;\;\tikzmark{c1} \;\;\;\, \tikzmark{d}\;\;\tikzmark{d1} \\[8pt] 
    $\bigl(\frac12,0\bigr)$ & \,\tikzmark{e}\;\;\tikzmark{e1} \;\;\; \tikzmark{f}\;\;\tikzmark{f1} & ${\blue\circlearrowleft}$ \;\;\; \tikzmark{g}\;\;\tikzmark{g1} \\[8pt] 
    $\bigl(\frac12,\frac12\bigr)$  & \,\tikzmark{h}\;\;\tikzmark{h1} \;\; ${\magenta\circlearrowleft}$ & $\,{\blue\circlearrowleft}$ \;\;\, ${\magenta\circlearrowleft}$ \\[8pt]\hline
  \end{tabular}
  \begin{tikzpicture}[overlay, remember picture, yshift=.25\baselineskip, shorten >=-2pt, shorten <=-2pt]
    \draw [blue,->] ({pic cs:a}) [bend right] to ({pic cs:c});
    \draw [blue,->] ({pic cs:c1}) [bend right] to ({pic cs:a1});
    \draw [magenta,->] ({pic cs:b}) [bend right] to ({pic cs:f});
    \draw [magenta,->] ({pic cs:f1}) [bend right] to ({pic cs:b1});
    \draw [magenta,->] ({pic cs:d}) [bend right] to ({pic cs:g});
    \draw [magenta,->] ({pic cs:g1}) [bend right] to ({pic cs:d1});
    \draw [blue,->] ({pic cs:e}) [bend right] to ({pic cs:h});
    \draw [blue,->] ({pic cs:h1}) [bend right] to ({pic cs:e1});
  \end{tikzpicture}
\vspace{.5cm}\\
{\small Figure 3.\, Modular group action on the states changes their spin-structure.}
\end{center}
\end{figure}

Note that the modular transformations in general mix between different spin structures, according to the table Fig.\ 3. While $S$ transformation preserves the parity of the spin structure (it maps states with odd spin structure to odd, and even to even), the $T$ transformation in general does not. According to Fig.\ 3 there exist several possibilities. For even number of particles $\lambda=0$ the $(0,0)$ spin structure is preserved by the full modular group, and for odd number of particles $\lambda=\frac12$ the odd $(\frac12,\frac12)$ spin structure is preserved by full modular group.
For the choices $\lambda=0$, $(\frac12,\frac12)$ and $\lambda=\frac12$, $(0,0)$, the invariant subgroup of the modular group is $\Gamma_\theta$, generated by $(T^2,S)$, which appeared before in Refs.\ \cite{L1} and \cite{KW,CZ} for Laughlin states, where the situation is completely equivalent, see \S \ref{modultr}. Finally, if we take into account $(0,\frac12)$ and $(\frac12,0)$ spin structures, then the minimal subgroup is the normal subgroup $\Gamma (2)$ generated by $(T^2,ST^2S)$, see e.g.\ Ref.\ \cite{Lang}. This subgroup is also called the modular group $\Lambda$, see Ref.\ \cite{Wein} and its fundamental domain is pictured in Fig.\ 4 (right). 

Let us focus on the states invariant under the full modular group: $(0,0)$ at $\lambda=0$ and  $(\frac12,\frac12)$ at $\lambda=\frac12$. In order to check that the action of $PSL(2,\mathbb Z)$ is unitary for these two choices of spin structure, we need to check the action of $S^2$ and $(ST)^3$ on the basis. In both cases we obtain
\begin{equation}
(U^SU^T)^3=e^{2\pi i\theta}C,\quad\quad (U^S)^2= C,
\end{equation} 
where the constant $\theta=\frac{1}8$. Matrix $C$ satisfies $C^2=1$, and for $\lambda=0$ it has a particularly simple form $C_{lm}=\delta_{l,k-m}$. Thus the basis of states transform by the projective unitary representation of $SL(2,\mathbb Z)$. For more details on projective unitary representations and modular tensor categories we refer to \cite{Kit}.

The fundamental domain for the action of $PSL(2,\mathbb Z)$ on the upper-half plane is the complex structure moduli space $\mathcal M_1=\mathbb H/PSL(2,\mathbb Z)$, see Fig.\ 4 (left). It has a cusp (at $\tau=i\infty$) and two orbifold points (at $\tau=i$ and $\tau=e^{\pi i/3}\sim e^{2\pi i/3}$). The fundamental domain $R_{\Gamma(2)}$ of $\Gamma (2)$ has three cusps at $ i\infty,0$ and $1\sim-1$.

\begin{figure}[t]
\begin{center}
\raisebox{-.12cm}{\begin{tikzpicture}
\draw[-] (-1.5,0) -- (1.5,0) node[below] {};
  \draw[-] (0,-.5) -- (0,3) node[left] {};
  \draw[scale=1,domain=0.832:3,smooth,variable=\x,blue,very thick] plot ({1/2},{\x});
    \draw[scale=1,domain=0.832:3,smooth,variable=\x,blue,very thick] plot ({-1/2},{\x});
  \draw[scale=1,domain=-1/2:1/2,smooth,variable=\x,blue,very thick]  plot ({\x},{2-(1-(\x)^2)^.5});
  \draw[black] (-.6,0) node[below]{$\scriptstyle-\frac12$};
    \draw[black] (1/2,0) node[below]{$\scriptstyle\frac12$};
        \draw[black] (-0.07,1.19) node[right]{${\scriptstyle 1}$};
                \draw[black] (-0.07,0.19) node[right]{${\scriptstyle 0}$};
  \end{tikzpicture}}\quad\quad\quad\quad\quad\quad
  \begin{tikzpicture}
\draw[-] (-1.5,0) -- (1.5,0) node[below] {};
  \draw[-] (0,-.5) -- (0,3) node[left] {};
  \draw[scale=1,domain=0:3,smooth,variable=\x,magenta,very thick] plot ({1},{\x});
    \draw[scale=1,domain=0:3,smooth,variable=\x,magenta,very thick] plot ({-1},{\x});
  \draw[scale=1,domain=0:1,smooth,variable=\x,magenta,very thick]  plot ({\x},{sqrt(1/4-(\x-.5)^2)});
    \draw[scale=1,domain=-1:0,smooth,variable=\x,magenta,very thick]  plot ({\x},{sqrt(1/4-(\x+.5)^2)});
    \draw[black] (-1,0) node[below]{$\scriptstyle-1$};
        \draw[black] (1,0) node[below]{$\scriptstyle1$};
    \draw[black] (0.15,0) node[below]{$\scriptstyle 0$};
  \end{tikzpicture}
  \vspace{0.5cm} \\
{\small Figure 4.\, Fundamental domains: $\mathcal M_1$ of $PSL(2,\mathbb Z)$ (left) and $R_{\Gamma(2)}$ of $\Gamma(2)$ (right).}
\end{center}
\end{figure}

Now we recall basic setup of the geometric adiabatic transport on the moduli space \cite{ASZ,ASZ1}. We  choose a smooth closed contour $\mathcal C[t],\;[0\leqslant t\leqslant 1]$ in the moduli space $Y=T_{[\varphi]}\times \mathcal M_1$, or $Y=T_{[\varphi]}\times R_{\Gamma(2)}$. We consider first the transport on $T_{[\varphi]}$ with the phases $\dot\varphi_a(t)$ varying along the contour. This creates the electric field and thus the current $I_b$ along the cycle $b$ on the torus, according to
\begin{equation}
I_b=i\dot\varphi_a\sigma_{ab},
\end{equation}
where $\sigma_{ab}$ is the conductance matrix. This is the coefficient matrix of the adiabatic curvature 2-form, traced over all LLL states $\tr\mathcal R$ \eqref{adcurv}. Here we take the trace because we would like to consider $k$ fermionic particles on the LLL, i.e., completely filled lowest Landau level (this point will become obvious in \S \ref{quillenano}, where we make the same calculation for the integer QH state). On the torus using \eqref{Zfactor} we immediately obtain 
\begin{align}\label{projphi}
\mathcal R_{lm}=\frac{2\pi}k\delta_{lm}\frac{d\varphi\wedge d\bar\varphi}{\tau-\bar\tau}=\frac{2\pi}k\delta_{lm}\;d\varphi_1\wedge d\varphi_2,\\\nonumber
\tr\mathcal R=2\pi\sigma_{12}\;d\varphi_1\wedge d\varphi_2=2\pi\sigma_{12}\;d\varphi_1\wedge d\varphi_2.
\end{align}
Recall that the latter is actually the first Chern class of the ground state vector bundle $E$, defined as $c_1(E)=\frac1{2\pi}\tr\mathcal R$. Therefore the quantization of the Hall conductance
\begin{equation}\nonumber
\sigma_H=\sigma_{12}=1
\end{equation} 
in the integer QHE follows from the integrality of $c_1(E)$. Indeed, we have  
\begin{equation}\label{chernnum}
\sigma_H=\int_{T_{[\varphi]}}c_1(E|_{T_{[\varphi]}})=1,
\end{equation}
where notation $E|_{T_{[\varphi]}}$ means that we restrict the vector bundle to the Jacobian torus $T_{[\varphi]}$.  

Using Eq.\ \eqref{Zfactor} we can derive the adiabatic curvature for the full moduli space. We obtain,
\begin{align}\label{c1E}
\frac1{2\pi}\tr\mathcal R=-\frac{ i}{\tau-\bar\tau}\, i d\varphi\wedge d\bar\varphi+\frac k{4\pi(\tau-\bar\tau)^2} i d\tau\wedge d\bar\tau,
\end{align}
and the mixed terms $d\varphi\wedge d\bar\tau$ vanish. The second term in the adiabatic curvature was related in Ref.\ \cite{ASZ1} to the non-dissipative component of the viscosity tensor, and gives rise to a new adiabatic transport coefficient $\eta_H=\frac k4$, called anomalous or Hall viscosity. By analogy with the Hall conductance calculation above \eqref{chernnum}, we can identify the coefficient $\eta_H$ computing the integral of the first Chern class $c_1(E)$ restricting to the moduli space $\mathcal M_1$,
\begin{equation} \nonumber
\int_{\mathcal M_1}c_1(E|_{\mathcal M_1})=\frac k{8\pi}\int_{\mathcal M_1}\frac{d\tau_1d\tau_2}{\tau_2^2}=\frac k{24}=\frac{\eta_H}6,
\end{equation}
since the volume of $\mathcal M_1$ in Poincar\'e metric equals $\pi/3$. This Chern number is a fraction, since $\mathcal M_1$ is an orbifold, but this is still a topological invariant of the vector bundle of LLL states. Finally, we note that one can get rid of $1/6$ by replacing the moduli space $\mathcal M_1$ by $R_{\Gamma(2)}$, which seems appropriate in view of action of the modular group. Since $\Gamma(2)$ is a congruence subgroup of index 6, the volume of $R_{\Gamma(2)}$ in Poincar\'e metric is six times bigger than the volume of $\mathcal M_1$ and thus $\int_{R_{\Gamma(2)}}c_1(E|_{R_{\Gamma(2)}})=\eta_H$. 

\section{Integer quantum Hall state}
\label{iqhe}

\subsection{Free fermions on a Riemann surface} 

We consider Hermitian line bundle $(L^k,h_0^k)$ on a compact Riemann surface $(\Sigma,g_0)$. The basis of the states on the LLL is given by the holomorphic sections $\{s_l\},\,l=1,..,N=\dim H^0(\Sigma,\rm L)$ of ${\rm L}=L^k\otimes K^s\otimes L_{\varphi}$. In the integer QHE we have a system of $N$ free fermions, occupying the lowest Landau level. Such a system is described by a multi-particle wave function, which is a completely antisymmetric combination of the one-particle ground states. Thus the (holomorphic part of) integer quantum Hall wave function on $\Sigma^N$ (more precisely on $\Sigma^N/S_N$ since the particles are identical) is given by the Slater determinant
\begin{equation}\label{slater}
\mathcal S(z_1,...,z_N)=\frac1{\sqrt{N!}}\det[s_l(z_m)]|_{l,m=1}^{N}.
\end{equation}
Similar to the one-particle case, the mod squared of the actual wave function is given by point-wise  Hermitian norm of \eqref{slater},
\begin{equation}\nonumber
|\Psi(z_1,...,z_N)|^2=||\mathcal S(z_1,...,z_N)||^2=\frac1{N!}|\det[s_l(z_m)]|^2\prod_{l=1}^{N}h_0^k(z_l,\bz_l) g_{0z\bz}^{-s}.
\end{equation}
Now, if the basis $\{s_l\}$ is chosen to be orthonormal with respect to $L^2$ norm,
\begin{equation}\label{normal}
\frac1{2\pi}\int_\Sigma\bs_l(\bz)s_m(z)h_0^k(z,\bz) g_{0z\bz}^{-s}\sqrt gd^2z=\delta_{lm},
\end{equation}
then the integer quantum Hall state is automatically normalized (with respect to the $L^2$ norm on $\Sigma^N$)
\begin{align}\label{detform}
&\frac1{(2\pi)^{N}}\int_{\Sigma^{N}}|\Psi(z_1,...,z_{N})|^2\prod_{j=1}^{N}\sqrt gd^2z_j=\\
=\frac1{(2\pi)^{N}N!}\int_{\Sigma^{N}}&|\det[s_l(z_m)]|^2\prod_{l=1}^{N}h_0^k(z_l,\bz_l) g_{0z\bz}^{-s}\sqrt{g_0}d^2z_l=\det \langle s_l,s_m\rangle_{L^2}=1\nonumber,
\end{align}
where in the last line we used a straightforward combinatorial identity, expressing the multiple integral as the determinant of one-particle $L^2$ norms \eqref{norm}.

{\it Plane}. On the complex plane $\mathbb C$ and for the constant perpendicular magnetic field $B$ the number of states on the LLL is infinite, so we take the first $N$ states $s_l=z^{l-1},l=1,...,N$, imposing a cut-off on the angular momentum $r^me^{2\pi i m\phi}, \,m<N$. Since the metric is flat, we can set the gravitational spin $s$ to zero. Using the Vandermonde determinant formula $\det z_l^{m-1}=\prod_{l<m}(z_l-z_m)$, the integer QH state can be written as
\begin{align}\label{intplane}
\Psi(z_1,...,z_N)=\frac{\mathcal N}{\sqrt{N!}}\det[s_l(z_m)] \prod_{l=1}^Nh_0^{B/2}(z_l,\bz_l)=\frac{\mathcal N}{\sqrt{N!}}\prod_{l<m}^{N}(z_l-z_m)\cdot e^{-\frac14B\sum_l|z_l|^2},
\end{align}
up to an overall normalization constant $\mathcal N$. 

{\it Sphere}. On the sphere with uniform constant magnetic field we can write the wave function \eqref{slater} explicitly, using the basis \eqref{spherebasis} and the Vandermonde identity 
\begin{align}\nonumber
\frac1{\sqrt{N!}}\det[s_l(z_m)]&=\frac{1}{\sqrt{N!}}\prod_{l=1}^{N} c_l\cdot\prod_{l<m}^{N}(z_l-z_m)\cdot (dz_1)^s\otimes(dz_2)^s\otimes\cdots\otimes(dz_{N})^s,
\end{align}
where number of particles $N=k-2s+1$ and normalization constants $c_l$ are given in Eq.\ \eqref{spherebasis}.

{\it Torus}. On the torus the analog of the Vandermonde formula above is known as the "bosonisation formula", see e.g.\ \cite[Eq.\ 5.33]{Fay}. For the constant magnetic field and flat metric the orthonormal basis of sections was constructed in \eqref{basis} and the number of particles is $N=k$. The following identity holds,
\begin{align}\label{ILtor}\nonumber
\mathcal S(z_1,..,z_k)=&\det[s_l^{\varepsilon,\delta}(z_m)]\\
&=e^{\pi i\epsilon}\,\eta(\tau)^{k-1}\vartheta\left[\begin{array}{c}\scriptstyle\varepsilon-\lambda+\frac12\\\scriptstyle\delta-\lambda+\frac12\end{array}\right](z_{\rm cm}+\varphi,\tau)\prod_{l<m}^k\frac{\vartheta_1(z_l-z_m,\tau)}{\eta(\tau)},
\end{align}
where $\epsilon$ is a constant depending only on $\varepsilon,\delta$ and $k$, $\lambda$ is the parity indicator for $k$ Eq.\ \eqref{lambda0}, and $\eta(\tau)$ is the Dedekind eta function. Also we introduced the following notation
\begin{equation}\label{com}
z_{\rm cm}=\sum_{l=1}^kz_l,
\end{equation} 
for the so called center-of-mass coordinate. For the proof of this identity we first note that both sides are sections of $L^k$ for each coordinate $z_m$.  This can be checked by comparing automorphy factors under lattice shifts, which on the lhs can be read off immediately from Eq.\ \eqref{torusmonod}, and on the rhs can be worked out from Eqns.\ (\ref{transtheta}, \ref{shift1}) in the Appendix.  Then we check that the zeroes on both sides coincide. For each $z_m$ there are manifestly $k-1$ zeroes at $z_m=z_1,..,z_{m-1},z_{m+1},..z_k$ on both sides: on the lhs because determinant vanishes due to coincident rows, and on the rhs because $\theta_1(0)=0$. Since lhs is a section of $L^k\otimes L_\varphi$ for $z_m$, there is an extra hidden zero, which is located exactly where the center-of-mass piece on the rhs vanishes. The indirect argument for the extra zero on the lhs relies upon the general correspondence between the line bundles and divisors \cite{GH,Fay}. After that the overall $z$-independent factor can be fixed by computing the $PSL(2,\mathbb Z)$ action on both sides with the help of Eqns.\ (\ref{Ttr}-\ref{Ttr1}) and concluding that it must be a modular form with a prescribed behavior at infinity, i.e., a certain power of the Dedekind eta function.

The action of the $T$ and $S$ transformation preserves the spin structure, when $\varepsilon=\delta=\lambda=0$ or $\varepsilon=\delta=\lambda=\frac12$, according to the table Fig.\ 3. Since the integer QH state is not degenerate, in both cases the action is given by the phase factors,
\begin{equation}\nonumber
T\circ\mathcal S=e^{\frac{\pi i}{12}(k^2+2)}\mathcal S,\quad S\circ\mathcal S=e^{-\frac{\pi i}4(k^2-k+2)}\mathcal S,
\end{equation}
which follows immediately from (\ref{T}, \ref{S}).

\subsection{Generating functional}

In quantum mechanics the main objects are wave functions, which have to be  $L^2$ normalized.
Usually one can explicitly find normalization constants for one-particle LLL states and for the IQHE state on the surface with constant scalar curvature metric, like e.g. in the examples of plane, sphere and torus above. However our goal is to define IQHE state for an arbitrary metric $g$ and inhomogeneous magnetic field $B$. Thus we have two sets of geometric data: background metric $g_0$ and Hermitian metric $h_0^k$ on $L^k$ defining background magnetic field $B_0$ (e.g. constant scalar curvature metric and constant magnetic field) and an arbitrary metric $g=g_0+\p\bp\phi$ and Hermitian metric $h^k=h_0^ke^{-k\psi}$, related to the background data exactly as in \eqref{gmetric} and \eqref{magneticp}, so that $g_0$ and $g$ and $F_0$ and $F$ are in the same \kahler class. The surface $\Sigma$ and the line bundle $L^k$ is the same in both cases, only the corresponding metrics are changed, so the number of LLL states does not change. 

Now, in order to construct the IQHE state for arbitrary $g$ and $B$ we start with a basis of one-particle states, normalized with respect to the background metrics, as in \eqref{normal}. Then we write the norm-squared of the IQHE state as
\begin{equation}\label{iqhewf}
|\Psi[g,B](z_1,...,z_N)|^2=\frac1{Z_k}\frac1{N!}|\det[s_l(z_m)]|^2\prod_{l=1}^{N}h^k(z_l,\bz_l) g_{z\bz}^{-s}\sqrt gd^2z_l,
\end{equation}
where the point-wise Hermitian norm of Slater determinant is taken with respect to the curved metric \eqref{gmetric} and the potential \eqref{magneticp}. In order to make this state $L^2$ normalized $\langle\Psi,\Psi\rangle_{L^2}=1$, we included the normalization factor $Z_k$, given by the following integral
\begin{equation}\label{Z1}
Z_k=\frac1{(2\pi)^{N}N!}\int_{\Sigma^{N}}|\det[s_l(z_m)]|^2\prod_{l=1}^{N}h^k(z_l,\bz_l) g_{z\bz}^{-s}\sqrt gd^2z_l,
\end{equation}
which we will call the partition function.
This is a functional of $Z_k=Z_k[h_0,g_0,\psi,\phi]$, or equivalently $Z_k=Z_k[B_0,g_0,B,g]$. However we note that the normalized IQHE state $|\Psi[g,B](z_1,...,z_N)|^2$ depends only on the metrics $g$ and inhomogeneous $B$ and not on $g_0$ and $B_0$. Indeed, under the change of the background metric $g_0\to g_0'$ the basis of sections transforms linearly $s\to s'=As$, where $A\in GL(N,\mathbb C)$. It follows immediately that dependence on $A$ cancels out between the numerator and denominator in \eqref{iqhewf}, hence the normalized IQHE state is independent of the choice of background metrics and of the choice of the basis of sections.

The logarithm of partition function $\log Z_k$ is called the generating functional and is the main object of our interest, since it contains a wealth of information. For example, it generates the density-density connected correlation functions, produced by variations wrt $\psi$,
\begin{equation}\label{dens}
\frac{\delta}{\delta\psi(w_1,\bar w_1)}\cdots 
\frac{\delta}{\delta\psi(w_m,\bar w_m)}\log Z_k
=(-k)^m\big\langle\rho(w_1,\bar w_1)...\rho(w_m,\bar w_m)
\big\rangle_{\rm conn},
\end{equation}
where the density of states operator reads $\rho(z,\bz)=\sum_{l=1}^{N} \delta(z,z_l)$. 

The remarkable property of the generating functional is the existence of the asymptotic expansion for large magnetic field, i.e., large $k$ expansion. This expansion can be derived as follows \cite{K,KMMW}. First, we write \eqref{Z1} in the determinant form
\begin{equation}\label{detfor}
Z_k=\det\int_{\Sigma}\bs_l(\bz)s_m(z)h^k(z,\bz) g_{z\bz}^{-s}\sqrt gd^2z,
\end{equation}
by the same combinatorial identity used previously in Eq.\ \eqref{detform}. Next, we use the following variational method (suggested in Ref.\ \cite{Don2} in a different context). Denoting the matrix inside $\det$ in \eqref{detfor} as $G_{lm}$, we can write for the variational derivative of $\log Z_k$ with respect to $\delta\phi$ and $\delta\psi$,
\begin{align}\label{freeen}
&\delta \log Z_k=\delta\,\tr\log G_{lm}\\\nonumber
&=-\frac1{2\pi}\sum_{l,m} (G^{-1})_{ml} 
\int_\Sigma\left(\frac{s-1}2(\Delta_g\delta\phi)
+k\delta\psi\right)\bs_ls_mh^kg_{z\bz}^{-s}\sqrt gd^2z\\\nonumber&
=-\frac1{2\pi}\int_\Sigma \left(\frac{s-1}2(\Delta_g B_k(z,\bz))\,
\delta\phi+kB_k(z,\bz)\,\delta\psi\right)\sqrt gd^2z.
\end{align}
The function $B_k(z,\bz)$ here
\begin{equation}\label{bergman1}
B_k(z,\bz)=\sum_{l,m} (G^{-1})_{ml}\bs_l(\bz)s_m(z)h^k(z,\bz)g_{z\bz}^{-s}(z,\bz),
\end{equation}
is called the Bergman kernel, which is a well known object in the theory of holomorphic line bundles.
Its physical meaning becomes apparent when with the help of \eqref{dens},\eqref{freeen} we obtain
\begin{equation}\nonumber
\langle\rho(z,\bz)\rangle=\frac{1}{2\pi}B_k(z,\bz).
\end{equation}
Hence the Bergman kernel is the average density of particles. The asymptotic expansion of $\log Z_k$ will follow from the asymptotic expansion of the Bergman kernel. Thus we now review the Bergman kernel expansion and then resume from Eq.\ \eqref{freeen} the derivation of the asymptotic expansion of $\log Z_k$ in \S \ref{agf}.
 
\subsection{Bergman kernel and density of states}

Bergman kernel on the diagonal  for the line bundle $L^k\otimes K^s$ has a straightforward interpretation as the density of states function on the LLL. Indeed, we write the Hermitian matrix in Eq. \eqref{bergman1} as $G^{-1}=AA^+$ for $A\in GL(N,\mathbb C)$, the basis $s'_l=A_{ml}s_m$ (up to $U(N)$ rotation) becomes orthonormal with respect to $h^k,g$, and we can write the Bergman kernel as the sum over the orthonormal LLL ground states
\begin{equation}\nonumber
B_k(z,\bz)=\sum_{l=1}^{N} \bs'_l(\bz)s'_l(z)h^k(z,\bz)g_{z\bz}^{-s}(z,\bz).
\end{equation}
They key fact about the Bergman kernel is the existence of the complete asymptotic expansion for large degree of the line bundle $k$ on \kahler manifold of any dimension, which was proven in \cite{Z,C}. In complex dimension, i.e. for the Riemann surfaces, the first few terms in the expansion can be read off from \cite[Eq.\ (44)]{KMMW},
\begin{align}\label{expansion}
B_k(z,\bz)=\,&B+\frac{1-2s}4R+\frac14\Delta_g\log B+
\frac{2-3s}{24}\Delta_g(B^{-1}R)\\\nonumber
&+\frac1{24}\Delta_g(B^{-1}
\Delta_g\log B)
+\mathcal O(1/k^2)\,.
\end{align}
The expansion involves only magnetic field and curvature invariants and their covariant derivatives. It goes in the inverse powers of the magnetic field $B$. For this reason we formally should require $B>0$ everywhere on $\Sigma$ and also that $B$  is of order $k$. Also the potentials $\phi,\psi$ here are $\mathcal C^\infty(\Sigma)$ functions on $\Sigma$. In particular, the expansion in the form of Eq.\ \eqref{expansion} in general breaks down near singularities of the curvature and the magnetic field. 

The asymptotic expansion Eq.\ \eqref{expansion} can be derived by quantum mechanical methods \cite{DK}. The density of states on the LLL can be represented as the path integral
\begin{equation}\nonumber
B_k(z,\bz)=\lim_{T\to\infty}\int_{x(0)=z}^{x(T)=z}e^{-\frac1\hbar\int_0^T(g_{a\bar b}\dot x^a\dot{\bar x}^{\bar b}+A_a\dot x^a+A_{\bar b}\dot{\bar x}^{\bar b})dt}\,\mathcal Dx(t)
\end{equation} 
for a particle in the magnetic field $F=dA$ on $\Sigma$ (more generally, on a \kahler manifold of complex dimension $n\geqslant1$), here at spin $s=0$.  The $T\to\infty$ limit projects the density of states to the lowest Landau level. The large $k$ expansion can be derived via perturbation theory techniques \cite{DK}, and the Planck constant $\hbar$ enters as the order-counting parameter in \eqref{expansion}.

There exists a closed formula for the coefficients of the Bergman kernel to all orders in $k$ in any complex dimension $n$, see Ref.\ \cite{Xu}. Let us illustrate this formula for the case when $s=0$ and magnetic field is constant $B=k$. We consider a local normal coordinate system around a point $z_0$, where 
\begin{equation}\nonumber
g_{i\bar j}(z_0)=\delta_{i\bar j},\quad g_{i\bar j_1...\bar j_m}=g_{i_1 \bar j i_2... i_m}=0.
\end{equation} 
The $m$th term $a_m$ in the expansion of the Bergman kernel
\begin{equation}\nonumber
B_k(z,\bz)=a_0(z)k^n+a_1(z)k^{n-1}+... a_m(z)k^{n-m}+...
\end{equation} 
involves exactly $2m$ derivatives of the metric $g_{i\bar j}$ in the local coordinate system. For example the order $4$ term will involve the structures as e.g. $g^{i_1\bar j_2}g^{i_2\bar j_1}g^{k_1\bar l_2}g^{k_2\bar l_1}g_{i_1\bar j_1 k_1\bar l_1}$ $g_{i_2\bar j_2 k_2\bar l_2}$. In general one can associate a directed graph $G$ to the structures of this kind, where the positions of $g_{i_1\bar j_1 k_1\bar l_1}$ and $g_{i_2\bar j_2 k_2\bar l_2}$ are represented by vertices and contractions with respect to $g^{i_1\bar j_2}$, etc., are represented by directed arrows between the vertices. At each vertex the number of incoming and outgoing vertices is at least $2$.
The local coefficient $a_m$ is then given by the sum
\begin{equation}\nonumber
a_m(z)=\sum_{G\in G(m)}z(G)\cdot G
\end{equation}
over the set of all such not necessarily connected graphs $G(m)$ at level $m$. The remarkable fact is that coefficients $z(G)$ are given by easily computable formulas.
For strongly connected graphs (when there exists a directed path from each vertex in $G$ to every other vertex)  
\begin{equation}\nonumber
z(G)=-\frac{\det (A-I)}{Aut(G)},
\end{equation}
where $A$ is the adjacency matrix of the graph. For connected but not strongly connected graphs $z(G)=0$ and for disconnected sum of $p$ subgraphs $G_j$ the coefficient is given by
\begin{equation}\nonumber
z(G)=\prod_{j=1}^pz(G_j)/|{\rm Sym}(G_1,...,G_p)|,
\end{equation}
where ${\rm Sym}(G_1,...,G_p)$ is the permutation group of these subgraphs. The hard part of the calculation (at high orders of $m$) is to transfer the expressions of the type $g^{i_1\bar j_2}g^{i_2\bar j_1}g^{k_1\bar l_2}g^{k_2\bar l_1}g_{i_1\bar j_1 k_1\bar l_1}g_{i_2\bar j_2 k_2\bar l_2}$ in the normal coordinate system back to the invariant form involving scalar curvature, Ricci and Riemann tensors and their derivatives, see Ref. \cite{XY} for the state of the art. 

\subsection{Anomalies and geometric functionals}
\label{agf}

Going back to the variational formula \eqref{freeen} and plugging the expansion Eq.\ \eqref{expansion} we can now integrate it, imposing the boundary condition $\log Z_k[\phi=0,\psi=0]=0$. The calculation was performed for the constant magnetic field in Ref.\ \cite{K} and generalized to inhomogeneous magnetic fields in Ref.\ \cite[Thm.\ 1]{KMMW}. We refer to these papers for more details and here we only state the result. The asymptotic expansion has the following general form
\begin{equation}\label{thm1exp}
\log Z_k=\log \frac{Z_H}{Z_{H0}}+\mathcal F-\mathcal F_0,
\end{equation}
where $\log Z_H$ is the "anomalous part" of the expansion and $\mathcal F$ is the "exact part". The former consists of only three terms
\begin{align}\label{ano}
\log Z_H-\log Z_{H0}=&-k^2S_2(g_0,B_0,\phi)+k\frac{1-2s}2 S_1(g_0,B_0,\phi,\psi)\\\nonumber&
-\left(\frac1{12}-\frac{(1-2s)^2}4\right)S_L(g_0,\phi),
\end{align}
where the following functionals appear
\begin{align}\label{func1}
S_2(g_0,B_0,\psi)=\frac1{2\pi}\int_\Sigma&\left(\frac14\psi\Delta_0\psi
+\frac1k B_0\psi\right)\sqrt{g_0}d^2z,\\\label{func2}
S_1(g_0,B_0,\phi,\psi)=\frac1{2\pi}\int_\Sigma&\left(-\frac12\psi R_0\right.\\\nonumber&
\left.+\bigl(\frac1k B_0+\frac12\Delta_0\psi\bigr)
\log\bigl(1+\frac12\Delta_0\phi\bigr)\right)\sqrt{g_0}d^2z,\\
\label{func3}
S_L(g_0,\phi)=\frac1{2\pi}\int_\Sigma&\left(-\frac14\log\bigl(1
+\frac12\Delta_0\phi\bigr)\,\Delta_0\log\bigl(1
+\frac12\Delta_0\phi\bigr)\right.\\\nonumber
&\left.+\frac12R_0\log\bigl(1
+\frac12\Delta_0\phi\bigr)\right)\sqrt{g_0}d^2z.
\end{align}
These are geometric functionals, which do not have a local expression in terms of the metric and magnetic field, and thus physically they correspond to anomaly terms, as we explain below. The last functional is the Liouville action $S_L$, and the first two are certain energy functionals well-known in \kahler geometry, see below.

The exact part $\mathcal F=\mathcal F[g,B]$ and $\mathcal F_0=\mathcal F[g_0,B_0]$ consists of infinitely many terms, which are local integrals of the magnetic field and curvature and their derivatives. Terms up to the order $\mathcal O(1)$ in $1/k$ read
\begin{align}\label{exact1}\nonumber
\mathcal F[g,B]=-\frac1{2\pi}\int_\Sigma &\left[\frac12B\log \frac{B}{2\pi}
+\frac{2-3s}{12}R\log\frac{B}{2\pi}\right.\\&\left.+\frac1{24}(\log B)\Delta_g(\log B)\right]
\sqrt gd^2z+\mathcal O(1/k).
\end{align}

Written in the form (\ref{func1}-\ref{func3}) the meaning of the anomalous action is not completely manifest. In order to make it transparent we now rewrite it in two equivalent forms. First, as the following double integral
\begin{align}\nonumber
\log Z_H=&-\frac1{2\pi}  \int_{\Sigma\times\Sigma} \left(B+\frac{1-2s}4R\right)\big|_z
\Delta_g^{-1}(z,y)\left(B+\frac{1-2s}4R\right)\big|_y\sqrt g d^2z\sqrt g d^2y\\&\label{di}
+\frac1{96\pi} \int_{\Sigma\times\Sigma} R(z)\Delta_g^{-1}(z,y)R(y)\sqrt g d^2z\sqrt g d^2y,
\end{align}
where the operator $\Delta_g^{-1}$ is formally defined as the inverse Laplacian $\Delta_g^z\Delta_g^{-1}(z,y)=\delta(z,y)$. The second term here is the gravitational anomaly in the form of the Polyakov effective action $\int_{\Sigma\times\Sigma} R\Delta^{-1}R$, see Ref.\ \cite{P1}. 

Yet another form uses the symmetric gauge \eqref{symgauge}, where by integration by parts we can rewrite $\log Z_H$ as
 \begin{align}\label{cs}
\log Z_H= \frac2{\pi}\int_\Sigma\left[A_zA_{\bz}
+\frac{1-2s}2(A_z\omega_{\bz}+
\omega_zA_{\bz})-\left(\frac1{12}-\frac{(1-2s)^2}4\right)
\omega_z\omega_{\bz}\right]d^2z.
\end{align}
Here we see, that the first term corresponds to 2d $U(1)$ gauge anomaly, the last term is the gravitational anomaly and the middle term is the mixed anomaly, known as Wen-Zee term \cite{WZ}, see also \cite{CF}. This form of the generating functional is a 2d avatar of the Chern-Simons action which will appear later \S \ref{cs-form}.

Two interesting special cases of Eq.\ \eqref{thm1exp} correspond to the two natural choices of magnetic field $B$ on $\Sigma$: $B=$const and $B={\rm const}\cdot R$. 

{\it Conformal regime.} For any $\Sigma$ of genus $\rm g \neq1$ we can choose magnetic field to be proportional to the scalar curvature $B=\frac{k}{4(1-{\rm g})}R$. For non-constant $R$, in order to keep $B$ positive we should require $R>0$ everywhere on $\Sigma$ for the $\Sigma=S^2$ and $R<0$ everywhere on $\Sigma$ for higher genus surfaces $\rm g>1$. In this case the anomalous part combines into one term
\begin{align}\nonumber
\log Z_H=&\frac1{96\pi}(1-3Q^2) \int_{\Sigma\times\Sigma} R(z)\Delta_g^{-1}(z,y)R(y)\sqrt g d^2z\sqrt g d^2y.
\end{align}
This formally corresponds to gravitational anomaly in a CFT with central charge $c=1-3Q^2$ with a very large background charge $Q=\frac k{1-{\rm g}}+1-2s$.

{\it K\"ahler regime}. This is the case of constant magnetic field $B=k$ and arbitrary metric $g$. As was pointed out  in Eq. \eqref{constmagn} this means the K\"ahler and magnetic potentials are equal $\phi=\psi$, possibly up to an irrelevant constant. This case was considered  in Ref.\ \cite{K}. In this case the first two terms in the expansion of the anomalous part 
\begin{align}\label{anok}
&\log Z_H-\log Z_{H0}=\\\nonumber&-kNS_{AY}(g_0,\phi)+k\frac{1-2s}2 S_M(g_0,\phi)
-\left(\frac1{12}-\frac{(1-2s)^2}4\right)S_L(g_0,\phi),
\end{align}
reduce to the Aubin-Yau and Mabuchi functionals, ubiquitous in \kahler geometry see e.g. \cite{PS} for review. These are defined by their variational formulas 
\begin{align}\label{func4}
&\delta S_{AY}(g_0,\phi)=\frac1{2\pi}\int \delta\phi\sqrt gd^2z,\\\nonumber&\delta S_{M}(g_0,\phi)=\frac1{4\pi}\int \delta\phi\,(2\chi(\Sigma)-R)\sqrt gd^2z,
\end{align}
and explicit formulas can be given 
\begin{align}\label{func5}
&S_{AY}(g_0,\phi)=S_2(g_0,kg_0,\phi),\\\nonumber&S_{M}(g_0,\phi)=\chi(\Sigma)S_2(g_0,kg_0,\phi,\phi)+S_1(g_0,kg_0,\phi,\phi),
\end{align}
in terms of functionals defined in Eq.\ \eqref{func1} and \eqref{func2}.

\subsection{Regularized spectral determinant}

Breaking up the generating functional into the anomalous and exact parts \eqref{thm1exp} has an interesting interpretation in terms of regularized determinants of spectral laplacian and Quillen metric, which we now recall following \cite{KMMW}. 

We consider the $\bp_{\rm L}$ operator \eqref{dbareq} and its adjoint $\bp^*_{\rm L}:\;\Omega^{0,1}(\Sigma,\rm L)\to \mathcal C^\infty(\Sigma,\rm L)$ under the $L^2$ inner product \eqref{norm}. Thus we can define the laplacian acting on sections of the line bundle $\rm L$ (Dolbeault laplacian):
\begin{equation}\label{dolb}
\Delta_{\rm L}=\bp^*_{\rm L}\bp_{\rm L}:\;\mathcal C^\infty(\Sigma,\rm L)\to \mathcal C^\infty(\Sigma,\rm L).
\end{equation} 
This operator, which is just the kinetic term in the one-particle hamiltonian \eqref{H}, is sometimes called the "magnetic laplacian". The regularized spectral determinant of this laplacian can be defined in the usual way. We consider the non-zero eigenvalues $\lambda$ of $\Delta_{\rm L}$ taken with multiplicities and define the zeta-function $\zeta(u)=\sum_\lambda\lambda^{-u}$. Then $\det'\Delta_{\rm L}=\exp(-\zeta'(0))$.

The relation to our setup is as follows. The holomorphic part of the integer QH state \eqref{slater} is a section $\mathcal S$ of the determinant line bundle $\mathcal L=\det H^0(\Sigma,L^k\otimes K^s)$ over the parameter space $Y=\mathcal M_{\rm g}\times Jac(\Sigma)$, Eq.\ \eqref{parsp}. Quillen defined a Hermitian metric on sections $\mathcal S$ of $\mathcal L$ as follows
\begin{equation}\label{quillenmet}
||\mathcal S||^2=\frac{Z_k}{\det'\Delta_{\rm L}}.
\end{equation}
Using the determinantal formula Eq.\ \eqref{detfor} the previous formula can also be written as 
\begin{equation}\nonumber
||\mathcal S||^2=\frac{\det\langle s_l,s_m\rangle_{L^2}}{\det'\Delta_{\rm L}}.
\end{equation}
Note that the Quillen metric \eqref{quillenmet} is defined with respect to a choice of $h^k$ and $g$.
Now we ask how $||\mathcal S||^2$ varies under the variations \eqref{gmetric} and \eqref{magneticp} of these metrics. The following exact formula holds 
\begin{equation}\label{thm21}
\log\frac{||\mathcal S||^2}{||\mathcal S||_0^2}=\log\frac{Z_H}{Z_{H0}},
\end{equation}
where $||\mathcal S||^2_0$ is defined for $(h_0^k,g_0)$. In other words, the anomalous part \eqref{ano} of the transformation formula for the partition function \eqref{thm1exp} is entirely due to the ratio of $Z_k$ and determinant of laplacian that enters the Quillen metric \eqref{quillenmet}. As an immediate consequence we see that the exact part $\mathcal F$ \eqref{exact1} corresponds to the regularized determinant of laplacian
\begin{equation}\label{thm22}
\mathcal F-\mathcal F_0=\log\frac{\det'\Delta_{\rm L}}{\det'\Delta_{\rm L0}},
\end{equation}
where $\Delta_{\rm L}$ is defined wrt $(h^k,g)$ and $\Delta_{\rm L0}$ wrt $(h_0^k,g_0)$.
The proof of \eqref{thm21} and \eqref{thm22} is a standard heat kernel calculation, see \cite[Thm. 2]{KMMW} for details. On the compact Riemann surfaces the regularized determinants are known explicitly. On the round sphere with the metric \eqref{roundsp} and magnetic potential \eqref{roundsp1} we have an exact formula,
\begin{align}\label{detsph}
 \log {\det}'\Delta_{L^k}
 &=2\sum_{j=1}^{k} (k-j) \log (j+1)-(k+1)\log(k+1)!
 \\ \nonumber
 &-4\zeta'(-1)+\frac{(k+1)^2}2=-\frac k2\log\frac k{2\pi}-\frac23\log k
 +\mathcal O(1),
\end{align}
see \cite[\S 4]{KMMW} and references therein.
This is in perfect agreement with Eq.\ \eqref{exact1} at $s=0$. On the flat torus with constant magnetic field we have 
\begin{align}\label{dettor}
\log {\det}'\Delta_{L^k}=-\frac k2\log\frac k{2\pi},
\end{align}
valid for any $k>0$, see \cite{Ber01}. This is also consistent with Eq.\ \eqref{exact1}.  
In \S  \ref{quillenano} we will consider the case of higher-genus surfaces, and the Quillen metric will turn out to be a useful tool for the study of the geometric adiabatic transport on the moduli space.

\section{Laughlin states on Riemann surfaces}
\label{lsrs}
\subsection{Definition of the Laughlin state}

We consider $N$ particles, labelled by their positions $z_1,...,z_N$, confined to the plane in the perpendicular constant magnetic field $B$. The Laughlin state in this setup was introduced in Ref. \cite{L},
\begin{equation}\label{LS}
\Psi(z_1,...,z_N)\sim\prod_{l<m}(z_l-z_m)^\beta \cdot e^{-\frac14B\sum_{l=1}^N|z_l|^2}, \quad\beta\in\mathbb Z_+,
\end{equation}
up to a normalization factor.
These states are associated with the Quantum Hall plateaux with the values of the Hall conductance  $\sigma_H=1/R_{xy}=1/\beta$. The graph Fig.\ 1 includes only one such state, labelled $1/3$. In this section we will  define and construct the Laughlin state \eqref{LS} on Riemann surfaces, and in the next section we come back to the relation between the Hall conductance and $\beta$ via geometric adiabatic transport.

At $\beta=1$ the Laughlin state \eqref{LS} reduces to the integer QH state \eqref{intplane}, which corresponds to free fermions. For $\beta>1$ the Laughlin state takes into account Coulomb interactions between the electrons. However, it is not an exact ground state of the full interacting Hamiltonian. Nevertheless, there exists a model Hamiltonian with the short-range interaction potential, see e.g.\ Refs. \cite{H} and \cite{Gir,Wen}, for which the state \eqref{LS} is an exact ground state with zero energy,
\begin{equation}\label{modelH}
H=\sum_{l=1}^ND_l\bar D_l+\sum_{l,m}V_\beta(z_l-z_m),\quad V_\beta(z)=\sum_{p=1}^{\beta-1}(-1)^pv_p\p_{\bz}^p\delta^2(z)\p_z^p,
\end{equation}
where $v_p$ are arbitrary positive constants and kinetic term is the sum over one particle operators \eqref{H}. 

In order to define the Laughlin state on a compact Riemann surface, we first note that \eqref{modelH} is already written in a covariant fashion, applicable to Riemann surfaces with arbitrary metric and magnetic field. The most important features of \eqref{LS} is that it consists of the holomorphic function of coordinates $F(z_1,...,z_N)$ and overall gaussian factor, depending on magnetic field. In order to minimize the kinetic energy in \eqref{modelH} written on a compact Riemann surface, for each coordinate $z_m$ the function $F(...,z_m,...)$ should transform as a holomorphic section of the line bundle $L^{\NPhi}$ of degree $\NPhi$ on $\Sigma$, where $\NPhi$ is the total flux of the magnetic field, as follows from Eq. \eqref{lll} and discussion in \S\ref{holosec}. Also due to compactness the number of particles and total flux are related,
\begin{equation}\label{frac}
\NPhi\approx\beta N. 
\end{equation}
Indeed, on a compact surface $\NPhi$ is integer and the allowed number of LLL states ($\dim H^0(\Sigma,L^{\NPhi})$) is of order $\NPhi$ \eqref{Nk}. Therefore the number of points in Eq.\ \eqref{LS} equals to $1/\beta$ times the number of allowed LLL states, while in the integer QH state \eqref{intplane} the number $N$ of particles was exactly equal to the number of LLL states. Thus one could say that only a $1/\beta$ fraction of all available LLL states is activated in the Laughlin state ("fractional" Quantum Hall effect). This interpretation is not a precise statement, but only an analogy, since in general the Laughlin state cannot be constructed from one-particle LLL states.

The magnetic field flux on a compact surface is quantized $\NPhi\in\mathbb Z$ \eqref{flux}, and without loss of generality we can write $\NPhi=\beta k+p,\; k\in\mathbb Z_+,\; p=0,...,\beta-1$,
where $p$ is the remainder of division of $\NPhi$ by $\beta$. The Laughlin state will be defined specifically at $p=0$. The wave functions corresponding to $p>0$ describe quasi-hole excitations over the Laughlin state \cite{L}, which are extremely important in physics of QHE, but we here we will focus only on  Laughlin states $p=0$. 

Now we have to choose the number of particles $N$ (of order $\sim \NPhi/\beta$), thus fixing the exact relation between $k$ and $N$. At this point in the discussion we can also turn on the power of canonical line bundle $K^{\ij}$, which we take to be quantized as $\ij=\beta s,\,s\in\frac12\mathbb Z$. To make the formulas consistent with the integer QH case we will choose the number of particles exactly as before \eqref{Nk} at $\beta=1$ (and $q=0$), 
\begin{equation}\label{N1}
N=\dim H^0(\Sigma,L^k\otimes K^s)=k+(1-{\rm g})(1-2s),\quad s=\frac{\ij}{\beta}.
\end{equation}
Thus on the Riemann surface the relation between the flux and number of particles is
\begin{equation}\label{numberpart}
\NPhi=\beta N-{\mathsf S},\quad {\rm where}\quad\mathsf S=(1-{\rm g})(\beta-2\ij),
\end{equation} 
generalizing the planar relation \eqref{frac}. Here $\mathsf S$ is usually called the shift in the QHE literature \cite{WZ}.  

Given this data we can now define a Laughlin state on a compact Riemann surface $\Sigma$. 
\begin{definition}\label{def1}{\it
Consider the holomorphic line bundle ${\rm L}=L^{\NPhi}\otimes K^\ij\otimes L_{\varphi}$ on a compact Riemann surface $\Sigma$. Let  $\NPhi=\beta k$ and $\ij=\beta s$ as in Eq. \eqref{N1}. Take $N$ points $z_1,...,z_{N}$ on $\Sigma$, where $N$ is given by Eq. \eqref{N1} and $\NPhi$ and $N$ are related as in Eq.\ \eqref{numberpart}. Then the (holomorphic part of the) Laughlin state on $\Sigma^N$ is $F(z_1,...,z_N)$, satisfying the following conditions
\begin{enumerate}
\item $F(...,z_m,...)$ with all coordinates, except $z_m$, fixed, transforms as a holomorphic section of $L^{\NPhi}\otimes K^\ij\otimes L_{\varphi}$. 
\item When all $z_l$'s are near the diagonal in $\Sigma^N$ at a generic point on $\Sigma$, $F(z_1,...,z_N)$ satisfies the vanishing condition
\begin{equation}\label{FL}
F(z_1,...,z_N)\sim\prod_{l<m}^N(z_l-z_m)^\beta.
\end{equation}
\end{enumerate}}
\end{definition}
As we have already mentioned, the first condition ensures the vanishing of the kinetic term, while the second condition guarantees minimization of the short-range pseudopotential in \eqref{modelH}. The Laughlin states on $\Sigma$ are degenerate for $\rm g>0$. This fact is usually referred to as the topological degeneracy. 

So far Def.\ \ref{def1} defines only the holomorphic part of the wave function. Thus we need to define the Hermitian norm, which is induced from the choice of the Hermitian metric on the line bundle. As in Eq.\ \eqref{F}, let $h^{\NPhi}(z,\bz)$ be an Hermitian metric on $L^{\NPhi}$ so that the magnetic field is given by
\begin{equation}\nonumber
F_{z\bz}=-(\p_z\p_{\bz}\log h^{\NPhi}), \quad B=g^{z\bz}F_{z\bz},\quad \NPhi=\frac1{2\pi}\int_\Sigma B\sqrt gd^2z.
\end{equation}
The natural Hermitian metric for the holomorphic part of the Laughlin state is the point-wise product of $h^{\NPhi}(z,\bz)$ on $\Sigma^N$,
\begin{equation}\label{hermnormL}
||F(z_1,...,z_N)||^2=|F(z_1,...,z_N)|^2 \prod_{l=1}^Nh^{\NPhi}(z_l,\bz_l)g_{z\bz}^{-\ij}(z_l,\bz_l).
\end{equation}
This is  a scalar function on $\Sigma^N$ and we can thus compute the $L^2$ norm and write the normalized Laughlin state $\Psi$ as follows
\begin{align}\nonumber\label{normLaugh}
&|\Psi(z_1,...,z_N)|^2=\frac1{\mathcal N}||F(z_1,...,z_N)||^2,\\
&\langle \Psi|\Psi\rangle_{L^2}=\frac1{\mathcal N}\frac1{(2\pi)^N}\int_{\Sigma^N} ||F(z_1,...,z_N)||^2\prod_{\l=1}^N\sqrt gd^2z_l=1.
\end{align}
The normalization factor $\mathcal N$ is a functional of the geometric data: the metric $g$, magnetic field $B$, complex structure moduli $J$ of $\Sigma$ and line bundle moduli $\varphi\in Jac(\Sigma)$, $\mathcal N=\mathcal N[g,B,J,\varphi]$.

\subsection{Examples}
\label{topdeg}

{\it Sphere}. The spherical Laughlin state was constructed in Ref.\ \cite{H}.
On the sphere the Laughlin state is unique. This is easy to see since, $\prod_{l<m}(z_l-z_m)^\beta$ is the only combination, which meets both conditions in Def.\ \ref{def1}. As we have already emphasized, the definition of the Laughlin state does not make any reference to the lowest Landau level wave functions. However, specifically for the sphere we can express the Laughlin state as a power of the Slater determinant, for the basis of LLL states Eq.\ \eqref{spherebasis}, as
\begin{align}\label{Laughsph}\nonumber
&F(z_1,...,z_N)=(\det s_l(z_m))^\beta,\\
&|\Psi(z_1,...,z_N)|^2=\frac 1{\mathcal N_0}\cdot |\det s_l(z_m)|^{2\beta} \prod_{l=1}^N h_0^{\NPhi}(z_l,\bz_l)g_0^{-\ij}(z_l,\bz_l)\\\nonumber
&=\frac 1{\mathcal N_0}\prod_{l=1}^Nc_l^{2\beta}\cdot\prod_{l<m}^N|z_l-z_m|^{2\beta}\prod_{l=1}^N\frac1{(1+|z_l|^2)^{\NPhi-2\ij}},
\end{align}
where the number of particles is $N=k+1-2s$, $c_l$ is given in Eq.\ \eqref{spherebasis} and constant $\mathcal N_0$ is such that the $L^2$ norm is one: $\langle\Psi|\Psi\rangle_{L^2}=1$. 

{\it Torus}. Laughlin states on the torus were constructed in Ref.\ \cite{HR}. We also refer to  \cite{L1989,KVW,R,Wen} for other excellent accounts. On the torus the canonical line bundle is trivial and for this reason we set $\rm j=0$. 
The first condition in Def.\ \ref{def1} implies that under the lattice shifts the wave function transforms with the same factors of automorphy as in Eq.\ \eqref{torusmonod} for each coordinate $z_m$,
\begin{align}\label{shift2}
&F(z_1,...,z_m+t_1+t_2\tau,...,z_N)\\\nonumber&=(-1)^{2t_2\delta+2t_1\varepsilon}e^{-2\pi i \NPhi t_2z_m-\pi i \NPhi t_2^2\tau-2\pi it_2\varphi}\cdot F(z_1,...,z_m,...,z_N).
\end{align}
Here $\varepsilon,\delta\in\{0,\frac12\}$ label the choice of spin structures, which are independent of $m$ since the particles are identical.
To fulfil the second condition Eq. \eqref{FL}, without any loss of generality we can assume the ansatz 
\begin{align}\nonumber
F(z_1,...,z_N)\sim f(z_1,...,z_N)\prod_{l<m}^N\left(\vartheta_1(z_l-z_m,\tau)\right)^\beta,
\end{align}
since $\vartheta_1(z)$ has only one simple zero at $z=0$.
From the lattice shift transformation formula Eq.\ \eqref{shift1} for the product of theta functions in the previous equation it follows that in order to be consistent with Eq.\ \eqref{shift2}, the function $f$ should transform as
\begin{align}\label{eqfr}
&f(z_1,...,z_m+t_1+t_2\tau,z_N)\\\nonumber&=(-1)^{t_2(2\delta-\NPhi+\beta)+t_1(2\varepsilon-\NPhi+\beta)}e^{-i\pi \beta\tau t_2^2-2\pi i t_2\beta z_{\rm cm}-2\pi it_2\varphi}\cdot f(z_1,...,z_N).
\end{align} 
Comparing with Eq. \eqref{torusmonod} this condition essentially means that $f(z_1,...,z_N)=f(z_{\rm cm})$. Moreover, with respect to the center-of-mass coordinate $z_{\rm cm}$ the function $f$ transforms as a section of the line bundle $L^{\beta}$. Since $\dim H^0(\Sigma,L^\beta)=\beta$ the degeneracy of the center-of-mass factor equals $\beta$. Thus we can write down the basis of solutions $f_r, r=1,..,\beta$ to \eqref{eqfr} explicitly using e.g.\ the basis of sections in Eq.\ \eqref{basis}. We consider first odd values of $\beta$ and, by analogy with \eqref{lambda0}, we shall introduce the parity indicator parameter for the number of particles:
\begin{align}\label{lambda}
\lambda=\frac N2-\left[\frac N2\right]=\begin{cases}
0, & \text{for }N\in{\rm even}\\
\frac12, & \text{for }N\in{\rm odd}.\\
\end{cases}
\end{align}
Then the following basis solves the condition \eqref{eqfr}
\begin{equation}\nonumber
\beta\in{\rm odd}:\quad f_r(z_{\rm cm})=\vartheta\left[\begin{array}{c}\scriptstyle\frac{r+\varepsilon}\beta-\lambda+\frac12 \\{\scriptstyle\delta-\beta\lambda+\frac\beta2}\end{array}\right](\beta z_{\rm cm}+\varphi,\beta\tau), \quad r=1,...,\beta.
\end{equation}
Then the basis of Laughlin states reads
\begin{align}\label{Ltorus}
F^{\scriptstyle\varepsilon,\delta}_r(z_1,...,z_N)=&\vartheta\left[\begin{array}{c}\scriptstyle\frac{r+\varepsilon}\beta-\lambda+\frac12 \\{\scriptstyle\delta-\beta\lambda+\frac\beta2}\end{array}\right](\beta z_{\rm cm}+\varphi,\beta\tau)\prod_{l<m}^N\left(\vartheta_1(z_l-z_m,\tau)\right)^\beta,\\\nonumber
&{\rm where}\;\;\beta\in{\rm odd}.
\end{align}
Here index $r=1,...,\beta$ labels the topological degeneracy and $\varepsilon,\delta$ label the spin structure constants, i.e., the choice of boundary conditions Eq.\ \eqref{shift2}. As a consistency check, note that for $\beta=1$ we recover the integer QH state on the torus Eq.\ \eqref{ILtor}, up to a normalization constant, to which we will come back further down the road.

For $\beta\in$ even (and thus $\NPhi\in$ even) the state is bosonic, i.e., completely symmetric under exchange of the coordinates, and spin structures are redundant. In this case we set
\begin{equation}\nonumber
\beta\in{\rm even}:\quad f_r(z_{\rm cm})=\vartheta\left[\begin{array}{c}\scriptstyle\frac{r}\beta \\{\scriptstyle0}\end{array}\right](\beta z_{\rm cm}+\varphi,\beta\tau), \quad r=1,...,\beta.
\end{equation}
We will mostly focus here on odd values of $\beta$. In order to see that the states Eq.\ \eqref{Ltorus} indeed form an orthogonal basis we consider flat torus and constant magnetic field, and rewrite the point-wise Hermitian norm on $F_r$ \eqref{hermnormL} as follows 
\begin{align}\label{Fr2}
&||F_r||^2=|F_r(z_1,...,z_N)|^2\prod_{l=1}^Nh_0^{\NPhi}(z_l,\bz_l)\\\nonumber
&=|f_r(z_{\rm cm})|^2h_0^{\beta}(z_{\rm cm},\bz_{\rm cm})\cdot\prod_{l<m}^N\left|\vartheta_1(z_l-z_m,\tau)\right|^{2\beta}\cdot e^{\frac{\pi i\beta}{\tau-\bar\tau}(z_l-z_m-(\bz_l-\bz_m))^2},
\end{align}
where we used the Hermitian metric $h_0$ corresponding to the constant magnetic field \eqref{hermtor}.
The $L^2$ inner product for the flat metric $g_0$ \eqref{torusmetr} then reads
\begin{align}\nonumber
\langle F_r,F_{r'}\rangle=\frac1{(2\pi)^{\NPhi}}\int_{\Sigma^N}& \bar f_r(\bz_{\rm cm})f_{r'}(z_{\rm cm})h_0^{\beta}(z_{\rm cm},\bz_{\rm cm})\\\nonumber
&\cdot\prod_{l<m}^N\left|\vartheta_1(z_l-z_m,\tau)\right|^{2\beta}\cdot e^{\frac{\pi i\beta}{\tau-\bar\tau}(z_l-z_m-(\bz_l-\bz_m))^2}\prod_{l=1}^N\sqrt{g_0}d^2z_l.
\end{align}
Note that the dependence on the center-of-mass coordinate in the integrand decouples from the relative distances $z_l-z_m$ between the points. Thus, we can pass to the integration over $z_{\rm cm}$ and $z_1-z_m,\,m=2,...,N$, and note that the latter is independent of $z_{\rm cm}$. Now, as we already noticed, $f_r(z_{\rm cm})$ is an orthonormal basis of holomorphic sections of the line bundle $L^\beta$ over the center-of-mass, as can be seen from \eqref{basis}, formally replacing $k\to\beta$ and $z\to z_{\rm cm}$. Since this basis is orthogonal
\eqref{L2s}, we conclude that the overlap matrix of $L^2$ norms
\begin{align}\label{torusortho}
\langle F_r,F_{r'}\rangle=\mathcal N_0(\tau,\bar\tau,\varphi,\bar\varphi)\delta_{rr'},
\end{align}
is a scalar matrix, where the constant $\mathcal N_0$ is independent of the index $r$ and is a function of only $\tau$ and $\varphi$. This is the normalization factor in Eq. \eqref{normLaugh} specified to the flat torus with the constant magnetic field $\mathcal N_0=\mathcal N[g_0,B_0,\tau,\varphi]$.

{\it Higher genus}.
The number of degenerate Laughlin states on a higher genus surface is $n_{\beta,{\rm g}}=\beta^{\rm g}$ \cite{WN}. For more details on higher genus states we refer to \cite{CMMN,IL} and \cite{K2016}.

\subsection{Vertex operator construction}\label{vertexop}

Vertex operator construction of Laughlin states was originally proposed in Ref.\ \cite{MR}. Here we review the construction of Laughlin states on a Riemann surface $(\Sigma,g)$ of genus $\rm g$ following Refs.\ \cite{FK,KW}, and consider the example on sphere and torus in full detail. We start with the gaussian free field $\sigma(z,\bz)$, compactified on a circle $\sigma\sim\sigma+2\pi R_c$, with radius $R_c=\sqrt \beta$ (``compactified boson''). The action functional has the form
\begin{equation}\label{action0}
S(g,B,\sigma)=\frac1{2\pi}\int_\Sigma\bigl(\p_z\sigma\p_{\bz}\sigma+\frac{ i q}4\sigma R\sqrt g\bigr)d^2z+\frac i{2\pi\sqrt\beta}\int_\Sigma A\wedge d\sigma,
\end{equation}
where we will specify the constant $q$ later in Eq.\ \eqref{q}. The second term here is the coupling of the field to the Riemannian metric on $\Sigma$ and the last term is the coupling to the magnetic field. The latter is written in this form to take into account a possible contribution from nontrivial flat connection part $A^{\varphi}$, see Eq.\ \eqref{flatconn}, of the gauge connection on surfaces of genus ${\rm g}>0$. We can write this contribution explicitly, writing the gauge connection as $A+A^{\varphi}$ and using product rule for derivative
\begin{equation}\label{action1}
S(g,B,\sigma)=\frac1{2\pi}\int_\Sigma\bigl(\p_z\sigma\p_{\bz}\sigma+\frac{ i q}4\sigma R\sqrt g+\frac i{\sqrt\beta}\sigma B\sqrt g\bigr)d^2z+\frac i{2\pi\sqrt\beta}\int_\Sigma A^{\varphi}\wedge d\sigma.
\end{equation}
For inhomogeneous magnetic field and curved metric this action was proposed in QHE context in Refs.\ \cite{Kv,FK,KW}. We emphasize that, while at the zero magnetic field this theory is a conformal field theory with background charge $q$ and central charge $c=1-3q^2$, the coupling to the magnetic field breaks conformal invariance, since it introduces the magnetic length scale $l^2_B\sim1/B$ to the theory.  

We consider now the (unnormalized) correlation function of a $N$ vertex operators at points $z_1,...,z_N$. 
\begin{equation}\label{PI}
\mathcal V\bigl(g,B,\{z_l\}\bigr)=\int e^{ i\sqrt\beta\sigma(z_1)}\cdots e^{ i\sqrt\beta\sigma(z_N)}e^{-S(g,B,\sigma)}\mathcal D_g\sigma.
\end{equation}
Its integral  over the coordinates will be denoted as
\begin{equation}\label{PI1}
e^{\mathcal F_\beta(g,B)}=\frac1{(2\pi)^N}\int \int_\Sigma e^{ i\sqrt\beta\sigma(z_1)}\sqrt gd^2z_1\cdots \int_\Sigma e^{ i\sqrt\beta\sigma(z_N)}\sqrt gd^2z_N\;e^{-S(g,B,\sigma)}\mathcal D_g\sigma.
\end{equation}

The key observation is that the correlation function \eqref{PI} gives the sum over all normalized degenerate Laughlin states $\Psi^{\varepsilon,\delta}_r,\,r=1,...,n_{\beta,{\rm g}}$ and over all $2^{2\rm g}$ choices of spin structures $\vec\varepsilon,\vec\delta\in\{0,\frac12\}^{\rm g}$ on the Riemann surface $\Sigma$, 
\begin{equation}\label{PI2}
\frac1{2^{\rm g}\cdot n_{\beta,{\rm g}}}\sum_{\vec\varepsilon,\vec\delta}c_{\varepsilon,\delta}\sum_{r=1}^{n_{\beta,{\rm g}}}|\Psi^{\varepsilon,\delta}_r(z_1,...,z_N)|^2=e^{-\mathcal F_\beta(g,B)}\mathcal V\bigl(g,B,\{z_l\}\bigr),
\end{equation}
where the constant $c_{\varepsilon,\delta}=\pm1$ depending on the parity of spin structure\footnote{This holds for odd number of particles, for even number of particles the constant is  slightly different, see Eq.\ \eqref{sumwave}} 
\begin{equation}\nonumber
c_{\varepsilon,\delta}=e^{4\pi i\vec\varepsilon\cdot\vec\delta}=\begin{cases}
+1, & \text{for } (\vec\varepsilon,\vec\delta)\in{\rm even}\\
-1, & \text{for } (\vec\varepsilon,\vec\delta)\in{\rm odd}.\\
\end{cases}
\end{equation}
The number of even spin structures $2^{\rm g-1}(2^{\rm g}+1)$ minus the number of odd spin structures $2^{\rm g-1}(2^{\rm g}-1)$ equals $2^{\rm g}$, which explains the overall normalization factor in \eqref{PI2}. The formula above also holds on the sphere where $n_{\beta,{\rm g}}=1$ and spin structures are absent.

We now follow the standard prescription for computing bosonic path integrals, see e.g. \cite{VV,DP}. The field $\sigma$ has a constant zero mode $\sigma_0$, defined as 
\begin{equation}\nonumber
\sigma(z,\bz)=\sigma_0+\tilde\sigma(z,\bz),\quad\int\tilde\sigma\sqrt gd^2z=0.
\end{equation}
Then integration over the zero mode yields the relation between the number of particles $N$ and the coefficients in the action
\begin{equation}\label{Nrel}
N=\frac1\beta \NPhi+\frac{q}{2\sqrt\beta}\chi(\Sigma).
\end{equation}
Hence to be in accordance with \eqref{numberpart} we fix the parameter $q$ as
\begin{equation}\label{q}
q=\sqrt\beta-\frac{2\ij}{\sqrt\beta}.
\end{equation}
Since the number of particles is always an integer, it follows that $\frac1\beta(\NPhi-\ij\chi(\Sigma))$ should be an integer. Therefore we assume $\frac1\beta\NPhi\in\mathbb Z$  and $\ij\in\frac\beta2\mathbb Z$.

We consider now the path integral \eqref{PI} on the sphere, where we can drop the flat connection part and compactification of the field does not play any role since all 1-cycles are contractible and there are no non-trivial winding configurations of the field. Then the integral over $\tilde\sigma$ can be computed according to standard rules for gaussian integrals. We introduce the standard Green function for scalar laplacian
\begin{align}
\label{green}
&\Delta_g G^{g}(z,y)=-2\pi\delta(z,y)+1,\\\label{intgreen}
&\int_MG^{g}(z,y)\sqrt{g}d^2y=0,
\end{align}
and the regularized Green function at coincident point,
\begin{equation}
\label{reggreen}
G^{g}_{\rm reg}(z)=\lim_{z\to y}\bigl(G^{g}(z,y)+\log d_{g}(z,y)\bigr),
\end{equation}
where $d_{g}(z,y)$ is the geodesic distance between the points in the metric $g$.

Next, we can make a linear shift of  $\tilde\sigma$ without changing the measure of integration \begin{align}
\nonumber
&\tilde\sigma(z)\to\tilde\sigma(z)+\int_MG^{g}(z,z')j(z')\sqrt{g}d^2z',\\\nonumber
&j(z)=2 i\sqrt\beta\sum_{j=1}^{N}\delta(z,z_j)-\frac{ i q}{4\pi} R(z)-\frac i{\sqrt\beta} B(z).
\end{align}
Then the integral \eqref{PI} becomes purely gaussian and can be written as
\begin{align}\label{LPI0}
&\mathcal V\bigl(g,B,\{z_j\}\bigr)=\left[\frac{\det'\Delta_g}{2\pi}\right]^{-1/2}\\\nonumber
\cdot\exp\left(-\frac1{4\pi^2}\right.&\left.\int_{\Sigma\times\Sigma}\left(\frac q4R+\frac1{\sqrt\beta}B\right)\big|_zG^{g
}(z,z')\left(\frac q4R+\frac1{\sqrt\beta}B\right)\big|_{z'}\sqrt{g}d^2z\,\sqrt{g}d^2z'\right)\\\nonumber
\cdot\exp\left(\frac{\sqrt\beta}{\pi}\right.&\left.\sum_{l=1}^{N}\int_\Sigma G^{g}(z_l,z)\left(\frac q4R+\frac1{\sqrt\beta}B\right)\big|_z\sqrt{g}d^2z\right.\\\nonumber&\left.-\beta\sum_{l\neq m}^{N}G^{g}(z_l,z_m)-\beta\sum_{l=1}^{N}G^{g}_{\rm reg}(z_l)\right).
\end{align}
Here the regularized Green function $G^{g}_{\rm reg}(z_l)$ replaces $G^g(z_l,z_l)$ on the diagonal, where the latter is infinite.

Let us consider now the round sphere $R_0=4$ and constant magnetic field $B_0=\NPhi$. The round metric is given by Eq.\ \eqref{roundsp} and the corresponding Green function reads
\begin{align}\nonumber
G^{g_0}(z,z')=-\log\frac{|z-z'|}{\sqrt{(1+|z|^2)(1+|z'|^2)}}-\frac12.
\end{align}
The regularized Green function is just a constant $G_{\rm reg}^{g_0}(z)=-\frac12$. Due to the property Eq.\ \eqref{intgreen} the integrals over $\Sigma$ in Eq. \eqref{LPI0} vanish and we arrive at
\begin{align}\nonumber
\mathcal V\bigl(g_0,B_0,\{z_l\}\bigr)=e^{2\zeta'(-1)-\frac14-\frac12\beta N^2}\prod_{l<m}^N|z_l-z_m|^{2\beta}\prod_{l=1}^N\frac1{(1+|z_l|^2)^{\NPhi-2\ij}},
\end{align}
where we used the value of the regularized determinant of the laplacian on the round sphere $\det'\Delta_0/2\pi=e^{\frac12-4\zeta'(-1)}$ \cite{OPS}. Comparing this equation to \eqref{Laughsph}, we conclude that the normalized Laughlin state can be expressed as
\begin{align}\nonumber
|\Psi(z_1,...,z_N)|^2=e^{-\mathcal F_\beta(g_0,B_0)}\mathcal V\bigl(g_0,B_0,\{z_l\}\bigr)
\end{align}
and it has norm one by definition \eqref{PI1}. We also see that the normalization constant in Eq.\ \eqref{Laughsph} is controlled by $e^{\mathcal F_\beta(g_0,B_0)}$ up to numerical factors.

Finally, let us comment on the $\beta=1$ case of the formula \eqref{PI2}. At $\beta=1$ this construction reduces to the bosonisation formula on Riemann surfaces \cite{ABMNV,VV}. Bosonisation formula is the statement that the correlation function \eqref{PI} equals to the correlator of $N$ insertions of $b\bar b$-operators in the theory of free fermions $b,c$ with spins ${\rm j},1-{\rm j}$, see Eq.\ \cite[Eq.\ (3.1)$'$]{ABMNV}. The construction of Refs.\ \cite{ABMNV,VV} applies to the case of the canonical line bundle and no magnetic field, but it can be straightforwardly generalized to the case of line bundle ${\rm L}=L^k\otimes K^s$ (recall that at $\beta=1$, ${\rm j}=s,\,\NPhi=k$). The main statement is that
\begin{equation}\nonumber
\mathcal V_{\beta=1}\bigl(g,B,\{z_j\}\bigr)=\langle b(z_1)\bar b(z_1)... b(z_N)\bar b(z_N)\rangle=\frac{\det'\Delta_{\rm L}}{Z_k}||\det s_l(z_m)||^2,
\end{equation}
cf.\ \cite[Eq.\ (4.15)]{ABMNV}, where on the right hand side we recognize the Hermitian norm of the integer QH state \eqref{slater} and the Quillen metric \eqref{quillenmet}. Then from \eqref{PI2} we it follows that $\mathcal F_\beta$ at $\beta=1$ reduces to the logarithm of the spectral determinant of laplacian \eqref{dolb} for the line bundle $\rm L$,
\begin{equation}\label{Fdet}
\mathcal F_{\beta=1}=\log{\det}' \Delta_{\rm L}.
\end{equation}
In this sense, the formula \eqref{PI2} for the Laughlin states can be thought of as a $\beta$-deformation of bosonisation formulas on Riemann surfaces.

\subsection{Laughlin states on the torus from free fields}

The computation of the correlation function Eq.\ \eqref{PI} on the torus is a version of the standard computation in CFT \cite[Ch.\ 10]{DMS}, slightly modified to include the magnetic field. We go over this calculation here in order to account for non-homogeneous magnetic field and curved metric. For CFT-type calculation for the Laughlin states on the flat torus and in constant magnetic field, we also refer to \cite{CMMN,CZ} and excellent recent accounts \cite{HS,FHS, DN1}, where also QH hierarchy states are constructed.

There are two nontrivial 1-cycles on the torus, hence there exist classical configurations $\sigma_{mm'}$ of the compactified boson, labelled by the integers $m,m'\in\mathbb Z$, 
\begin{align}\label{config0}
&\sigma=\sigma_0+\sigma_{mm'}(z)+\tilde\sigma(z),\\
&\sigma_{mm'}(z+a\tau+b)=\sigma_{mm'}(z)+2\pi R_c(mb+m'a),
\end{align}
winding $m,m'$ times around each of the cycles, with $\tilde\sigma(z)$ being a single-valued scalar function. The last equation can be solved by
\begin{equation}\label{config}
\sigma_{mm'}(z)=2\pi R_c\left(\frac{m'-m\bar\tau}{\tau-\bar\tau}(z-z_0)-\frac{m'-m\tau}{\tau-\bar\tau}(\bz-\bz_0)\right),
\end{equation}
where $z_0$ is so far an arbitrary point on the torus.
After the zero-mode integration, which fixes the number of particles Eq.\ \eqref{Nrel}, the integral decomposes into the product of the classical part $Z_{\rm cl}$ due to $\sigma_{mm'}$ and quantum part $Z_{\rm qu}$ due to integration over $\tilde\sigma$
\begin{equation}\label{ZZ}
\mathcal V=Z_{\rm cl}Z_{\rm qu}.
\end{equation}
Here $Z_{\rm cl}$ is the sum over the sectors with different $m,m'$. 

In order to define $Z_{\rm cl}$ we need to compute the action Eq.\ \eqref{action1} on the field configuration \eqref{config}. There is a certain subtlety arising from multi-valuedness of the compactified boson, which manifests itself in ambiguity in the choice of base-point $z_0$ in Eq.\ \eqref{config}. This has no effect on the first and the last terms in the action \eqref{action1}, but the second and third term need to be defined more carefully. Proper definition should ensure modular invariance of the correlation functions. For our purposes it suffices to choose the base-point as $z_0=(\tau+1)/2$, for which the second and third terms in the action \eqref{action1} vanish. However, we shall note that various other prescriptions are possible for the terms of this type, on the torus \cite{DSZ} and on higher-genus surfaces \cite{VV}, that also preserve the modular invariance.

Then the value of the action Eq.\ \eqref{action1} on the configuration \eqref{config} is easily computed
\begin{equation}\nonumber
S(g_0,\NPhi,\sigma_{mm'})=\frac{\pi i R_c^2}{(\tau-\bar\tau)}|m'-m\tau|^2-
\frac{2\pi iR_c}{\sqrt\beta}(m\varphi_2+m'\varphi_1),
\end{equation}
where the first term comes form the kinetic term and the second term is the contribution of the flat connections Eq.\ \eqref{Atorus}. Taking into account the contribution from the vertex operators $e^{i\sqrt\beta\sigma_{mm'}(z_l)}$, $Z_{\rm cl}$ reads
\begin{align}\nonumber
Z_{\rm cl}=\frac{R_c}{\sqrt 2}\sum_{m,m'\in \mathbb Z}\exp\left(-\frac{\pi i R_c^2}{(\tau-\bar\tau)}|m'-m\tau|^2+
\frac{2\pi iR_c}{\sqrt\beta}(m\varphi_2+m'\varphi_1)\right.\\\nonumber
\left.+2\pi iR_c\sqrt\beta\left(\frac{m'-m\bar\tau}{\tau-\bar\tau}z_{\rm cm}-\frac{m'-m\tau}{\tau-\bar\tau}\bz_{\rm cm}\right)\right).
\end{align}
Now we apply Poisson summation formula Eq.\ \eqref{poisson} to the sum over $m'$
\begin{align}\label{sum10}\nonumber
Z_{\rm cl}=\sqrt{{\rm Im}\,\tau}&\cdot e^{\frac{\pi i\beta}{\tau-\bar\tau}\left(z_{\rm cm}^\varphi-\bz_{\rm cm}^\varphi\right)^2}\sum_{m,n\in\mathbb Z}\exp\left(i\pi\tau\left(\frac n{R_c}-\frac{mR_c}2\right)^2-i\pi\bar\tau\left(\frac n{R_c}+\frac{mR_c}2\right)^2\right.\\
&\left.-2\pi i\sqrt\beta\tau\left(\frac n{R_c}-\frac{mR_c}2\right)z_{\rm cm}^\varphi+2\pi i\sqrt\beta\bar\tau\left(\frac n{R_c}+\frac{mR_c}2\right)\bz_{\rm cm}^\varphi\right),
\end{align}
where we introduced the short-hand notation 
\begin{equation}\nonumber
z_{\rm cm}^\varphi=z_{\rm cm}+\frac\varphi\beta,\quad
\bz_{\rm cm}^\varphi=\bz_{\rm cm}+\frac{\bar\varphi}\beta.
\end{equation} 
Now we set $R_c=\sqrt\beta$ and change the summation indices $m,n\to p,r,\varepsilon$ according to: $n=\beta p+r,\;p\in\mathbb Z,\;r=1,..,\beta;\;m=2(q+\varepsilon),\;q\in\mathbb Z,\;\varepsilon=\{0,\frac12\}$. Then the sum over $m,n$ in Eq.\ \eqref{sum10} reads 
\begin{align}\nonumber
\sum_{m,n\in\mathbb Z}...=\sum_{\varepsilon={0,\frac12}}\sum_{r=1}^\beta\sum_{p,q\in\mathbb Z}\exp\left(i\pi\beta\tau\left(p-q+\frac r\beta-\varepsilon\right)^2+2\pi i\left(p-q+\frac r\beta-\varepsilon\right)\beta z_{\rm cm}^\varphi\right.\\\nonumber
\left.+i\pi\beta\bar\tau\left(p+q+\frac r\beta+\varepsilon\right)^2-2\pi i\left(p+q+\frac r\beta+\varepsilon\right)\beta \bz_{\rm cm}^\varphi\right).
\end{align}
Next we redefine the summation variables as follows, $n=p-q,\,m=p+q,\,n,m\in\mathbb Z$, which implies the constraint on $n+m$ being even, 
\begin{align}\nonumber
\sum_{m,n\in\mathbb Z}...=\sum_{\varepsilon={0,\frac12}}\sum_{r=1}^\beta\sum_{m,n\in\mathbb Z}\exp\left(i\pi\beta\tau\left(n+\frac r\beta-\varepsilon\right)^2+2\pi i\left(n+\frac r\beta-\varepsilon\right)\beta z_{\rm cm}^\varphi\right.\\\nonumber
\left.+i\pi\beta\bar\tau\left(m+\frac r\beta+\varepsilon\right)^2-2\pi i\left(m+\frac r\beta+\varepsilon\right)\beta \bz_{\rm cm}^\varphi\right)\left(\frac{1+e^{\pi i(n+m)}}2\right),
\end{align}
where the last term enforces this constraint. Writing 
\begin{equation}\nonumber
(1+e^{\pi i(n+m)})=\sum_{\delta=\{0,\frac12\}}e^{2\pi i\delta(n+m)},
\end{equation}
we can finally recast $Z_{\rm cl}$ in the form of the sum of absolute values squared of theta functions 
\begin{align}\label{sum1}
Z_{\rm cl}=\frac12\sqrt{{\rm Im}\,\tau}\cdot e^{\frac{\pi i\beta}{\tau-\bar\tau}\left(z_{\rm cm}-\bz_{\rm cm}+\frac{\varphi-\bar\varphi}\beta\right)^2}\sum_{\varepsilon,\delta=\{0,\frac12\}}\sum_{r=1}^\beta\, e^{4\pi i\varepsilon\delta}\left|\vartheta\left[\begin{array}{c}\scriptstyle\frac r\beta+\varepsilon \\\scriptstyle\delta\end{array}\right]\bigl(\beta z_{\rm cm}+\varphi,\beta\tau\bigr)\right|^2.
\end{align}
The sum over  $\varepsilon,\delta$ is nothing but the sum over four different spin structures, by analogy with Eq.\ \eqref{basis}. However, we note that for $\beta$ even the sum over $\varepsilon,\delta$ collapses into one term:
\begin{align}\nonumber
\beta\in{\rm even}:\quad Z_{\rm cl}=\sqrt{{\rm Im}\,\tau}\cdot e^{\frac{\pi i\beta}{\tau-\bar\tau}\left(z_{\rm cm}-\bz_{\rm cm}+\frac{\varphi-\bar\varphi}\beta\right)^2}\sum_{r=1}^\beta\left|\vartheta\left[\begin{array}{c}\scriptstyle\frac r\beta \\\scriptstyle 0\end{array}\right]\bigl(\beta z_{\rm cm}+\varphi,\beta\tau\bigr)\right|^2.
\end{align}
From now on we will only consider odd values of $\beta$. In order to be consistent with Eq.\ \eqref{Ltorus} we rewrite the expression Eq.\ \eqref{sum1} in the equivalent form, using the parity indicator $\lambda$ \eqref{lambda} for the number of particles,
\begin{align}\nonumber
Z_{\rm cl}=&\frac12\sqrt{{\rm Im}\,\tau}\cdot e^{\frac{\pi i\beta}{\tau-\bar\tau}\left(z_{\rm cm}-\bz_{\rm cm}+\frac{\varphi-\bar\varphi}\beta\right)^2}\\\nonumber&\cdot\sum_{\varepsilon,\delta=\{0,\frac12\}}\sum_{r=1}^\beta\, e^{4\pi i(\varepsilon-\lambda+\frac12)(\delta-\lambda+\frac12)}\left|\vartheta\left[\begin{array}{c}\scriptstyle\frac{r+\varepsilon}\beta-\lambda+\frac12 \\{\scriptstyle\delta-\beta\lambda+\frac\beta2}\end{array}\right]\bigl(\beta z_{\rm cm}+\varphi,\beta\tau\bigr)\right|^2.
\end{align}
Next we compute the quantum part $Z_{\rm qu}$ in Eq.\ \eqref{ZZ}. This is given by Eq.\ \eqref{LPI0}, where the Green function \eqref{green} and regularized Green function \eqref{reggreen} on the flat torus read
\begin{align}\nonumber
&G^{g_0}(z,z')=\frac12\frac{\pi i}{\tau-\bar\tau}\bigl(z-z-(\bz-\bz')\bigr)^2+\log\left|\frac{\theta_1(z-z',\tau)}{\eta(\tau)}\right|,\\\nonumber
&G^{g_0}_{\rm reg}(z)=-\log\bigl(\sqrt{2\pi{\rm Im}\,\tau}\,|\eta(\tau)|^2\bigr).
\end{align}
Plugging this to \eqref{LPI0}, and observing that the integrals over $\Sigma$ vanish due to \eqref{intgreen}, we obtain
\begin{align}\label{Zqu}
&Z_{\rm qu}=\left[\frac{\det'\Delta_{g_0}}{2\pi}\right]^{-1/2}\cdot\exp\left(-\beta\sum_{l\neq m}^{N}G^{g}(z_l,z_m)-\beta\sum_{l=1}^{N}G^{g}_{\rm reg}(z_l)\right)\\\nonumber
=\sqrt{2\pi}\bigl(&2\pi{\rm Im\,}\tau|\eta(\tau)|^4\bigr)^{\frac{\NPhi-1}2}\prod_{l<m}^N\left|\frac{\theta_1(z_l-z_m,\tau)}{\eta(\tau)}\right|^{2\beta}\cdot e^{\frac{\pi i\NPhi}{\tau-\bar\tau}\sum_l(z_l-\bz_l)^2-\frac{\pi i\beta}{\tau-\bar\tau}(z_{\rm cm}-\bz_{\rm cm})^2},
\end{align}
where we used the formula for the regularized determinant of laplacian on the torus 
\begin{equation}\nonumber
{\det}'\Delta_{g_0}=2\pi|\eta(\tau)|^4\,{\rm Im\,}\tau,
\end{equation} 
see e.g.\ Ref.\ \cite{DMS}. Putting together $Z_{\rm cl}$ and $Z_{\rm qu}$ we arrive at 
\begin{align}\nonumber
\mathcal V=&\frac12\bigl(2\pi{\rm Im\,}\tau|\eta(\tau)|^4\bigr)^{\frac{\NPhi}2}\cdot e^{\frac{\pi i}{\beta(\tau-\bar\tau)}(\varphi-\bar\varphi)^2}\\\nonumber
&\cdot\frac1{|\eta(\tau)|^2}\sum_{\varepsilon,\delta=\{0,\frac12\}}\sum_{r=1}^\beta\, e^{4\pi i(\varepsilon-\lambda+\frac12)(\delta-\lambda+\frac12)}\left|\vartheta\left[\begin{array}{c}\scriptstyle\frac{r+\varepsilon}\beta-\lambda+\frac12 \\{\scriptstyle\delta-\beta\lambda+\frac\beta2}\end{array}\right]\bigl(\beta z_{\rm cm}+\varphi,\beta\tau\bigr)\right|^2\\\nonumber&\cdot
\prod_{l<m}^N\left|\frac{\theta_1(z_l-z_m,\tau)}{\eta(\tau)}\right|^{2\beta}\cdot \prod_{l=1}^Ne^{\frac{\pi i\NPhi}{\tau-\bar\tau}(z_l-\bz_l)^2+\frac{2\pi i}{\tau-\bar\tau}(z_l-\bz_l)(\varphi-\bar\varphi)}.
\end{align}
This is the final result for the correlation function of vertex operators Eqns.\ (\ref{PI}, \ref{ZZ}) on the torus.

\subsection{Holomorphic structure and modular group action}
\label{modultr}
Comparing with the (holomorphic parts) of the Laughlin state given in Eq.\ \eqref{Ltorus}, and taking into account Eq.\ \eqref{Fr2} we see that the sum above is in the form of Eq.\ \eqref{PI2}, 
\begin{align}\label{sumwave}
\frac1{2\beta}\sum_{\varepsilon,\delta=\{0,\frac12\}}c_{\varepsilon,\delta}\sum_{r=1}^\beta |\Psi^{\varepsilon,\delta}_r|^2=&\frac{e^{-\mathcal F_\beta(g_0,B_0)} }{Z_\beta(\tau,\bar\tau,\varphi,\bar\varphi)}\\\nonumber&\cdot\sum_{\varepsilon,\delta=\{0,\frac12\}}\sum_{r=1}^\beta c_{\varepsilon,\delta}|F^{\varepsilon,\delta}_r|^2\prod_{l=1}^{N}h_0^{\NPhi}(z_l,\bz_l),
\end{align}
with $c_{\varepsilon,\delta}=e^{4\pi i(\varepsilon-\lambda+\frac12)(\delta-\lambda+\frac12)}$. Here we introduced the the $Z$-factor and redefined the holomorphic part of the Laughlin state to include the $\eta$ functions exactly as they appear from the path integral calculation
\begin{align}\label{Ltorus1}
F^{\varepsilon,\delta}_r(\{z_l\})=&\eta(\tau)^{\NPhi-1}\vartheta\left[\begin{array}{c}\scriptstyle\frac{r+\varepsilon}\beta-\lambda+\frac12 \\{\scriptstyle\delta-\beta\lambda+\frac\beta2}\end{array}\right]\bigl(\beta z_{\rm cm}+\varphi,\beta\tau\bigr)\prod_{l<m}^N\left(\frac{\theta_1(z_l-z_m,\tau)}{\eta(\tau)}\right)^{\beta},\\\label{LZ}
&Z_\beta(\tau,\bar\tau,\varphi,\bar\varphi)=\bigl(2\pi{\rm Im\,}\tau\bigr)^{-\frac{\NPhi}2}\cdot e^{-\frac{\pi i}{\beta(\tau-\bar\tau)}(\varphi-\bar\varphi)^2}.
\end{align}

Comparing Eqns.\ \eqref{torusortho} and \eqref{PI2} we can write the relation between the normalization factors $Z_\beta$ and $\mathcal N_0$ as follows
\begin{equation}\nonumber
\mathcal N_0[g_0,B_0,\tau,\varphi]=e^{\mathcal F_\beta(g_0,B_0)}\cdot Z_\beta(\tau,\bar\tau,\varphi,\bar\varphi).
\end{equation}
{\it Remark.} There is some ambiguity in the choice of the holomorphic part and of the $Z_\beta$-factor. Namely, one can redefine $Z_\beta\to|f(\tau,\varphi)|^2Z_\beta$ and correspondingly $F_r\to f(\tau,\varphi)F_r$ by a holomorphic function of the moduli. Since $Jac(\Sigma)$ is compact $f$ is a function of $\tau$ only. We can also take $f(\tau)$ to be non-vanishing on an open set of $\mathcal M_1$. In particular, $f(\tau)$ can be a power of $\eta(\tau)$, since the latter is non-vanishing in the upper half plane $\tau\in\mathbb H$ with a zero as $\tau\to i\infty$, where $\eta(\tau)\sim q^{1/24},\, q=e^{2\pi i\tau}$. This
will modify the adiabatic connection \eqref{adconn} and add the delta-function term, localized at $i\infty$, to the adiabatic curvature, but also modify the monodromies of the Laughlin states on the moduli space (which we review below), while preserving the adiabatic phases. In general, it should be possible to study the behavior of the normalized Laughlin states near the boundary of the moduli space of complex structures, by applying techniques of Ref.\ \cite{BK1}.

Next we note that the wave functions $\Psi_r^{\varepsilon,\delta}$ have the same general form as \eqref{wf}, so the relations \eqref{adconn} and \eqref{adcurv} apply to the Laughlin states, with the substitution $Z(y,\bar y)=e^{\mathcal F_\beta}\cdot Z_\beta$. In particular, the adiabatic connection is projectively flat and adiabatic curvature is a scalar matrix. The factor $e^{\mathcal F_\beta}$ is given by a nontrivial path integral expression Eq.\ \eqref{PI1} and we will study it in the next section. 

Let us now discuss the action of lattice shifts in the Jacobian $T_{[\varphi]}$ and the modular transformations. This is completely analogous to the action on one-particle states, worked out in \S \ref{monodrom}. The group of lattice shifts $\varphi\to\varphi+t_1+t_t\tau$ acts in the unitary representation
\begin{align}\nonumber
&F_r^{\varepsilon,\delta}(\{z_l\}|\varphi+t_1+t_2\tau,\tau)=e^{-\frac{\pi i}{\beta}t_2^2\tau-\frac{2\pi i}{\beta} t_2(\beta z_{\rm cm}+\varphi)}\cdot \sum_{r'=1}^\beta U_{rr'} F_{r'}^{\varepsilon,\delta}(\{z_l\}|\varphi,\tau),\\\nonumber
&{\rm where}\quad U_{rr'}=e^{\frac{2\pi i}{\beta}\bigl(t_1r+t_1(\varepsilon+\beta(\frac12-\lambda))-t_2(\delta+(\frac12-\lambda))\bigr)}\delta_{r,r'-t_2},\\\nonumber
& \prod_{l=1}^Nh_0^{\NPhi}(z_l,\bz_l|\varphi+t_1+t_2\tau,\tau)=e^{2\pi it_2(z_{\rm cm}-\bz_{\rm cm})} \prod_{l=1}^Nh_0^{\NPhi}(z_l,\bz_l|\varphi,\tau),
\\
\nonumber
&Z_\beta(\varphi+t_1+t_2\tau,\bar\varphi+t_1+t_2\bar\tau,\tau,\bar\tau)=e^{-\frac{\pi i}{\beta} t_2^2(\tau-\bar\tau)-\frac{2\pi i}\beta t_2(\varphi-\bar\varphi)}\cdot Z_\beta(\varphi,\bar\varphi,\tau,\bar\tau),
\end{align}
and, as was already the case for the LLL states on the torus, $t_1$-shifts act diagonally and $t_2$-shift action is non-diagonal.

The formulas for the action of the modular group, for $\beta\in$ odd, are listed in the Appendix. The action of the modular group on the basis of Laughlin states is very similar the action on the basis of one particle states (\ref{T}, \ref{S}), formally interchanging $\beta$ and $k$. In particular the action on spin-structures is the same as in Fig.\ 3. For even number of particles $\lambda=0$ the $(0,0)$ spin structure is conserved and for $\lambda=\frac12$ the $(\frac12,\frac12)$ spin structure is conserved. In these cases we have 
$$(U^SU^T)^3=e^{2\pi i\theta\NPhi}C,\;(U^S)^2=C,$$ 
where $C^2=1$ and $\theta=\frac18$, and thus Laughlin states transform in projective unitary representation of the modular group.

\section{Geometric adiabatic transport and anomaly formulas}

\subsection{Generating functional for Laughlin states}

The generating functional, which was defined for integer QH state \eqref{Z1}, can be defined for the Laughlin states as well. The definition is analogous to the one given in \eqref{Z1}. We start with the background configuration $(\Sigma,g_0)$ and $(L^{\NPhi},h_0^{\NPhi})$ and choose the corresponding $L^2$ normalized basis of holomorphic states $F_{0r}^{\varepsilon,\delta}$ \eqref{hermnormL}, \eqref{normLaugh},
\begin{align}\label{hermnormL1}
&||F_{0r}^{\varepsilon,\delta}(z_1,...,z_N)||^2=|F_{0r}^{\varepsilon,\delta}(z_1,...,z_N)|^2 \prod_{l=1}^Nh_0^{\NPhi}(z_l,\bz_l)g_{0z\bz}^{-\ij}(z_l,\bz_l),\\\nonumber
&\langle \Psi_{0r}^{\varepsilon,\delta}|\Psi_{0r}^{\varepsilon,\delta}\rangle_{L^2}=\frac1{\mathcal N_{0}}\frac1{(2\pi)^N}\int_{\Sigma^N} ||F_{0r}^{\varepsilon,\delta}(z_1,...,z_N)||^2\prod_{l=1}^N\sqrt{g_0}d^2z_l=1,
\end{align}
where the normalization factor $\mathcal N_{0}$ is $r$-independent, which is the case e.g., for the flat torus with constant magnetic field \eqref{torusortho}.
Next, we consider the curved metric $(\Sigma,g)$ and magnetic field $(L^{\NPhi},h^{\NPhi})$, where the $g$ and $F$ are in the same \kahler class
\begin{align}\label{tr1}
&g=g_0+\p_z\p_{\bz}\phi,\\\label{tr2}
&h^{\NPhi}=h_0^{\NPhi}e^{-\NPhi\psi},\\\label{tr3}
&F_{z\bz}=F_{0z\bz}+\NPhi\p_z\bp_{\bz}\psi.
\end{align}
The partition function is then defined as
\begin{align}\label{betaZ}
Z_{\NPhi}[g_0,B_0,g,B]=\frac1{\mathcal N_0}\frac1{(2\pi)^N}\int_{\Sigma^N} &\frac1{2^{\rm g}n_{\beta,\rm g}}\sum_{r,\varepsilon,\delta} c_{\varepsilon,\delta}||F_{0r}^{\varepsilon,\delta}(z_1,...,z_N)||^2\\\nonumber&\cdot\prod_{l=1}^Nh^{\NPhi}(z_l,\bz_l)g_{z\bz}^{-\ij}(z_l,\bz_l)\sqrt{g}d^2z_l.
\end{align}
In other words, we change the Hermitian metric on the line bundle and the metric on the surface, staying in the same \kahler class, and compute the sum of the norms of the wave functions in the new metric. It follows that the partition function is normalized as
\begin{equation}\nonumber
Z_{\NPhi}[g_0,B_0,g_0,B_0]=1.
\end{equation}

The expression \eqref{betaZ} is written on the surface of genus ${\rm g}>0$ and includes sum over all degenerate Laughlin states and also over the spin-structures. On the sphere this formula simplifies, since there is only one Laughlin state. In the notations of Eq. \eqref{Laughsph}, we can write on the sphere
\begin{equation}\label{betaZ1}
Z_{\NPhi}[g_0,B_0,g,B]=\frac1{\mathcal N_0}\frac1{(2\pi)^N}\int_{(S^2)^N} |\det s_l(z_m)|^{2\beta}\prod_{l=1}^Nh^{\NPhi}(z_l,\bz_l)g_{z\bz}^{-\ij}(z_l,\bz_l)\sqrt{g}d^2z_l.
\end{equation}
 Taking into account \eqref{PI2}, we can rewrite \eqref{betaZ} via the correlator of vertex operators
\begin{align}\label{Z}
Z_{\NPhi}[g_0,B_0,g,B]=&\frac1{(2\pi)^N}\int_{\Sigma^N} \frac1{2^{\rm g}n_{\beta,\rm g}}\sum_{r,\varepsilon,\delta} c_{\varepsilon,\delta}\bigl|\Psi_{0r}^{\varepsilon,\delta}(z_1,...,z_N)\bigr|^2\\\nonumber&\cdot e^{-\sum_{l=1}^N\bigl(\NPhi\psi(z_l,\bz_l)+{\rm j}\log\frac{\sqrt{g}}{\sqrt{g_0}}|_{z_l}\bigr)}\prod_{l=1}^N\sqrt{g}d^2z_l\\\nonumber
=\frac{1}{e^{\mathcal F_\beta(g_0,B_0)}}\frac1{(2\pi)^N}\int_{\Sigma^N}&\mathcal V\bigl(g_0,B_0,\{z_l\}\bigr)e^{-\sum_{l=1}^N\bigl(\NPhi\psi(z_l,\bz_l)+{\rm j}\log\frac{\sqrt{g}}{\sqrt{g_0}}|_{z_l}\bigr)}\prod_{l=1}^N\sqrt{g}d^2z_l.
\end{align}
As usual the logarithm of the partition function is called the generating functional. In this section we show that $\log Z_{\NPhi}$ admits an expansion for large magnetic field, which is analogous to the expansion of the generating functional for the integer QH state 
\begin{align}\nonumber
\log Z_{\NPhi}[g_0,B_0,g,B]=\log \frac{Z_{H,\beta}}{Z_{H0,\beta}}+\mathcal F_\beta[g,B]-\mathcal F_{\beta}[g_0,B_0],
\end{align}
where $\mathcal F_\beta$ is local functional of $g$ and $B$ and $\log Z_{H,\beta}$ is non-local functional representing anomaly, where all the terms depend nontrivially on $\beta$. 

In the integer QHE case we were able to compute the asymptotic expansion of $\log Z_{\NPhi}$ due to the determinantal representation of partition function \eqref{detfor}, which allowed us to reduce the computation to the Bergman kernel expansion for high powers of line bundle. However, the partition function \eqref{betaZ} does not admit the determinantal representation and novel methods are required. Here we review the path integral derivation of the asymptotic expansion following \cite{FK}; another derivation of this result can be found in \cite{CLW,CLW1}, where the Ward identity method of Refs.\ \cite{WZ1,ZW,Zab} was employed.

\subsection{Effective action and gravitational anomaly}

The calculation of asymptotic expansion of $Z_{\NPhi}$ is performed in two steps. At the first step we start from the path integral expression for $\mathcal V\bigl(g,B,\{z_l\}\bigr)$ in Eq.\ \eqref{PI} and compute its transformation formula under the change of metrics $(g,h)$ to $(g_0,h_0)$ in the same \kahler class, Eqns.\ (\ref{tr1}, \ref{tr2}). As we have seen in Eq.\ \eqref{ZZ}, for the surfaces of genus ${\rm g}>0$ the path integral is the product of the classical and quantum parts $\mathcal V=Z_{\rm cl}Z_{\rm qu}$, where the classical part $Z_{\rm cl}$ essentially depends only on the \kahler class of the metric and not on a particular choice of the metric in that class. Therefore it suffices to derive the transformation formulas for the change of metrics in the quantum part of the path integral $Z_{\rm qu}$. The latter is given by the same formal expression Eq.\ \eqref{LPI0} for the surfaces of any genus, including sphere. It remains to compute the transformation rules for different objects in that expression. The scalar curvatures and magnetic fields in background and curved metrics are related as
\begin{align}\label{scb}
R\sqrt g= R_0\sqrt{g_0}-\sqrt{g_0}\;\Delta_0\log\frac{\sqrt g}{\sqrt{g_0}},\\\nonumber
B\sqrt g=B_0\sqrt{g_0}+\frac12\NPhi\sqrt{g_0}\;\Delta_0\psi.
\end{align} 
For the metrics $g=g_0+\p\bp\phi$ in the same \kahler class the regularized determinant of the laplacian transforms according to the Polyakov gravitational anomaly formula \cite{P},
\begin{equation}
\label{trans1}
\frac{\det'\Delta_g}{\det'\Delta_0}=e^{-\frac16S_L(g_0,\phi)},
\end{equation}
where $S_L(g_0,\phi)$ is the Liouville action Eq.\ \eqref{func3}. 

The transformation formulas for other terms in \eqref{LPI0} can be found using the identities for the transformation of Green functions and their integrals, derived in \cite[\S 3]{FKZ3} and \cite[\S 4]{FK}. After a tedious but straightforward calculation, we arrive at
\begin{align}\label{VV0}
&\log\frac{\mathcal V\bigl(g,B,\{z_l\}\bigr)}{\mathcal V\bigl(g_0,B_0,\{z_l\}\bigr)}=-\NPhi\sum_{l=1}^N\psi(z_l,\bz_l)-{\rm j}\sum_{l=1}^N\log\frac{\sqrt g}{\sqrt{g_0}}\big|_{z_l}\\\nonumber
&+\frac1\beta\NPhi^2S_2(g_0,B_0,\psi)-\frac q{2\sqrt\beta}\NPhi S_1(g_0,B_0,\psi,\phi)+\frac1{12}(1-3q^2)S_L(g_0,\phi).
\end{align} 
Recall that the constant $q=\sqrt\beta-2{\rm j}/\sqrt\beta$ here is defined in Eq.\ \eqref{q}, and the functionals $S_1$ and $S_2$ are defined exactly as in Eqns.\ \eqref{func1}, \eqref{func2} with $k\to\NPhi$, namely
\begin{align}\nonumber
&S_2(g_0,B_0,\psi)=\frac1{2\pi}\int_\Sigma\left(\frac14\psi\Delta_0\psi
+\frac1{\NPhi} B_0\psi\right)\sqrt{g_0}d^2z,\\\nonumber
&S_1(g_0,B_0,\phi,\psi)=\frac1{2\pi}\int_\Sigma\left(-\frac12\psi R_0
+\left(\frac1{\NPhi} B_0+\frac12\Delta_0\psi\right)
\log\bigl(1+\frac12\Delta_0\phi\bigr)\right)\sqrt{g_0}d^2z.
\end{align}
Using Eq.\ \eqref{VV0} we can express the generating functional Eq.\ \eqref{Z} as follows
\begin{align}\label{expZbeta}
\log Z_{\NPhi}[g_0,B_0,g,B]=&-\frac1\beta\NPhi^2S_2(g_0,B_0,\psi)+\frac q{2\sqrt\beta}\NPhi S_1(g_0,B_0,\psi,\phi)\\\nonumber
&-\frac1{12}(1-3q^2)S_L(g_0,\phi)+\mathcal F_\beta(g,B)-\mathcal F_\beta(g_0,B_0).
\end{align}
The exact terms $\mathcal F_\beta$ are formally defined by the path integral Eq.\ \eqref{PI1} and we will come back to them shortly. The first three terms on the rhs in Eq.\ \eqref{expZbeta} contribute to the anomalous part of the generating functional $\log Z_H$. By analogy with Eqns.\ (\ref{di}, \ref{cs}) for the integer QH state, these can be written in two equivalent forms: as a double integral,
\begin{align}\label{di1}
&\log Z_{H,\beta}=\\\nonumber
-\frac1{2\pi\beta}  &\int_{\Sigma\times\Sigma} \left(B+\frac{\beta-2{\rm j}}4R\right)\big|_z
\Delta_g^{-1}(z,y)\left(B+\frac{\beta-2{\rm j}}4R\right)\big|_y\sqrt g d^2z\sqrt g d^2y\\\nonumber
+\frac1{96\pi} &\int_{\Sigma\times\Sigma} R(z)\Delta_g^{-1}(z,y)R(y)\sqrt g d^2z\sqrt g d^2y,
\end{align}
and as a quadratic form in gauge and spin connections,
\begin{equation}\label{logzh}
\log Z_{H,\beta}= \frac2{\pi}\int_\Sigma\left[\sigma_H A_zA_{\bz}
+2\varsigma_H(A_z\omega_{\bz}+
\omega_zA_{\bz})-\frac1{12}c_H
\omega_z\omega_{\bz}\right]d^2z,
\end{equation}
where we introduced the following constants
\begin{equation}\label{constants}
\sigma_H=\frac1\beta,\quad\varsigma_H=\frac{q}{4\sqrt\beta},\quad c_H=1-3q^2,
\end{equation}
and $q=\sqrt\beta-2\ij/\sqrt\beta$.
Recall that Eq.\ \eqref{logzh} assumes symmetric gauge Eq.\ \eqref{symgauge} for the gauge and spin connections. As a consistency check, at $\beta=1$ we have ${\rm j}=s$ and $q=1-2s$ and the expressions above agree with Eqns.\ (\ref{di}, \ref{cs}).

At the moment Eq.\ \eqref{expZbeta} is a formal expression, valid for any $\NPhi\geqslant0$. It turns out that the terms $\mathcal F_\beta$ \eqref{PI1} can be better understood for $\NPhi$ large. We can already see that the first three terms in Eq.\ \eqref{expZbeta} are written in the form of the large $\NPhi$ expansion, similar to anomalous terms \eqref{ano} in the integer case. We argue that $\mathcal F_\beta$ admits asymptotic expansion for large $\NPhi$ with coefficients given by local functional of the metric and the magnetic field, which is similar to Eq.\ \eqref{exact1} in the integer case, but now $\beta$-deformed. The argument is as follows \cite{FK}. Starting from the representation \eqref{PI1} we can rewrite the remainder term in the form
\begin{equation}\nonumber
e^{\mathcal F_\beta(g,B)}=\int e^{-S(g,B,\sigma)+N\log\frac1{2\pi}\int_\Sigma e^{i\sqrt\beta\sigma(z)}\sqrt gd^2z}\mathcal D_g\sigma.
\end{equation}
This is the path integral of the interacting scalar field, where the number of particles $N$ plays the role of a large parameter. Therefore one can apply the stationary phase method. The standard analysis of perturbation theory in $1/N$ reveals \cite{FK} that the contributions from the Feynman diagrams reduce to local integrals of polynomials in curvature and magnetic field and their derivatives. In particular, the leading term in the large magnetic field expansion of $\mathcal F_\beta$ has the form \cite{K2016},
\begin{equation}\label{asexpf}
\mathcal F_\beta=\frac{\beta-2}{4\pi\beta}\int_\Sigma B\left(\log \frac{B}{2\pi}\right)\sqrt gd^2z+\mathcal O(\log B).
\end{equation}
At $\beta=1$ this coincides with the integer QH result for $\mathcal F$, cf. first term in Eq.\ \eqref{exact1}.

\subsection{Quillen metric and geometric adiabatic transport}
\label{quillenano}

First we discuss the geometric adiabatic transport in the integer QH state and then turn to the Laughlin states. We consider how the wave functions vary over the parameter space $Y=\mathcal M_{\rm g}\times Jac(\Sigma)$. Following the discussion in \S \ref{at} it is especially convenient to put the wave function in the form Eq.\ \eqref{wf}, which emphasizes the holomorphic structure. For the integer QH state we can always choose the basis $s_l(z|y)$ in the space of holomorphic sections $H_0(\Sigma, L^k\otimes K^s)$, so that it depends holomorphically on local complex coordinate $y\in Y$. Then the integer QH state $\mathcal S$ \eqref{slater} transforms as a holomorphic section of the Quillen's determinant line bundle $\mathcal L=\det H^0(\Sigma_y,L^k\otimes K^s_y)$ over $Y$. The point-wise Hermitian norm of the section $\det s_l(z_m)$ and the $L^2$ norm is defined as before in Eq.\ \eqref{Z1}. The adiabatic connection and adiabatic curvature are then given by Eqns.\ \eqref{aconn}, \eqref{acurv} and thus the adiabatic curvature can be expressed in terms of the $L^2$ norm of the IQHE state as
\begin{equation}\label{omega}
\mathcal R=-(\p_y\p_{\bar y}\log Z_k) idy\wedge d\bar y.
\end{equation} 
Here $\mathcal R$ carries no indices, since the integer QH state is not degenerate and is a section of the  line bundle. We can now compute \eqref{omega} using the following observation originally due to Avron-Seiler-Zograf \cite{ASZ}. We rewrite $\mathcal R$ as
\begin{equation}\label{omega1}
\mathcal R=-\left(\p_y\p_{\bar y}\log\frac{Z_k}{\det'\Delta_{\rm L}}\right)idy\wedge d\bar y -\bigl(\p_y\p_{\bar y}\log{\det}'\Delta_{\rm L}\bigr)\,idy\wedge d\bar y,
\end{equation} 
and note that the first term here is the curvature $\mathcal R^{\mathcal L}$ of the Quillen metric on $\mathcal L$, where the latter is defined in Eq.\ \eqref{quillenmet}. The formula for the curvature of the Quillen metric is known in physics literature as the Quillen anomaly formula, and it was first computed as part of the proof of the holomorphic factorisation of string theory integration measure in Ref.\ \cite{BK1}, see also \cite{Q,ZT} for mathematical references. For the Riemann surfaces the curvature of Quillen metric can be written explicitly, 
\begin{equation}\label{omegaL}
\mathcal R^{\mathcal L}=2\pi d\varphi\wedge (\Omega-\bar\Omega)^{-1}d\bar\varphi-\left(\frac k4(1-2s)-\frac1{12}\bigl(1-3(1-2s)^2\bigr)\chi(\Sigma)\right)\Omega_{WP}.
\end{equation}
The first term here is the 2-form on $Jac(\Sigma)$ written in complex coordinates \eqref{varphic} and summation over indices labelling 1-cycles is understood. In real coordinates on the Jacobian \eqref{flatconn} the first term reads 
\begin{equation}\nonumber
2\pi d\varphi\wedge (\Omega-\bar\Omega)^{-1}d\bar\varphi=2\pi\sum_{a=1}^{\rm g}d\varphi_1^a\wedge\varphi_2^a,
\end{equation}
and thus corresponds to the flat Euclidean metric on the $2{\rm g}$ dimensional torus. 
The Weil-Petersson form $\Omega_{WP}$ on the moduli space of complex structures $\mathcal M_{\rm g}$, for constant scalar curvature metrics on $\Sigma$, enters the second term in Eq.\ \eqref{omegaL} and is defined as follows. The deformations of the metric, preserving the area of $\Sigma$, along the moduli space have the form
\begin{equation}\nonumber
\delta(g_{z\bz}dzd\bz)=\frac1{1-|\delta\mu|^2}g_{z\bz}|dz+\delta\bar\mu d\bz|^2-g_{z\bz}dzd\bz,
\end{equation} 
where $\delta\mu={\delta\mu^z}_{\bz}d\bz(dz)^{-1}$ is the Beltrami differential with the weight $(-1,1)$. There exists \cite{DP} $3{\rm g}-3$ independent holomorphic quadratic differentials $\eta$ on a surface of a genus ${\rm g}>1$ and the corresponding Beltrami differential  $\delta\mu=g^{z\bz}\sum_{\nu=1}^{3{\rm g}-3}\bar\eta_\nu d y_\nu$ is characterized by $3{\rm g}-3$ local complex coordinates $y_{\nu}$. The K\"ahler $(1,1)$ form on $\mathcal M_{\rm g}$ corresponding to the Weil-Petersson metric can be written as 
\begin{equation}\nonumber
\Omega_{WP}=\frac1{2\pi}\int_\Sigma(i\delta\mu\wedge \delta\bar\mu)\sqrt gd^2z=\frac1{\pi}\int_\Sigma g^{z\bz}\bar\eta_\nu\eta_\mu d^2z\; idy^\nu\wedge d\bar y^\mu.
\end{equation}  
We note that in the standard definition the scalar curvature $R$ enters the integrand, which is constant in our case $R=2\chi(\Sigma)$, and the factor of $\chi(\Sigma)$ is already present in the formula \eqref{omegaL}.
On the sphere there are no solenoid phases and the choice of complex structure is unique, so the corresponding moduli space is just a point. On the torus Eq.\ \eqref{omegaL} is still valid with the replacement of Weil-Petersson form $\Omega_{WP}$ by the Poincar\'e metric on $\mathcal M_1$, and setting $s=0$. Indeed, using the determinantal formula \eqref{detfor} we immediately obtain for the flat torus and constant magnetic field, 
\begin{equation}\nonumber
Z_k=(Z)^k,
\end{equation}
where $Z=Z(\tau,\bar\tau,\varphi,\bar\varphi)$ is the normalization $Z$-factor for one particle LLL states on the torus Eq.\ \eqref{Zfactor}. Next, the determinant of the laplacian for the line bundle $L^k$ is moduli-independent constant \eqref{dettor}, so the adiabatic curvature $\mathcal R$ and Quillen curvature \eqref{omegaL} coincide. Hence in the case of torus we obtain 
\begin{equation}\label{omegaL1}\nonumber
\mathcal R_{T^2}=\mathcal R^{\mathcal L}_{T^2}=(-\p_y\p_{\bar y}\log Z_k)\,idy\wedge d\bar y=2\pi \frac{d\varphi\wedge d\bar\varphi}{\tau-\bar\tau}+\frac k4\frac{2id\tau\wedge d\bar\tau}{(\tau-\bar\tau)^2},
\end{equation}
in complete agreement with \eqref{c1E}. 

On higher genus surfaces the determinant of the laplacian depends on the moduli in a nontrivial way. Namely for the constant scalar curvature metric on $\Sigma_{\rm g}$ with $\rm g>1$ and for canonical line bundle $L=K$ we have, according to Refs.\ \cite{DP1,ASZ},
\begin{equation}\label{magndet}
{\det}'\Delta_{L^k}=e^{-c_k\chi(\Sigma)}\prod_{\gamma}\prod_{j=1}^\infty
\bigl[1-e^{i\sum_{a=1}^{2{\rm g}}\varphi_an_a(\gamma)}e^{-(j+k)l(\gamma)}\bigr].
\end{equation} 
 Here the surface $\Sigma$ is realized as an orbit space for a discrete subgroup
$\Gamma$ of $SL(2,\mathbb R)$ acting on upper half plane and $\gamma\in\Gamma$ are primitive 
hyperbolic elements of $\Gamma$ representing conjugacy 
classes corresponding to closed geodesics on $\Sigma$. Next, $n_a$ counts the number of times the closed geodesic goes around the $a$th fundamental loop, 
$l(\gamma)$ is the length of geodesic, and  $c_k$ is a constant. In the large $k$ limit the leading term (apart form the constant, irrelevant for the computation of curvature) in 
$\log{\det}'\Delta_{L^k}$ decays exponentially as $e^{-kl(\gamma_{\rm min})}$, 
where $\gamma_{\rm min}$ is the length of the shortest geodesic. When the latter is bounded from zero, i.e., away from the boundary of the moduli space, the second term in \eqref{omega} represents small fluctuations and $\mathcal R\approx\mathcal R^{\mathcal L}+\mathcal O(e^{-kl(\gamma_{\rm min})})$ 
with exponential precision. However, the exponential asymptotic of determinant changes to polynomial near the boundary of the moduli space, see \cite[\S V.F]{DP}, where the second term in \eqref{omega} starts to play a more prominent role. This regime, where the Riemann surface becomes singular, deserves to be understood in more detail. For recent work in QH states on singular surfaces we refer to \cite{CCLW,Gr}, see also \cite{AKPS}.

The formula \eqref{omegaL} is can be read off directly from the anomalous part of the generating functional written as the quadratic form \eqref{cs}. We write the variation of the gauge connection along the moduli space as
\begin{equation}\nonumber
\delta (A_zdz)=2\pi\delta\varphi(\Omega-\bar\Omega)^{-1}\bar\omega,\quad \delta (A_{\bz}d\bz)=-2\pi\delta\bar\varphi(\Omega-\bar\Omega)^{-1}\omega,
\end{equation}
and the second variation of the spin connection (the first variation vanishes) as
\begin{equation}\nonumber
\delta\wedge\bar\delta (\omega_zdz)=\frac 12\p_z (\delta\mu\wedge\delta\bar\mu),\quad \delta\wedge\bar\delta (\omega_{\bz}d\bz)dz=-\frac 12\p_{\bz} (\delta\mu\wedge\delta\bar\mu)d\bz.
\end{equation}
Computing the second variation of $\log Z_H$ and plugging the above result is another way to obtain Eq.\ \eqref{omegaL},
\begin{align}\label{genfcal}
&-(\p_y\p_{\bar y}\log Z_H)idy\wedge d\bar y\\\nonumber
=&\frac 1\pi\int_\Sigma\left(\delta (A_zdz)\wedge\delta(A_{\bz}d\bz)+\frac{(1-2s)}2\bigl(A_zdz\,\delta\wedge\bar\delta(\omega_{\bz}d\bz)+c.c.\bigr)\right.\\\nonumber
&\left.-\frac1{12}\bigl(1-3(1-2s)^2\bigr)(\omega_zdz\,\delta\wedge\bar\delta(\omega_{\bz}d\bz)+c.c.)\right)=\mathcal R^{\mathcal L}.
\end{align}
Here in the second and third term we applied integration by parts and then projected the result onto the constant magnetic field and constant scalar curvature metric.  

\subsection{Geometric adiabatic transport for Laughlin states}

Now we turn to the adiabatic curvature for the Laughlin states. On a Riemann surface $\Sigma$ of genus ${\rm g}>0$ the number of Laughlin states is $n_{\beta,\rm g}=\beta^{\rm g}$, as was mentioned in \S \ref{topdeg}. Thus Laughlin states transform as the sections of a vector bundle of degree $n_{\beta,\rm g}$ over the parameter space $Y$. The adiabatic connection on this vector bundle is projectively flat \eqref{adconn}, which was demonstrated for the Laughlin states on the torus in \S \ref{topdeg} (this is assumed to be the case for the higher-genus surfaces as well). Then the adiabatic curvature can be determined from the norm of the wave functions as in Eq.\ \eqref{adcurv}, or by analogy with the integer case, from the generating functional
\begin{equation}\label{Zbe}
\mathcal R_{rr'}=\mathcal R\delta_{rr'}=-\delta_{rr'}(\p_y\p_{\bar y}\log Z_{\NPhi}) idy\wedge d\bar y.
\end{equation}
Now we can determine adiabatic curvature $\mathcal R$ applying the variational method of Eq.\ \eqref{genfcal} to the anomalous part of the generating functional \eqref{logzh}. We immediately obtain
\begin{align}\label{ombeta}
\mathcal R=\,&2\pi\sigma_H d\varphi\wedge (\Omega-\bar\Omega)^{-1}d\bar\varphi\\\nonumber&-\left(\varsigma_H\NPhi-\frac1{12}c_H\chi(\Sigma)\right)\Omega_{WP}-(\p_y\p_{\bar y}\mathcal F_\beta)\, idy\wedge d\bar y,
\end{align}
where the constants $\sigma_H,\varsigma_H,c_H$ are given in Eq.\ \eqref{constants}. The first two terms here differ from Eq.\ \eqref{omegaL} only in overall coefficients. The last term in \eqref{ombeta} reduces to the logarithm of the regularized determinant $\mathcal F_{\beta=1}=\log{\det}'\Delta_{\rm L}$ due to the bosonisation formula, as was discussed before Eq.\ \eqref{Fdet}. The quantization argument for the Hall conductance in the integer case Eq.\ \eqref{chernnum} will go through for the Laughlin states if we can show that the last term is an exact $(1,1)$ from, corresponding to small fluctuations at large $\NPhi$. We have already seen from the $1/N$ perturbation theory arguments \cite{FK} that $\mathcal F_\beta$ admits asymptotic expansion in large magnetic field with coefficients given by local curvature invariants \eqref{asexpf}. Since local terms are moduli-independent, $\p_y\p_{\bar y}\mathcal F_\beta$ is zero perturbatively, i.e., for all terms in asymptotic $1/\NPhi$ expansion. However exponential corrections of the form $e^{-\NPhi f}$ are possible, where $f$ can be a nontrivial function of the moduli (in fact they appear already at $\beta=1$ in the log determinant on higher-genus surfaces\eqref{magndet}). Since $f$ is a function $\p_y\p_{\bar y}\mathcal F_\beta$ is exact, and the exponential suppression means that the last term in \eqref{ombeta} represents exponentially small fluctuations of the adiabatic curvature. It would be interesting to check this indirect argument, e.g.\ by computing $\mathcal F_\beta$ from its path integral representation \eqref{PI1}. 

We can now apply the general formula Eq.\ \eqref{ombeta} to the torus, where we worked out explicit expression for $Z_\beta$ in Eq.\ \eqref{LZ}. Plugging $Z_{\NPhi}=e^{\mathcal F_\beta}Z_\beta$ in \eqref{Zbe} we obtain
\begin{equation}\label{att}
\mathcal R=2\pi\sigma_H \frac{d\varphi\wedge d\bar\varphi}{\tau-\bar\tau}+\frac \NPhi4\frac{2id\tau\wedge d\bar\tau}{(\tau-\bar\tau)^2}-(\p_y\p_{\bar y}\mathcal F_\beta)\, idy\wedge d\bar y,
\end{equation} 
in agreement with \eqref{ombeta} at $\chi(\Sigma)=0$ and $\rm j=0$. Now, by analogy with \eqref{chernnum} the first Chern class of the vector bundle of Laughlin states restricted to the Jacobian $E|_{T_{[\varphi]}}$ equals one,
\begin{equation}
\int_{T_{[\varphi]}}c_1(E|_{T_{[\varphi]}})=\int_{T_{[\varphi]}}\frac1{2\pi}\tr \mathcal R_{rr'}=\beta\sigma_H=1,
\end{equation}
and thus the Hall conductance $\sigma_H=1/\beta$ is a fraction in this case. This argument in the fractional QHE was suggested in \cite{TW,T2}.

As before in the integer case, the adiabatic transport on the moduli space of complex structure of the torus gives rise to the anomalous viscosity and the Hall viscosity coefficient is also proportional to the magnetic field flux $\eta_H=\NPhi/4$ \eqref{att}, see Refs.\ \cite{TV1,TV2} and \cite{R}. On the surfaces of genus $\rm g>1$ the corresponding coefficient acquires the finite-size correction \eqref{ombeta}, proportional to $c_H$, for "Hall central charge", since it appears also as the coefficient in front of the Liouville action \eqref{expZbeta}.

\subsection{Adiabatic phase and Chern-Simons action}
\label{cs-form}

We have already noticed the resemblance of the generating functional for the integer QH state in the form  Eq.\ \eqref{cs} to the 2+1d Chern-Simons action. Effective long-distance description of the quantum Hall effect in terms of Chern-Simons theory goes back to \cite{WZ,FKer,FS}. The gravitational Chern-Simons term in 2+1d, corresponding to the 2d gravitational anomaly term in Eq.\ \eqref{cs}, was derived only recently \cite{AG1,AG2,AG4}. Following Ref.\ \cite{KMMW} we recall how Chern-Simons functional arises from the adiabatic curvature \eqref{omega}, and more precisely from the Quillen anomaly part of the adiabatic curvature $\mathcal R^{\mathcal L}$ \eqref{omegaL}.

We shall now consider the family of surfaces $\Sigma_y$ parameterized by $y\in Y$ and the space $M$ which is the union of all $\Sigma_y$ over $Y$ (sometimes $M$ is called "the universal curve"). In particular, the dimension of $M$ equals $\dim M=\dim Y+2$ where $2$ is the dimension of the Riemann surface. We consider also the family of line bundles $L^k_y\otimes K_y^s\to\Sigma_y,\,y\in Y$. The union of all such line bundles extends to the holomorphic line bundle $E$ over $M$, and Hermitian metric $h^k$ extends to the Hermitian metric $h^E$ on $E$. We denote the curvature of the metric $h^E$ as $F^E$. We also consider the extension of the union of tangent bundles $T\Sigma_y$ to the bundle $TM|Y$, which is still a line bundle (as opposed to the usual tangent bundle $TM$). Let $g^{TM|Y}$ be a smooth Hermitian metric on $TM|Y$ and $R_{TM|Y}$ be its curvature 2-form.  

The following formula for the curvature of the Quillen determinant line bundle $\mathcal R^{\mathcal L}$ is due to Bismut-Gillet-Soul\'e \cite[Thm. 1.27]{BGS3}, see also \cite{BF},
\begin{equation}\label{bgs}
\mathcal R^{\mathcal L}
=-2\pi i\int_{M|Y}\bigl[{\rm Ch}(E){\rm Td}(TM|Y)\bigr]_{(4)}.
\end{equation}
The notation $M|Y$ means that the integration goes over the fibers in the fibration $\sigma:M\to Y$, i.e., over the spaces $\Sigma_y$ at  $y$ fixed. The expression in the brackets is a form of mixed degree on $M$, $\rm Ch$ is the Chern character an $\rm Td$ is the Todd class, see e.g.\ \cite{GH} for standard definitions. The subscript $(4)$ means that only the $4$-form component of the integrand is retained, and after the integration we end up with the 2-form on $Y$. In order to apply the results of Ref.\ \cite{BGS3} in our context we need to check that the fibration $\sigma:M\to Y$ is locally K\"ahler, which turns out to be the case, as explained in Ref.\ \cite{KMMW}. 

Next, we choose an adiabatic process, which is a smooth closed contour $\mathcal C\in Y$. When the wave function (in this context we are talking only about the integer QH state) is transported around the contour, we can compute the geometric part of the adiabatic phase (corresponding to $\mathcal R^{\mathcal L}$) as $\int_{\mathcal C} \mathcal A^{\mathcal L}$, where the connection 1-form $\mathcal A^{\mathcal L}$ on $\mathcal L$ can be computed using the formula \eqref{bgs} locally as $\mathcal R^{\mathcal L}=d\mathcal A$. 

First, we need to write down the integrand in Eq.\ \eqref{bgs}  explicitly in our case. Since we are interested only in the 4-form part in the integrand, we expand the Chern character form ${\rm Ch}(E)$ and the Todd form ${\rm Td}(TM|Y)$ up to the 4-form order and restrict to the line bundle case (i.e., setting $c_2(E)=c_2(TM|Y)=0$),
\begin{align}\nonumber
&{\rm Ch}(E)=1+c_1(E)+\frac12c_1^2(E)
+...,\\\nonumber
&{\rm Td}(TM|Y)=1+\frac12c_1(TM|Y)+\frac1{12}c_1^2(TM|Y)+...,
\end{align}
where the forms representing first Chern classes read
\begin{align}\nonumber
c_1(E)=\frac{i}{2\pi}\tr  \mathrm{F}^E,\quad c_1(TM)=\frac{i}{2\pi}\tr \mathrm{R}^{TM}
\end{align}
Also we split the curvature 2-form of the bundle $E$ as:
$\mathrm{F}^{E}=\mathrm{F}-s\mathrm{R}_{TM|Y}$
where $\mathrm{F}$ now refers to the part of the curvature
2-form corresponding to the line bundle
$\tilde L^k\to M$, which is the union of all bundles
$L^k_y\to \Sigma_y$. Using the composition property
${\rm Ch}(E\otimes E')={\rm Ch}(E)\cdot {\rm Ch}(E')$ for the product of two bundles $E,E'$, we obtain
\begin{align}
\nonumber
{\rm Ch}(L^k\otimes K^s)
=\,&1+c_1(L^k)-sc_1(TM|Y)-sc_1(L^k)c_1(TM|Y)\\\nonumber&
+\frac12\bigl(c_1^2(L^k)+s^2c_1^2(TM|Y)\bigr)+...
\end{align} 
Then the curvature formula \eqref{bgs} specified to our case reads
\begin{align}\label{BF}
\mathcal R^{\mathcal L}=\frac{i}{4\pi}
\int_{M|Y}&\left[\mathrm{F}\wedge \mathrm{F}
+(1-2s)\, \mathrm{F}\wedge \mathrm{R}_{TM|Y}
\right.\\\nonumber&\left.+\left(\frac{(1-2s)^2}4-\frac1{12}\right)\,
\mathrm{R}_{TM|Y}\wedge \mathrm{R}_{TM|Y}\right].
\end{align} 
Now we introduce notations for the one forms $F=dA_{(M)}$ and $R_{TM|Y}=d\omega_{(M)}$ on $M$. Locally we can write the integrand as the derivative of the Chern-Simons term  
\begin{equation}\nonumber
\mathcal R^{\mathcal L}=\frac{i}{4\pi}
\int_{M|Y}d_MCS(A_{(M)},\omega_{(M)}),
\end{equation}
and use the formula for the commutation of the exterior derivative with the integral along the fiber \cite[Eq.\ (1.17)]{GH}, $d_Y\int_{M|Y}\alpha=d_M\alpha$, in order to show that
\begin{align}\label{gcs}
&\int_{\mathcal C}\mathcal A^{\mathcal L}=\frac1{4\pi}\int_{\sigma^{-1}(\mathcal{C})}CS(A_{(M)},\omega_{(M)})
\\\nonumber=\frac1{4\pi}\int_{\sigma^{-1}(\mathcal{C})} &A\wedge dA
+\frac{1-2s}2(A\wedge d\omega+dA\wedge\omega)+
\left(\frac{(1-2s)^2}4-\frac1{12}\right)\omega\wedge d\omega\,.
\end{align}
The notation $\sigma^{-1}(\mathcal{C})$ means that the integration goes over the $2+1d$ space with one dimension along the contour and two dimensions along the fiber $\Sigma_y$ at the point $y\in Y$.  Here $A$ and $\omega$ are $2+1$d connection 1-forms with components along the fiber retained and third component $A_0$ is the projection on the contour $A_0dt=A_ydy+A_{\bar y}d\bar y$ where $t$ is a parameter along the contour. 

Thus Eq.\ \eqref{gcs} is a formal derivation of the Chern-Simons action, which here has the meaning of the adiabatic phase acquired upon the transport of integer QH state along a contour $\mathcal C\in$ in the parameter space. We note that the full formula for adiabatic curvature Eq.\ \eqref{omega1} has also the exact form contribution due to determinant, which will also lead to a contribution to the phase, which is not reflected in Eq.\ \eqref{gcs} (by analogy with the discussion around Eq.\ \eqref{magndet}, we can argue that this part is exponentially small for large magnetic fields).

\section{Appendix}

Out notations for theta functions follow Mumford \cite{M}. Theta function with characteristics
\begin{equation}\label{theta}
\vartheta\left[\begin{array}{c}a \\b\end{array}\right](z,\tau)=\sum_{n\in\mathbb Z}\exp\left(\pi i(n+a)^2\tau+2\pi i(n+a)(z+b)\right),
\end{equation}
where $a,b\in\mathbb R$.
Their transformation property under the lattice shifts
\begin{equation}
\label{transtheta}
\vartheta\left[\begin{array}{c}a \\b\end{array}\right](z+t_1+t_2\tau,\tau)=e^{-i\pi t_2^2\tau-2\pi it_2z+2\pi i(at_1-bt_2)}\vartheta\left[\begin{array}{c}a \\b\end{array}\right](z,\tau),\quad t_1,t_2\in\mathbb Z.
\end{equation}
We use the standard notation
\begin{equation} \nonumber
\vartheta_1(z,\tau)=\vartheta\left[\begin{array}{c}\scriptstyle\frac12 \\\scriptstyle\frac12\end{array}\right](z,\tau).
\end{equation}
Another useful lattice shift formula
\begin{align}\nonumber \label{shift1}
&\prod_{j<l}^N\bigl(\vartheta_1(z_j-z_l,\tau)\bigr)^\beta\big|_{z_m\to z_m+t_1+t_2\tau}\\
&=e^{-i\pi\tau t_2^2\beta(N-1)+\pi i(t_1+t_2)\beta(N+1)-2\pi i t_2\beta Nz_m+2\pi i t_2\beta z_{\rm cm}}\prod_{j<l}^N\bigl(\vartheta_1(z_j-z_l,\tau)\bigr)^\beta
\end{align}

Poisson summation formula
\begin{equation}
\label{poisson}
\frac1{\sqrt A}\sum_{m'\in\mathbb Z}e^{-\frac{\pi}{A}\bigl(m'+\frac{B}{2\pi i}\bigr)^2}=\sum_{n\in\mathbb Z}e^{-\pi An^2+Bn}.
\end{equation}

Modular transformation formulas:
\begin{align}
\label{Ttr}
&\vartheta\left[\begin{array}{c}a \\b\end{array}\right](mz,m(\tau+1))=e^{-\pi ima(a+1)}\vartheta\left[\begin{array}{c}a \\b+m(a+{\scriptstyle\frac12})\end{array}\right](mz,m\tau),\\
\label{Str}
&\vartheta\left[\begin{array}{c}a \\b\end{array}\right]\left(m\frac z\tau,-m\frac1\tau\right)=\sqrt{\frac{-i\tau}m}\,e^{\pi im\frac{z^2}\tau+2\pi iab}\sum_{c=1}^{m}\vartheta\left[\begin{array}{c}\frac{b+c-1}m \\\scriptstyle-ma\end{array}\right](mz,m\tau),\\
&\eta(\tau+1) =e^{\frac{\pi i}{12}}\eta(\tau),\\
&\eta(-1/\tau)=\sqrt{-i\tau}\,\eta(\tau),\\
&\prod_{j<l}^N\bigl(\vartheta_1(z_j-z_l,\tau+1)\bigr)^\beta=e^{\frac{\pi i}8 \beta N(N-1)}\prod_{j<l}^N\bigl(\vartheta_1(z_j-z_l,\tau)\bigr)^\beta,\\\label{Ttr1}
&\prod_{j<l}^N\left(\vartheta_1\left(\frac{z_j-z_l}\tau,-\frac1\tau\right)\right)^\beta\\\nonumber&=\bigl(\sqrt{-i\tau}\bigr)^{\beta \frac{N(N-1)}2}e^{-\frac{\pi i}4 \beta N(N-1)+\frac{\pi i \NPhi}\tau\sum_lz_l^2-\frac{\pi i\beta}\tau z_{\rm cm}^2}\prod_{j<l}^N\left(\vartheta_1(z_j-z_l,\tau)\right)^\beta.
\end{align}

Modular group action on Laughlin states (\ref{sumwave}-\ref{LZ}), 
\begin{align}\nonumber
&T\circ F_r^{\varepsilon,\delta}(\{z_l\}|\varphi,\tau)=U^T_{rr'} F_{r'}^{\varepsilon,\delta+\varepsilon-\lambda}(\{z_l\}|\varphi,\tau),\\\nonumber
&{\rm where}\quad U^T_{rr'}=\delta_{rr'}e^{\frac{\pi i}{12}(N\NPhi-1)+\frac{\pi i}{\beta}(r+\varepsilon-\beta\lambda+\frac\beta2)(r-\varepsilon+(2-\beta)\lambda+\frac\beta2)},\\\nonumber
&T\circ \prod_{l=1}^Nh_0^{\NPhi}(z_l,\bz_l)=\prod_{l=1}^Nh_0^{\NPhi}(z_l,\bz_l),
\\\nonumber
&T\circ Z_\beta(\varphi,\bar\varphi,\tau,\bar\tau)=Z_\beta(\varphi,\bar\varphi,\tau,\bar\tau),\\\nonumber
&S\circ F^{\varepsilon,\delta}_r(\{z_l\}|\varphi,\tau)=(\sqrt{-i\tau})^{\NPhi}\cdot e^{\frac{\pi i \NPhi}\tau\sum_lz_l^2+\frac{2\pi i}\tau z_{\rm zm}\varphi+\frac{\pi i}\beta\frac{\varphi^2}{\tau}} \sum_{r'=1}^\beta U^S_{rr'} F^{\delta,\varepsilon}_{r'}(\{z_l\}|\varphi,\tau),\\\nonumber
&{\rm where}\quad U^S_{rr'}=\frac1{\sqrt \beta}e^{-\frac{\pi i}4 \NPhi(N-1)-\frac{2\pi i}{\beta}\bigl(\varepsilon+\beta(\frac12-\lambda)\bigr)\bigl(\delta+\beta(\frac12-\lambda)\bigr)-\frac{2\pi i}{\beta}r'(r+2\varepsilon)},\\\nonumber
&S\circ \prod_{l=1}^Nh_0^{\NPhi}(z_l,\bz_l)=e^{-\frac{\pi i\NPhi}{\tau}\sum_lz_l^2+\frac{\pi i\NPhi}{\bar\tau}\sum_l\bz_l^2-\frac{2\pi i}{\tau}z_{\rm cm}\varphi+\frac{2\pi i}{\bar\tau}\bz_{\rm cm}\bar\varphi}\prod_{l=1}^Nh_0^{\NPhi}(z_l,\bz_l),\\\nonumber
&S\circ Z_\beta(\varphi,\bar\varphi,\tau,\bar\tau)=(\sqrt{\tau\bar\tau})^{\NPhi}\cdot e^{\frac{\pi i}{\beta}\frac{\varphi^2}{\tau}-\frac{\pi i}{\beta}\frac{\bar\varphi^2}{\bar\tau}}\cdot Z_\beta(\varphi,\bar\varphi,\tau,\bar\tau).
\end{align}


\label{lastpage}

\begin{thebibliography}{0}


\bibitem{AG1} A.~G.~Abanov and A.~Gromov, {\it Electromagnetic and gravitational responses of two-dimensional non-interacting electrons in background magnetic field,} Phys.\ Rev.\ B {\bf 90} (2014) 014435, \href{http://arxiv.org/abs/1401.3703}{\tt arXiv:1401.3703 [cond-mat.str-el]}.

\bibitem{AMV} L.~Alvarez-Gaume, G.~Moore and C. Vafa,
{\it Theta functions, modular invariance, and strings,} Commun.\ Math.\ Phys.\ {\bf 106} (1986) 1--40.

\bibitem{ABMNV} L.~Alvarez-Gaume, J.-B.~Bost, G.~Moore, P.~Nelson and C. Vafa,
{\it Bosonization on higher genus Riemann surfaces,} Commun.\ Math.\ Phys.\ {\bf 112} (1987) 503--552.

\bibitem{AKPS} J.~E.~Avron, M.~Klein, A.~Pnueli and L.~Sadun, {\it Adiabatic Charge Transport and
Persistent Currents of Leaky Tori,} Phys.\ Rev.\ Lett.\ {\bf 69} (1992) 128.

\bibitem{ASS} J.~E.~Avron, R.~Seiler and B.~Simon, {\it Homotopy and quantization in condensed matter physics,} Phys.\ Rev.\ Lett.\ {\bf 51} (1983) 51.

\bibitem{AS} J.~E.~Avron and R.~Seiler, {\it Quantization of the Hall conductance for general, multiparticle Schr\"odinger hamiltonians,} Phys.\ Rev.\ Lett.\ {\bf 54} (1985) 259.

\bibitem{ASZ} J.~E.~Avron, R.~Seiler and P.~G.~Zograf, {\it Adiabatic quantum transport: quantization and fluctuations,} Phys.\ Rev.\ Lett.\ {\bf 73} no.\ 24 (1994) 3255--3257.

\bibitem{ASZ1} J.~E.~Avron, R.~Seiler and P.~G.~Zograf, {\it Viscosity of quantum Hall fluids,} Phys.\ Rev.\ Lett.\ {\bf 75} no.\ 4 (1995) 697--700, \href{http://arxiv.org/abs/cond-mat/9502011}{\tt arXiv:cond-mat/9502011}.

\bibitem{Av} J.~Avron, {\it Adiabatic quantum transport}, in Les Houches LXI, 1994,
"Mesoscopic Quantum Physics", E. Akkermans and G. Montambaux, J.L. Pichard and J. Zinn-Justin Eds., North-Holland (1995).

\bibitem{BJQ} M.~Barkeshli, C.-M.~Jian, and X.-L.~Qi, {\it Twist defects and projective non-Abelian braiding statistics,} Phys.\ Rev.\ B {\bf 87} (2013) 045130.

\bibitem{BK1} A.~Belavin and V.~Knizhnik, {\it Algebraic geometry and the geometry of quantum strings,} Phys.\ Lett.\ B {\bf 168} (1986) 201--206; {\it Complex geometry and the theory of quantum strings,} Sov.\ Phys.\ JETP {\bf 64} (1986) 215--228.

\bibitem{BES} J.~Bellissard, A.~van~Elst and H.~Schulz-Baldes, {\it The noncommutative geometry of the quantum Hall effect,} J.\ Math.\ Phys.\ {\bf 35} (1994) 5373.

\bibitem{B2} R.~Berman, {\it Determinantal point processes and fermions on complex manifolds: large deviations and bosonization,} Commun.\ Math.\ Phys.\ {\bf 327} (2014) 1--47, \href{http://arxiv.org/abs/0812.4224}{\tt arXiv:0812.4224 [math.CV]}.

\bibitem{Berry} M.~V.~Berry, {\it Quantal phase factors accompanying adiabatic changes,} Proc.\ R.\ Soc.\ Lond.\ A {\bf 392} (1984) 45--57.

\bibitem{Ber01} A.~Berthomieu, {\it Analytic torsion of all vector
bundles over an elliptic curve}.
 J. Math. Phys. {\bf 42} (2001), no. 9, 4466--4487.

\bibitem{BF} J.-M.~Bismut and D.~Freed, {\it The analysis of elliptic families. I-II,} Commun.\ Math.\ Phys.\ {\bf 106} no.\ 1 (1986) 159--176; {\bf 107} no.\ 1 (1987) 103--163.

\bibitem{BGS3} J.-M.~Bismut, H.~Gillet and C.~Soul\'e, {\it Analytic torsion and holomorphic determinant bundles. III,} Commun.\ Math.\ Phys.\ {\bf 115} (1988) 301--351.

\bibitem{BGN} P.~Bonderson, V.~Gurarie and C.~Nayak, {\it Plasma Analogy and Non-Abelian Statistics for Ising-type Quantum Hall States,} Phys.\ Rev.\ B {\bf 83} (2011) 075303, \href{http://arxiv.org/abs/1008.5194}{\tt arXiv:1008.5194 [cond-mat.str-el]}.

\bibitem{BR} B.~Bradlyn and N.~Read, {\it Low-energy effective theory in the bulk for transport in a topological phase,} Phys.\ Rev.\ B{\bf 91} (2015) 125303, \href{http://arxiv.org/abs/1407.2911}{\tt arXiv:1407.2911 [cond-mat.mes-hall]}.

\bibitem{BR1} B.~Bradlyn and N.~Read, {\it Topological central charge from Berry curvature: Gravitational anomalies in trial wave functions for topological phases,} Phys.\ Rev.\ B {\bf 91} (2015) 165306, \href{http://arxiv.org/abs/1502.04126}{\tt arXiv:1502.04126 [cond-mat.mes-hall]}.

\bibitem{CLW} T.~Can, M.~Laskin and P.~Wiegmann, {\it Fractional quantum Hall effect in a curved space: gravitational anomaly and electromagnetic response}, Phys.\ Rev.\ Lett.\ {\bf 113} (2014) 046803, \href{http://arxiv.org/abs/1402.1531}{\tt arXiv:1402.1531 [cond-mat.str-el]}.

\bibitem{CLW1} T.~Can, M.~Laskin and P.~Wiegmann, {\it Geometry of quantum Hall states: gravitational anomaly and transport coefficients}, Ann.\ Phys.\ {\bf 362} (2015) 752--794, \href{http://arxiv.org/abs/1411.3105}{\tt arXiv:1411.3105 [cond-mat.str-el]}.

\bibitem{CZ} A.~Cappelli and G.~Zemba, {\it Modular invariant partition functions in the quantum Hall effect,} Nucl.\ Phys.\  B{\bf 490}  no.\ 3 (1997) 595--632, \href{http://arxiv.org/abs/hep-th/9605127}{\tt arXiv:hep-th/9605127}.

\bibitem{CR} A.~Cappelli, E.~Randellini, {\it Multipole expansion in the quantum hall effect,} J.\ High Energ.\ Phys.\ {\bf 03} (2016) 105, \href{http://arxiv.org/abs/1512.02147}{\tt arXiv:1512.02147 [cond-mat.str-el]}.

\bibitem{C} D.~Catlin, {\it The Bergman kernel and a theorem of Tian,}
in "Analysis and geometry in several complex variables" (Katata, 1997), 
Trends Math., Birkh\"auser, Boston, MA, 1999, pp. 1--23.

\bibitem{CF} A.~H.~Chamseddine and J.~Fr\"ohlich, {\it Two dimensional Lorentz-Weyl anomaly and gravitational Chern-Simons theory,} Commun.\ Math.\ Phys.\ {\bf 147} (1992) 549--562.

\bibitem{CMMN} G.~Cristofano, G.~Maiella, R.~Musto and F.~Nicodemi, {\it Topological order in Quantum Hall Effect and two-dimensional conformal field theory,} Nucl.\ Phys.\ B (Proc. Suppl.) {\bf 33} C (1993) 119--133.

\bibitem{DP} E.~D'Hoker and D.~H.~Phong, {\it The geometry of string perturbation theory,} Rev.\ Mod.\ Phys.\  {\bf 60} (1988) 917.

\bibitem{DP1} E.~D'Hoker and D.~H.~Phong,
{\it On determinants of laplacians on riemann surfaces,}
Commun.\ Math.\ Phys.\ {\bf 104} (1986) 537--545.

\bibitem{DMS} P.~Di Francesco, P.~Mathieu and D.~Senechal, {\it Conformal Field Theory,} Springer (1996).

\bibitem{DSZ} P.~Di Francesco, H.~Saleur, J.~B.~Zuber, {\it Relations between the Coulomb gas picture and conformal invariance of two-dimensional critical models,} J.\ Stat.\ Phys.\ {\bf 49} (1987) 57.

\bibitem{DN1} A.~Deshpande and A.~Nielsen, {\it Lattice Laughlin states on the torus from conformal field theory,} J.\ Stat.\ Mech.\ {\bf 2016} (2016) 013102, \href{http://arxiv.org/abs/1507.04335}{\tt arXiv:1507.04335 [cond-mat.str-el]}.

\bibitem{Don2} S.~K.~Donaldson, {\it Scalar curvature and projective embeddings. II,} Q.\ J.\ Math.\ {\bf 56} no. 3 (2005) 345-356, \href{http://arxiv.org/abs/math/0407534}{\tt arXiv:math/0407534 [math.DG]}.

\bibitem{DK} M.~R.~Douglas and S.~Klevtsov,
{\it Bergman kernel from path integral,} Commun.\ Math.\ Phys.\ {\bf 293} no.\ 1 (2010)
205-230, \href{http://arxiv.org/abs/0808.2451}{\tt arXiv:0808.2451 [hep-th]}.

\bibitem{DN} B.~A.~Dubrovin and S.~P.~Novikov, {\it Ground state of a two dimensional electron
in a periodic magnetic field,} Sov.\ Phys.\ JETP {\bf 52} (1980) 511--516; {\it Ground
state in a periodic field, magnetic Bloch functions and vector bundles,} Sov.\ Math.\
Dokl.\ {\bf 22} (1980) 240--244.

\bibitem{Fay} J.~Fay, {\it Kernel functions, analytic torsion and moduli spaces,} Memoirs of AMS, {\bf 96} no.\ 464, Providence RI (1992).

\bibitem{FK} F.~Ferrari and S.~Klevtsov, {\it FQHE on curved backgrounds, free fields and large N,} J.\ High Energ.\ Phys.\ {\bf 12} (2014) 086, \href{http://arxiv.org/abs/1410.6802}{\tt arXiv:1410.6802 [hep-th]}.

\bibitem{FKZ3} F.~Ferrari, S.~Klevtsov and S.~Zelditch, {\it Gravitational actions in two dimensions and the Mabuchi functional,} Nucl.\ Phys.\ B {\bf 859} no.\ 3 (2012) 341--369, \href{http://arxiv.org/abs/1112.1352}{\tt arXiv:1112.1352 [hep-th]}.

\bibitem{FHS} M.~Fremling, T.~H.~Hansson and J.~Suorsa, {\it Hall viscosity of hierarchical quantum Hall states,} Phys.\ Rev.\ B {\bf 89} (2014) 125303, \href{http://arxiv.org/abs/1312.6038}{\tt arXiv:1312.6038 [cond-mat.str-el]}.

\bibitem{FKer} J.~Fr\"ohlich and T.~Kerler, {\it Universality in quantum Hall systems,} Nucl.\ Phys.\ B {\bf 354} (1991) 369.

\bibitem{FS} J.~Fr\"ohlich and U.~M.~Studer,
{\it $U(1)\times SU(2)$-gauge invariance
of non-relativistic quantum mechanics, and generalized Hall effects,}
Comm.\ Math.\ Phys.\ {\bf 148} (1992) 553--600.

\bibitem{Gir} S.~Girvin, {\it The quantum Hall effect: novel excitations and broken symmetries,} in"Topological Aspects of Low Dimensional Systems", Addison Wesley (2000), \href{http://arxiv.org/abs/cond-mat/9907002}{\tt arXiv:cond-mat/9907002 [cond-mat.mes-hall]}.

\bibitem{GH}
P.~Griffiths and J.~Harris, \emph{{Principles of Algebraic Geometry}},
John Wiley and Sons, New York, 1978.

\bibitem{AG2} A.~Gromov and A.~G.~Abanov, {\it Density-curvature response and gravitational anomaly,} Phys.\ Rev.\ Lett.\ {\bf 113} (2014) 266802, \href{http://arxiv.org/abs/1403.5809}{\tt arXiv:1403.5809 [cond-mat.str-el]}.

\bibitem{AG3} A.~Gromov and A.~G.~Abanov, {\it Thermal Hall effect and geometry with torsion,} Phys.\ Rev.\ Lett.\ {\bf 114} (2015) 016802, \href{http://arxiv.org/abs/1407.2908}{\tt arXiv:1407.2908 [cond-mat.str-el]}.

\bibitem{AG4} A.~Gromov, G.~Y.~Cho, Y.~You, A.~G.~Abanov and E.~Fradkin, {\it Framing anomaly in the effective theory of fractional quantum Hall effect,} Phys.\ Rev.\ Lett.\ {\bf 114} (2015) 016805, \href{http://arxiv.org/abs/1410.6812}{\tt arXiv:1410.6812 [cond-mat.str-el]}.

\bibitem{Gr} A.~Gromov, {\it Geometric defects in quantum Hall states,} \href{http://arxiv.org/abs/1604.03988}{\tt arXiv:1604.03988 [cond-mat.str-el]}.

\bibitem{GN} V.~Gurarie and C.~Nayak, {\it A plasma analogy and Berry matrices for
non-abelian quantum Hall states,} Nucl.\ Phys.\ B {\bf 506} (1997) 685--694, \href{http://arxiv.org/abs/cond-mat/9706227}{\tt arXiv:cond-mat/9706227}.

\bibitem{H} F.~D.~M.~Haldane, {\it Fractional quantization of the Hall effect: a hierarchy of incompressible quantum fluid states}, Phys.\ Rev.\ Lett.\ {\bf 51} no.\ 7 (1983) 605--608.

\bibitem{HR} F.~D.~M.~Haldane and E.~H.~Rezayi, {\it Periodic Laughlin-Jastrow wave functions for the fractional quantized Hall effect}, Phys.\ Rev.\ B{\bf 31} no.\ 4 (1985) 2529--2531.

\bibitem{H1} F.~D.~M.~Haldane, {\it Geometrical description of the fractional quantum Hall effect,} 
Phys.\ Rev.\ Lett.\ {\bf 107} (2011) 116801, \href{http://arxiv.org/abs/1106.3375}{\tt arXiv:1106.3375 [cond-mat.mes-hall]}.

\bibitem{Hal} B.~I.~Halperin, {\it Statistics of quasiparticles and the hierarchy of fractional quantized Hall states,} Phys.\ Rev.\ Lett.\ {\bf 52} (1984) 1583--1586.

\bibitem{HS} M.~Hermanns, J.~Suorsa, E.~J.~Bergholtz, T.~H.~Hansson and A.~Karlhede, {\it Quantum Hall wave functions on the torus,} Phys.\ Rev.\ B {\bf 77} (2008) 125321, \href{http://arxiv.org/abs/0711.4684}{\tt arXiv:0711.4684 [cond-mat.mes-hall]}.

\bibitem{Ho} C.~Hoyos, {\it Hall viscosity, topological states and effective theories,} Int.\ J.\ Mod.\ Phys.\ B {\bf 28} (2014) 1430007, \href{http://arxiv.org/abs/1403.4739}{\tt arXiv:1403.4739 [cond-mat.mes-hall]}.

\bibitem{Zh} J.~P.~Hu and S.~C.~Zhang, {\it Collective excitations at the boundary of a four-dimensional quantum Hall droplet,} Phys.\ Rev.\ B {\bf 66} (2002) 125301, \href{http://arxiv.org/abs/cond-mat/0112432}{\tt cond-mat/0112432}.

\bibitem{IL} R.~Iengo and D.~Li, {\it Quantum mechanics and quantum Hall effect on Riemann surfaces,} Nucl.\ Phys.\ B {\bf 413} (1994) 735, \href{http://arxiv.org/abs/hep-th/9307011}{\tt arXiv:hep-th/9307011}.

\bibitem{Jain} J.~K.~Jain, {\it Composite-fermion approach for the fractional quantum Hall effect,} Phys.\ Rev.\ Lett.\ {\bf 63} (1989) 199.

\bibitem{H2} S.~Johri, Z.~Papic, P.~Schmitteckert, R.~N.~Bhatt, F.~D.~M.~Haldane, {\it Probing the geometry of the Laughlin state,} \href{http://arxiv.org/abs/1512.08698}{\tt arXiv:1512.08698 [cond-mat.str-el]}.

\bibitem{KL} V.~Kalmeyer and R.~B.~Laughlin, {\it Equivalence of the resonating-valence-bond and fractional quantum Hall states,} Phys.\ Rev.\ Lett.\ {\bf 59} (1987) 2095.

\bibitem{KN} D.~Karabali and V.~P.~Nair, {\it Quantum Hall effect in higher dimensions,} Nucl.\ Phys.\  B {\bf 641} (2002) 533, \href{http://arxiv.org/abs/hep-th/0203264}{\tt arXiv:hep-th/0203264}.

\bibitem{KN1} D.~Karabali and V.~P.~Nair, {\it The geometry of quantum Hall effect: an effective action for all dimensions,} \href{http://arxiv.org/abs/1604.00722}{\tt arXiv:1604.00722 [hep-th]}.

\bibitem{KVW} E.~Keski-Vakkuri and X.~G.~Wen, {\it The ground state structure and modular transformations of fractional quantum Hall states on a torus,} Int.\ J.\ Mod.\ Phys.\ A {\bf 7} (1993) 4227--4259, \href{http://arxiv.org/abs/hep-th/9303155}{\tt arXiv:hep-th/9303155}.

\bibitem{Kir} W.~Kirwin, {\it Quantizing the geodesic flow via adapted complex structures,} \href{http://arxiv.org/abs/1408.1527}{\tt arXiv:1408.1527 [math.SG]}.

\bibitem{Kit} A.~Kitaev, {\it Anyons in an exactly solved model and beyond,} Ann.\ Phys.\ {\bf 321} (2006) 2--111, \href{http://arxiv.org/abs/cond-mat/0506438}{\tt arXiv:cond-mat/0506438 [cond-mat.mes-hall]}.

\bibitem{K} S.~Klevtsov, {\it Random normal matrices, Bergman kernel and projective embeddings,} J.\ High Energ.\ Phys.\  {\bf 1401} (2014) 133, \href{http://arxiv.org/abs/1309.7333}{\tt arXiv:1309.7333 [hep-th]}.

\bibitem{KMMW}  S.~Klevtsov, X.~Ma, G.~Marinescu and P.~Wiegmann, {\it Quantum Hall effect and Quillen metric,} \href{http://arxiv.org/abs/1510.06720}{\tt arXiv:1510.06720 [hep-th]}.

\bibitem{KW}  S.~Klevtsov and P.~Wiegmann, {\it Geometric adiabatic transport in Quantum Hall states,} Phys.\ Rev.\ Lett.\ {\bf 115} (2015) 086801, \href{http://arxiv.org/abs/1504.07198}{\tt arXiv:1504.07198 [cond-mat.str-el]}.

\bibitem{K2016} S.~Klevtsov, {\it in preparation}. 

\bibitem{Kv} T.~Kvorning, {\it Quantum Hall hierarchy in a spherical geometry,} Phys.\ Rev.\ B {\bf 87} (2013) 195131, \href{http://arxiv.org/abs/1302.3808}{\tt arXiv:1302.3808 [cond-mat.str-el]}.

\bibitem{Lang} S.~Lang, {\it Introduction to Modular Forms}, Grundlehren der mathematischen Wissenschaften, {\bf 222} (1976) Springer-Verlag.

\bibitem{LCW} M.~Laskin, T.~Can and P.~Wiegmann, {\it Collective field theory for quantum Hall states}, Phys.\ Rev.\ B {\bf 92} (2015) 235141, \href{http://arxiv.org/abs/1412.8716}{\tt arXiv:1412.8716 [cond-mat.str-el]}.

\bibitem{CCLW} M.~Laskin, Y.~H.~Chiu, T.~Can and P.~Wiegmann, {\it Emergent conformal symmetry of quantum Hall states on singular surfaces}, \href{http://arxiv.org/abs/1602.04802}{\tt arXiv:1602.04802 [cond-mat.str-el]}.

\bibitem{L1981} R.~B.~Laughlin, {\it Quantized Hall conductivity in two dimensions}, Phys.\ Rev.\ B {\bf 23} no.\ 10 (1981) 5632.

\bibitem{L} R.~B.~Laughlin, {\it Anomalous quantum Hall effect: an incompressible quantum fluid with fractionally charged excitations}, Phys.\ Rev.\ Lett.\ {\bf 50} no.\ 18 (1983) 1395.

\bibitem{L1989} R.~Laughlin, {\it Spin hamiltonian for which quantum Hall wave function is exact,} Ann.\ Phys.\ {\bf 191} no.\ 1 (1989) 163--202.

\bibitem{L1} P.~L\'evay, {\it Berry phases for Landau Hamiltonians on deformed tori,} J.\ Math.\ Phys.\ {\bf 36}  (1995) 2792. 

\bibitem{L2} P.~L\'evay, {\it Berry's phase, chaos, and the deformations of Riemann surfaces,} Phys.\ Rev.\ E {\bf 56} no.\ 5 (1997) 6173--6176.

\bibitem{MM} X.~Ma and G.~Marinescu, {\it Holomorphic Morse inequalities and Bergman kernels,} Progress in Mathematics, Birkh\"auser, Vol. {\bf 254} (2006).

\bibitem{MR} G.~Moore and N.~Read, {\it Nonabelions in the fractional quantum Hall effect}, Nucl.\ Phys.\ B {\bf 360} (1991) 362--396.

\bibitem{M} D.~Mumford, {\it Tata lectures on theta I,} Birkh\"auser, Boston (1983).

\bibitem{T2} Q.~Niu, D.~J.~Thouless and Y.-S.~ Wu, {\it Quantized Hall conductance as a topological invariant,} Phys.\ Rev.\ B{\bf 31}  (1985) 3372.

\bibitem{Nov} K.~S.~Novoselov, {\it et al,} {\it Room-Temperature Quantum Hall Effect in Graphene,} Science {\bf 315} (5817), (2007) 1379, arXiv: \href{http://arxiv.org/abs/cond-mat/0702408}{\tt arXiv:cond-mat/0702408}.

\bibitem{OPS}
B.~Osgood, R.~Phillips and P.~Sarnak,
{\it Extremals of determinants of laplacians,} J. Funct. Anal. {\bf 80}, 148--211 (1988).

\bibitem{PS} D.~H.~Phong and J.~Sturm, {\it Lectures on stability and constant scalar curvature,} Current developments in mathematics,  (2007) 101-176, Int. Press, Somerville MA (2009), \href{http://arxiv.org/abs/0801.4179}{\tt arXiv:0801.4179 [math.DG]}.

\bibitem{P} A.~M.~Polyakov, {\it Quantum geometry of bosonic strings,} Phys.\ Lett.\ B {\bf 103} (1981) 207--210.

\bibitem{P1} A.~M.~Polyakov, {\it Quantum gravity in two dimensions,} Mod.\ Phys.\ Lett.\ A {\bf 2} no.\ 11 (1987) 893--898.

\bibitem{Q} D.~Quillen, {\it Determinants of Cauchy-Riemann operators over a Riemann surface,} Funct.\ Anal.\ Appl.\ {\bf 19} no.\ 1 (1985) 37--41.

\bibitem{RG} N.~Read and D.~Green, {\it Paired states of fermions in two dimensions with breaking of parity and time-reversal symmetries, and the fractional quantum Hall effect,} Phys.\ Rev.\ B {\bf 61} (2000) 10267, \href{http://arxiv.org/abs/cond-mat/9906453}{\tt arXiv:cond-mat/9906453 [cond-mat.mes-hall]}.

\bibitem{R} N.~Read, {\it Non-Abelian adiabatic statistics and Hall viscosity in quantum Hall states and $p_x+ip_y$ paired superfluids,} Phys.\ Rev.\ B {\bf 79} no.\ 4 (2009) 045308, 
\href{http://arxiv.org/abs/0805.2507}{\tt arXiv:0805.2507 [cond-mat.mes-hall]}.

\bibitem{RR1999} N. Read and E. Rezayi, {\it Beyond paired quantum Hall states: parafermions and incompressible states in the first excited Landau level,} Phys.\ Rev.\ B {\bf 59} (1999) 8084, \href{http://arxiv.org/abs/cond-mat/9809384}{\tt arXiv:cond-mat/9809384 [cond-mat.mes-hall]}.

\bibitem{RR} N.~Read and E.~H.~Rezayi, {\it Hall viscosity, orbital spin, and geometry: Paired superfluids and quantum Hall systems,} Phys.\ Rev.\ B {\bf 84} no.\ 4 (2009) 085316,  
\href{http://arxiv.org/abs/1008.0210}{\tt arXiv:1008.0210 [cond-mat.mes-hall]}.

\bibitem{JSimon} N.~Schine, A.~Ryou, A.~Gromov, A.~Sommer and J.~Simon, {\it Synthetic Landau levels for photons,} Nature {\bf 534} (2016) 671--675, \href{http://arxiv.org/abs/1511.07381}{\tt arXiv:1511.07381 [cond-mat.mes-hall]}.

\bibitem{Simon} B.~Simon, {\it Holonomy, the Quantum Adiabatic Theorem, and Berry's Phase,} Phys.\ Rev.\ Lett.\ {\bf 51} (1983) 2167. 

\bibitem{Son} D.~T.~Son,
{\it Newton-Cartan Geometry and the Quantum Hall Effect,}
\href{http://arxiv.org/abs/1306.0638}
{\tt arXiv:1306.0638 [cond-mat.mes-hall]}.

\bibitem{TW} R.~Tao and Y.-S.~Wu, {\it Gauge invariance and fractional quantum Hall effect,} Phys. Rev. B {\bf 30} (1984) 1097.

\bibitem{TP06} C.~Tejero~Prieto,
{\it Fourier-Mukai transform and adiabatic curvature of spectral
 bundles for Landau Hamiltonians on Riemann surfaces},
 Comm.\ Math.\ Phys. \textbf{265} (2006) 373--396.
 
\bibitem{T1} D.~J.~Thouless, M.~Kohmoto, M.~P.~Nightingale and
M.~den Nijs, {\it Quantized Hall conductance in a two-dimensional periodic potential,} Phys.\ Rev.\ Lett.\ {\bf 49} (1982) 405.

\bibitem{TV1} I.~V.~Tokatly and G.~Vignale, {\it Lorentz shear modulus of a two-dimensional electron gas at high magnetic field,} Phys.\ Rev.\ B{\bf 76} 
(2007) 161305, \href{http://arxiv.org/abs/0706.2454}{\tt arXiv:0706.2454 [cond-mat.mes-hall]}.

\bibitem{TV2} I.~Tokatly and G.~Vignale, {\it Lorentz shear modulus of fractional quantum Hall states,} J.\ Phys. C {\bf 21} 
(2009) 275603, \href{http://arxiv.org/abs/0812.4331}{\tt arXiv:0812.4331 [cond-mat.mes-hall]}.

\bibitem{TSG1} D.~C.~Tsui, H.~L.~St\"ormer and A.~C.~Gossard, {\it The fractional quantum Hall effect,} Rev.\ Mod.\ Phys. {\bf 71} No. 2 (1999) S298.

\bibitem{VV}
E.~P.~Verlinde and H.~L.~Verlinde,
{\it Chiral bosonization, determinants and the string partition function,}
Nucl.\ Phys.\ {\bf B288} (1987) 357--396.

\bibitem{Xu}
H.~Xu, {\it A closed formula for the asymptotic expansion of the Bergman kernel, } Commun.\ Math.\ Phys.\ {\bf 314} (2012) 555-585, \href{http://arxiv.org/abs/1103.3060}{\tt arXiv:1103.3060 [math.DG]}.

\bibitem{XY} H.~Xu and S.-T.~Yau, {\it Trees and tensors on Kahler manifolds,} Ann.\ Global Anal.\ Geom.\ {\bf 44} (2013) 151--168.

\bibitem{Wein} E.~Weisstein, \href{http://mathworld.wolfram.com/ModularGroupLambda.html}{http://mathworld.wolfram.com/ModularGroupLambda.html}.

\bibitem{WN} X.~G.~Wen and Q.~Niu, {\it Ground-state degeneracy of the fractional quantum Hall states in the presence of a random potential and on high-genus Riemann surfaces}, Phys.\ Rev.\ B{\bf 41} no.\ 13 (1990) 9377--9396.

\bibitem{WZ} X.~G.~Wen and A.~Zee, {\it Shift and spin vector: New topological quantum numbers for the Hall fluids,} Phys.\ Rev.\ Lett.\ {\bf 69} (1992) 953.

\bibitem{Wen} X.~G.~Wen, {\it Modular transformation and bosonic/fermionic topological orders
in Abelian fractional quantum Hall states,} \href{http://arxiv.org/abs/1212.5121}{\tt arXiv:1212.5121 [cond-mat.str-el]}.

\bibitem{WZ1}
P.~Wiegmann and A.~Zabrodin, {\it Large $N$ expansion for normal and complex matrix ensembles,} Proc. of Les Houches Spring School (2003), \href{http://arxiv.org/abs/hep-th/0401165}{\tt hep-th/0401165}.

\bibitem{Zab}  A.~Zabrodin,
  {\it Matrix models and growth processes: From viscous flows to the quantum Hall effect,} Applications of random matrices in physics, Springer (2006), \href{http://arxiv.org/abs/hep-th/0412219}{\tt arXiv:hep-th/0412219}.
  
\bibitem{ZW} A.~Zabrodin and P.~Wiegmann, {\it Large $N$ expansion for the 2D Dyson gas,} J.\ Phys.\ A {\bf 39} (2006) 8933--8963, \href{http://arxiv.org/abs/hep-th/0601009}{\tt arXiv:hep-th/0601009}.

\bibitem{ZMP} M.~Zaletel, R.~Mong and F.~Pollmann, {\it Topological characterization of fractional quantum Hall ground states from microscopic hamiltonians,} Phys.\ Rev.\ Lett.\ {\bf 110} (2013) 236801, \href{http://arxiv.org/abs/1211.3733}{\tt arXiv:1211.3733 [cond-mat.str-el]}.

\bibitem{Z} S.~Zelditch, {\it \szego kernels and a theorem of Tian},
IMRN  {\bf 1998} no.\ 6 (1998) 317-331, \href{http://arxiv.org/abs/math-ph/0002009}{\tt arXiv:math-ph/0002009}.

\bibitem{ZT} P.~G.~Zograf and L.~A.~Takhtadzhyan,
{\it A local index theorem for families of $\bp$-operators on
Riemann surfaces,} Russian Math. Surveys {\bf 42} (1987) 169--190.

\end{thebibliography}
\end{document}